\documentclass[11pt,a4paper]{article}
\usepackage{jheppub}
\usepackage{amsmath}
\usepackage{multicol,bbm}
\usepackage{enumerate}
\usepackage{bibentry}

\graphicspath{{./Figs/}} 

\newcommand{\bea}{\begin{equation}\begin{aligned}}
\newcommand{\eea}{\end{aligned}\end{equation}}
\newcommand{\beq}{\begin{eqnarray}}
\newcommand{\eeq}{\end{eqnarray}}
\newcommand{\eq}{\begin{equation}\begin{aligned}}
\newcommand{\eqe}{\end{aligned}\end{equation}}
\newcommand{\eqa}{\begin{eqnarray}}
\newcommand{\eqae}{\end{eqnarray}}

\title{The Positive orthogonal Grassmannian and loop amplitudes of ABJM}

\author[a,c]{Yu-tin Huang,}
\author[b]{Congkao Wen,}
\author[a]{Dan Xie}

\affiliation[a]{School of Natural Sciences, Institute for Advanced
Study, Princeton, NJ 08540, USA} 
\affiliation[b]{Centre for Research in String Theory, Department of Physics, Queen Mary University of London, Mile End Road, London E1 4NS, UK}
\affiliation[c]{Department of Physics and Astronomy, National Taiwan University, Taipei 10617, Taiwan, ROC}

\abstract{In this paper we study the combinatorics associated with the positive orthogonal Grassmannian $OG_k$ and its connection to ABJM scattering amplitudes. We present a canonical embedding of $OG_k$ into the Grassmannian $Gr(k,2k)$, from which we deduce the canonical volume form that is invariant under equivalence moves. Remarkably the canonical forms of all reducible graphs can be converted into irreducible ones with products of $d\log$ forms.  Unlike $\mathcal{N}=4$ super Yang-Mills, here the Jacobian plays a crucial role to ensure the $d\log$ form of the reduced representation. Furthermore in the positive region, we identify the functional map that arises from the triangle equivalence move as a $3$-string scattering S-matrix which satisfies the tetrahedron equations by Zamolodchikov, implying $(2+1)$-dimensional integrability. We study the solution to the BCFW recursion relation for loop amplitudes, and demonstrate the presence of all physical singularities as well as the absence of all spurious ones. The on-shell diagram solution to the loop recursion relation exhibits manifest two-site cyclic symmetry and reveals that, to all loop, four and six-point amplitudes only have logarithmic singularities. }
\preprint{QMUL-PH-14-02}

\begin{document}

\maketitle 

\section{Introduction and conclusion}
An amazing aspect of recent studies in scattering amplitudes is the realization that the answer can often be reinterpreted as the solution to a complete different set of questions, the latter of which are unrelated to the observable being a description of a scattering process. Furthermore, in such a setup the usual physical requirement of locality and unitarity becomes an emergent property, transplanted by other principles. That this is possible can be seen by the continuous march of uncovering new formulation of scattering amplitudes in gauge theories (see e.g. \cite{Elvang:2013cua} for a review). 

In the study of scattering amplitudes in $\mathcal{N}=4$ super Yang-Mills theory (SYM), a new approach was developed by Arkani-Hamed et al~\cite{NimaBigBook}, where the scattering amplitudes are constructed by simply iteratively gluing together fundamental three-point on-shell amplitudes. Thus any amplitude is given by a sum of such ``on-shell" diagrams, which are diagrams with trivalent vertices taking black or white color depending on the helicity configuration (bi-partite network). Note that in this approach, all scattering process are interpreted as on-shell processes without the invocation of any ``off-shell" physics, and an appealing aspect of this approach is that it is amendable to applications for $\mathcal{N}<4$ SYM. Perhaps what is the most remarkable realization in the work of~\cite{NimaBigBook}, is that in this form, the iterative gluing of on-shell diagrams can be translated into the iterative parameterization of positive cells in the Grassmannian $Gr(k,n)_{+}$(see~\cite{Franco} for related work). The relation between  $\mathcal{N}=4$ SYM and integrals over Grassmannian integral was known (relatively) long ago~\cite{ArkaniHamed:2009dn}, and the on-shell diagram gives a ``first-principle" microscopic derivation of this relationship. Moreover, it shows that the amplitudes can be thought of as the collection of cells in the positive Grassmannian for which the boundaries correspond to physical singularities. 

There are two obvious generalizations, both of which were briefly discussed in~\cite{NimaBigBook}: extensions to theories other than four-dimensions, and non-maximal supersymmetry. The three-dimensional $\mathcal{N}=6$ supersymmetric Chern-Simons matter theory~\cite{ABJM1, ABJM2}, commonly known as ABJM theory, fits both of the bills. In particular, as already discussed in~\cite{NimaBigBook}, the scattering amplitudes for ABJM theory can be thought of as iteratively gluing together the fundamental four-point amplitude, the simplest non-trivial amplitude in the theory. The resulting four-valent diagrams, which we will refer to as {\it medial graph}, can be thought of as parameterizing the orthogonal Grassmannian $OG_{k}$, which is also long known to be related to the leading singularities of ABJM theory~\cite{LeeOG, Gang}.  This proposal was studied further by two of the authors in~\cite{HW}, where on-shell diagram representation of tree-level amplitudes was derived. Furthermore it was found that with suitable definition of the bilinear form for the orthogonal Grassmannian, one can similarly define positivity and identify the on-shell diagrams as constructing a parameterization of cells in the positive orthogonal Grassmannian.

In this paper we continue the study on the relation between $OG_{k+}$ and the scattering amplitudes of ABJM theory. We begin by noting that the combinatorics of $OG_{k+}$, as well as its canonical coordinates, can be most easily understood by considering its image on $Gr(k,2k)_{+}$\footnote{For the case of $k=2$, it was already realized in \cite{nimastring}}. To facilitate the construction of its image we present a simple way of constructing the face variables of $Gr(k,2k)_{+}$ in terms of vertex variables in $OG_{k+}$. Through this embedding, all equivalence moves within $OG_{k+}$ can be understood as a consequence of the equivalence moves of its image in $Gr(k,2k)_{+}$. This verifies that the reduction of  $Gr(k,2k)_{+}$ to $OG_{k+}$ is  stratification preserving. Note that only special cells in $Gr(k,2k)_{+}$ contains the images of $OG_{k+}$. Studying the boundary structures of $OG_{k+}$ allows one to construct a poset for the cells. Remarkably we find that, up to $k=4$, all cells in $OG_{k+}$ forms an Eulerian poset, i.e. the number of even dimensional cells, is greater than the odd-dimension cells by 1. This is exactly the combinatorics of a the face lattice of a convex polytope. Thus the cell structure of $OG_{k+}$ is combinatorially a polytope. Recently Kim and Lee~\cite{SangminNew} have presented an efficient method of enumerating cells, and have shown that the top-cell of $OG_{k+}$ does indeed form a Eulerian poset. Whilst combinatorially, $OG_{k+}$ can be considered as simply a descendent of $Gr(k,2k)_{+}$, when making connection to the on-shell diagram of ABJM theory, important differences emerge:

\begin{itemize}
  \item First is the canonical volume form, which is defined as the measure that contains only logarithmic singularity and is preserved under equivalence moves, modulo a sign. This form is important as the amplitudes for $\mathcal{N}=4$ SYM (ABJM) is given by the integration over $Gr(k,n)_{+}$ ($OG_{k+}$), with the canonical volume form as the measure. For $Gr(k,n)_{+}$, using the bipartite network one can identify the volume form as simply the product of $d\log f_i$, where $f_i$ are the face variables for the faces in the graph. Under the equivalence move, the face variables $f_i$ transforms in such a way that leaves the volume form invariant.\footnote{From here on we refer the volume form as invariant if it is invariant modulo a sign.} Here we find that the canonical volume form for $OG_{k+}$ is given by 
\eq
\mathcal{J} \times \prod_{i=1}^{n_v} d\log ({\rm tan}\theta_{i})
\eqe
where $n_v$ is the number of vertices, $\theta_{i}$ is the variable associated with the degree of freedom on each vertex, and $\mathcal{J}$ is a Jacobian factor that is present whenever the medial graph contains closed loops. The positive region is defined as $0\leq\theta_i\leq\pi/2$. Note that the form $d\log ({\rm tan}\theta_{i})$ has manifest logarithmic singularity on the boundaries of the positive region, $\theta_i=0$ and $\theta_i=\pi/2$. We give a simple rule of deriving $\mathcal{J}$ for an arbitrary medial graph. In the positive region, the Jacobian factors do not introduce new singularity, thus preserving the fact that the volume form only has logarithmic singularity. Interpreted in terms of the gluing of on-shell amplitudes, this Jacobian factor follows from the fact that there is a mismatch between bosonic ($\lambda^a$) and fermionic on-shell variables ($\eta^I$), that is unique to $\mathcal{N}=6$ supersymmetry. The presence of $\mathcal{J}$ is crucial in two ways: firstly the canonical volume form with right Jacobian factor is invariant under equivalence moves; secondly for reducible diagrams, it ensures that, by reduction, the dependency on the removed degrees of freedom appears only as an overall $d\log$ factor.
  \item Second, the amplitudes must live on both $OG_{k+}$ and $OG_{k-}$, where the later is defined by the constraint that the ratio of the an ordered minor and its image is $-1$. Remarkably, this requirement is a reflection of the fact that the on-shell massless kinematics for three-dimensions is separated into disconnected chambers~\cite{2LoopABJM2}. 
\end{itemize}

But even with these subtleties, one might still ask the following question: if we view $OG_{k+}$ as a lower-dimensional parameterization of $Gr(k,2k)_{+}$, is there anything special to this as oppose to any other parameterization? Remarkably, as we will show, $OG_{k+}$ is special in that the map that is induced from the triangle equivalence move for $OG_{3+}$, can actually be interpreted as the scattering S-matrix of infinite straight strings in $(2+1)$-dimensions. In particular, the map $R$ which maps the three vertex variables $\theta_i$ to the new variables $\theta'_i$, satisfies the ``tetrahedron equation", the $(2+1)$-dimensional generalization of Yang-Baxter equation. This equation was first introduced by Zamolodchikov~\cite{ZZ1, ZZ2} as the criteria for the $(2+1)$-dimensional integrability of straight strings scattering.

Equipped with the canonical volume form, we study the solution to the loop-level recursion relations proposed in~\cite{NimaBigBook}. We show how the known one-loop four-point amplitude is reconstructed from the recursion. For higher-points and higher-loops, we demonstrate that all physical singularities are present where as all spurious singularity cancels. Unlike the on-shell diagrams for $\mathcal{N}=4$ SYM, the solution to the loop level recursion manifests the required two-site cyclic symmetry similar to that of the tree-level solution~\cite{HW}. Another new feature appears where only half of the single-cut singularities are associated with external vertices, we identify the other half as associated to the internal vertices. Furthermore, we find that all singularities are covered by one coordinate chart of $OG_{k}$. We also find that via bubble reductions, all four and six-point multi-loop amplitudes can be reduced to a product of $d\log$s times the leading singularity, implying to all loop orders, the four- and six-point amplitudes only have logarithmic singularities. 

Since the solution to loop-level recursion relations include all physical singularities, they obviously include all leading singularities, which are associated with the reduced medial graphs. Since we know that distinct cells in $OG_{k+}$ correspond to inequivalent medial graphs, the leading singularities are then associated with cells in $OG_{k+}$. Each cell in $OG_{k+}$ encodes distinct linear dependency between consecutive columns in the orthogonal Grassmannian, which implies the vanishing of sets of ordered minors. These are precisely the singularities of which the orthogonal Grassmannian integral localizes on, and hence all residues of the integral indeed correspond to leading singularities of ABJM, as conjectured long ago~\cite{LeeOG}. 

As mentioned in the beginning of this introduction, it is possible to relegate locality and unitarity to emergent properties. In the recent works of Arkani-Hamed and Trnka \cite{Arkani-Hamed:2013jha, Arkani-Hamed:2013kca}, such a possibility was realized by transplanting the above physical properties by positivity. In particular, they showed that the scattering amplitudes of $\mathcal{N}=4$ SYM can be identified with a ``volume" of the space defined by the union of the positive Grassmannian $Gr(k,n)_{+}$ and  $Gr(4+k,n)_{+}$, where the latter is the Grassmannian associated with external kinematics data. The space is bounded by boundaries that are associated with physical singularities, thus ensuring locality. Unitarity is an emergent property due to positivity. An obvious question is whether such an object exists for ABJM theory. The current missing information lies in the optimal way of parameterizing the external data such that locality can be easily translated into properties of some orthogonal Grassmannian. An interesting application to such a formulation would be to verify the conjecture that IR-divergences of ABJM theory exponentiates in the same way as $\mathcal{N}=4$ SYM~\cite{BDSABJM}. 

ABJM is a theory with bi-fundamental matter fields with gauge group SU(N)$\times$SU(N). In principle there is nothing stopping us to consider on-shell diagrams for ABJ theory which has SU(N)$\times$SU(M)~\cite{ABJ}. The distinct gauge groups can be reflected in associating different weights to the faces of the medial graph. It will be interesting to consider more exotic gauge groups as its existence has been inferred by the recently found twistor string theory for $\mathcal{N}=6$ Chern-Simons matter theory~\cite{RaduString}.

This paper is organized as follows. In section \ref{section:Combinatorics}, after a brief introduction on positive orthogonal Grassmannian, we then extensively study its mathematical structures. In particular transformation rules for equivalent moves are understood and derived by embedding $OG_{k+}$ into $Gr(k,2k)_{+}$. This embedding also allows us to assign a face variables to medial graphs. A simple but general rule for obtaining Jacobian factors of diagrams containing closed loops is proposed and tested with non-trivial examples. We prove that with such a Jacobian factor, the volume form associated with a medial graph is invariant under equivalent moves. We also find that any cells $OG_{k+}$ form a Eulerian poset. 

Section \ref{section:Scattering} and section \ref{section:loop} are devoted to the study on the connection between positive orthogonal Grassmannian and the scattering amplitudes of ABJM theory. In section \ref{section:Scattering}, we review some of results of \cite{HW}, and then establish the fact that the scattering amplitudes in ABJM theory is the sum of two branches of $OG$, namely $OG_{k+}$ and $OG_{k {}-{}}$. At tree level, these two branches can be nicely combined into one, and leads to the usual BCFW representation of tree-level amplitudes. In section \ref{section:loop}, we study the all loop recursion relations of ABJM amplitudes in a great detail. We find the solutions to recursion relations have many interesting and nice properties. 

\section{Combinatorics of positive orthogonal Grassmannian} \label{section:Combinatorics}

\subsection{Definition of positive orthogonal Grassmannian}
Grassmannian $Gr(k, n)$ can be regarded as the space of $k$-dimensional subspaces inside $n$-dimensional complex space, $\mathbb{C}^n$. Take a basis $e_1,\ldots, e_n$ of  $\mathbb{C}^n$, then 
each point can be represented by a $(k\times n)$ matrix up to a $GL(k)$ transformation. For a canonical ordering of $e_1,\ldots, e_n$, there is a natural notion of positivity, namely the determinants of all
ordered minors $\Delta_I$ are non-negative. This notion of positivity is the same as the one defined by Lusztig on partial flag variety $(G/P)_{\geq 0}$ \cite{DanStuff}, as the Grassmannian can be 
identified with the partial flag manifold $GL(\mathbb{C})/P$, where $P$ is a parabolic subgroup of $GL(\mathbb{C})$. 

To define orthogonal Grassmannian $OG(k,n)$, one needs to first define a symmetric bilinear form on $\mathbb{C}^n$:
\begin{equation}
H[x,y]=\eta_{ij} x_i y_j,
\end{equation}
where $x_i, y_i$ are vectors in $\mathbb{C}^n$. In this paper, the bilinear $\eta_{ij}$ will be taken to be diagonal: i.e. $\eta_{ij}=\delta_{ij}$. Notice that in the complex case, different signatures of $\eta$ are isomorphic. The orthogonal Grassmannian is then defined as the space of $k$-planes inside $\mathbb{C}^{n}$, such that $H[x,y]=0$ on this subspace. Again each 
point of orthogonal Grassmannian can be represented by a $k\times n$ matrix, $C_{ai}$ with $a=1,\ldots,k,$ and $ i=1,\ldots,n$. The constraints $H[x,y]=0$ can be written as:
\begin{equation}
\eta^{ij}C_{ai} C_{bj} =0 \, .
\end{equation}
Note that there are $k\times(k+1)/2$ independent constraints. 

In this paper, we restrict ourselves to the orthogonal Grassmannian with $n=2k$, namely $OG_k\equiv OG(k,2k)$. The dimension of $OG_k$ is given by
\eq
k\times2k-k^2-\frac{k(k+1)}{2}=\frac{k(k-1)}{2}
\eqe
where we have removed $k^2$ degrees of freedom representing the $GL(k)$ redundancy of linearly recombining the $k$ $n$-dimensional vectors that span the $k$-dimensional planes. Let's order the column of $C_{ai}$ by $(1,2,\ldots, 2k)$ and denote $\mathsf{ I}$ an ordered $k$ subset. For example, choosing $\mathsf{I}=(1,2,3)$, we can use Gauss elimination algorithm to put the C matrix into an $\mathsf{ I}$-echelon form, i.e.  
\begin{equation}
C=\begin{pmatrix} 
1& 0&0&*&*&*\\
0 & 1 &0&*&*&* \\
0&0&1&*&*&*
\end{pmatrix}=(\mathsf{ I}_{3\times 3}, c_{3\times 3})\,.
\end{equation}
In the representation, the orthogonal Grassmannian is given by an $(9-6)=3$-dimensional parametrization of $ c_{3\times 3}$ such that, with $\eta^{ij}=\delta^{ij}$, $cc^T+1=0$. As a consequence, the minors constructed from the set of columns $C_{\mathsf{ I}}$, denoted by $M_{\mathsf{ I}}$ satisfying ${M_{\mathsf{ I}}/M_{\bar{\mathsf{ I}}}}=\pm(i)^{k}$, where $\bar{\mathsf{I}}$ is the ordered complement of $\mathsf{I}$. For example for $k=3$, if $\mathsf{ I}=(1,3,4)$, then $\bar{\mathsf{I}}=(2,5,6)$.

As discussed in \cite{HW}, the building blocks for scattering amplitudes of ABJM theory can be associated with the positive part of the orthogonal Grassmannian $OG_{k+}$. To define positive part of orthogonal Grassmannian, we need to first define a real subspace of $OG_k$, and in this case, the signature of $\eta_{ij}$ is important and 
different signature defines different subspace. As was discussed in \cite{HW}, to consistently define $OG_{k+}$ for arbitrary $k$, it is advantageous to use alternative signature $\eta_{ij}=(+,-,+,\ldots, -)$. Note that in this signature, the orthogonal condition becomes $1-cc^T=0$, and one instead has ${M_{\mathsf{ I}}/M_{\bar{\mathsf{ I}}}}=\pm1$. For $OG_{k+}$, one always has ${M_{\mathsf{ I}}/M_{\bar{\mathsf{ I}}}}=1$ for both $\mathsf{ I}$ and $\bar{\mathsf{I}}$ being ordered. 

Using the above notion of positivity, we can define a cell decomposition of non-negative part of $OG_{k+}$. We consider all possible linear dependency among consecutive chains of columns. The classification of all such linear dependency for the positive Grassmannian $Gr(k,n)_{+}$ is referred to as positroid stratification~\cite{PostGrass, TLam}. Each cell is then an element in the stratification. In terms of ordered minors it is defined as 
\begin{equation}
S_{\cal M}=\{A|M_\sigma(A)>0~~\text{for}~~\sigma\in {\cal M},~\text{and}~~M_\sigma(A)=0~~\text{for}~\sigma~\text{not}\in  {\cal M}\}\,.
\end{equation}

Let us begin with the first non-trivial example, $OG_{2+}$.\footnote{$OG_{1+}$ is zero-dimensional and given by $(1,1)$.} Taking $\mathsf{ I}=(1,2)$, the C-matrix is given by:
\begin{equation}
C=\begin{pmatrix} 
1& 0 &-\cot \theta&-\csc\theta\\
0 & 1 &\csc \theta&\cot \theta 
\end{pmatrix} \, .
\label{CanonicalGauge}
\end{equation}
On the other hand for $\mathsf{ I}=(1,3)$ we instead have:
\begin{equation}
C=\begin{pmatrix} 
1& \cos\theta &0&-\sin\theta\\
0 & \sin\theta &1&\cos\theta 
\end{pmatrix} \, .
\label{AlterGauge}
\end{equation}
As one can easily verify, the ordered minors of the above $C$-matrix are always positive for $0<\theta<{\pi\over 2}$. In this chapter, we will constrain ourself to this region, and use the abbreviation $s_i\equiv\sin\theta_i$, $c_i\equiv\cos\theta_i$.

\subsection{Medial graph and cells of positive orthogonal Grassmannian }
Cells of positive Grassmannian $Gr(k,n )_{+}$ can be represented nicely by bipartite network~\cite{PostGrass}, and one can find
the so-called cluster coordinates using the network.  The combinatorics of the bipartite network is very rich and is 
an extremely useful tool to study the geometry of the positive Grassmannian. One important ingredient is the equivalence 
moves for the network. Using those equivalence moves, Postnikov has established a one-to-one correspondence between 
the positive cells, and the bipartite networks that are distinct up to equivalence moves.

In the work of~\cite{NimaBigBook}, it was realized that the bipartite networks can be understood as on-shell diagrams of the scattering amplitudes in $\mathcal{N}=4$ SYM,  
and the aforementioned equivalence moves are given a physical interpretation as either the equivalence of kinematic constraints, or the equivalence of distinct ways to recursively construct the four-point amplitude. Therefore, we have the following triple:
\begin{equation}
\text{Cells of $Gr(k,n)_{+}$}\rightarrow \text{planar bi-partite network} = \text{on-shell diagram of $\mathcal{N}=4$ SYM}\nonumber
\end{equation}
It was soon realized that similar story can be established for the positive cell of orthogonal Grassmannian:
\begin{equation}
\text{Cells of $OG_{k+}$}\rightarrow \text{planar medial graph} = \text{on-shell diagram of ABJM theory} \nonumber
\end{equation}
We have reviewed the definition of cells of $OG_{k+}$ in last subsection, here we are going to explain the rules for constructing medial graphs and their equivalence moves. 
\begin{center}
\begin{figure}[htbp]
\small
\centering
\includegraphics[scale=0.6]{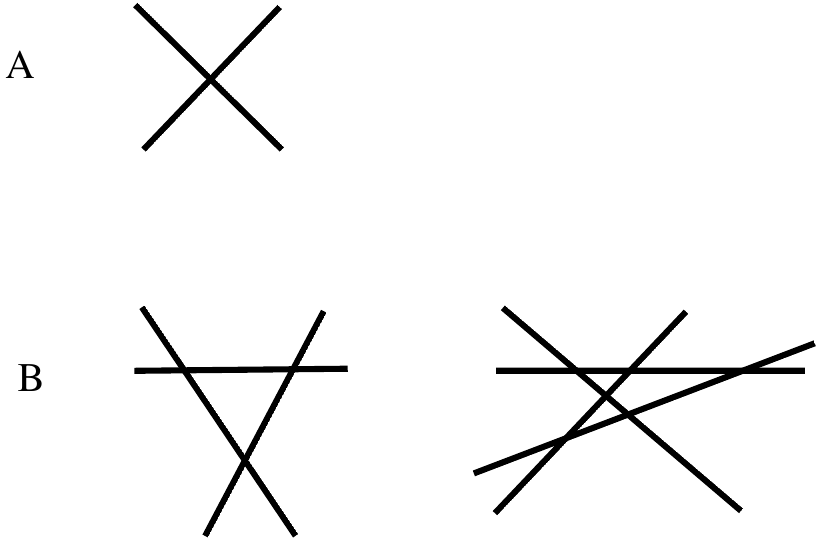}
\caption{A. The fundamental four-vertex for OG$_k$. B. Two medial graphs for OG$_3$ and OG$_4$, respectively.}
\label{vertex}
\end{figure}
\end{center}
The basic element for constructing medial graphs is a four-point vertex, see figure \ref{vertex}A. One can build 
complicated medial graphs by simply gluing the four-point vertices together as indicated in figure \ref{vertex}B. 
We limit ourselves to planar medial graphs, i.e. graphs defined on the disk, and the number of boundary vertices is always even ($2k$). We will associate one degree of freedom, $\theta$, to each vertex. Such graphs also appears in the study of electric networks~\cite{CIM, dVGV}, which is where the term medial graph came from, and the variable at each vertex is associated with the conductance of the resistors in the electric network.\footnote{ Recently it has been realized that the combinatorics can also be related to an orthogonal Grassmannian, albeit with different bilinear~\cite{Thomas}.} 

\begin{center}
\begin{figure}[htbp]
\small
\centering
\includegraphics[scale=0.7]{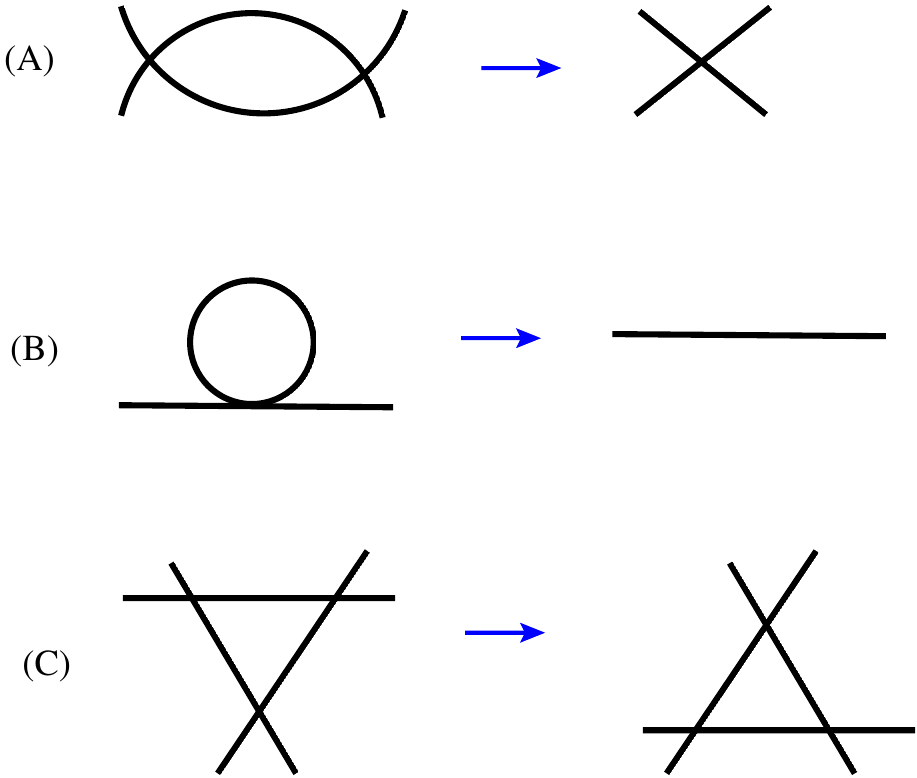}
\caption{}
\label{abjmmoves}
\end{figure}
\end{center}

As discussed in~\cite{NimaBigBook, HW}, distinct medial graphs represent distinct cells of $OG_{k+}$, and as the bipartite network, distinct diagrams are defined modulo equivalence moves. The equivalence moves associated with the medial graphs are shown in figure \ref{abjmmoves}: move A $\&$ B are called 
bubble reduction, and move C is called triangle move. Note that for the bubble reduction, the dimension of the diagram is reduced by one. An important notion is the irreducible graph, which is defined as:
\begin{itemize}
\item A medial graph is called irreducible if no bubble can be formed after doing any sequence of triangle moves.
\item Two irreducible medial graphs are equivalent and representing the same cell if they can be related by triangle moves.
\end{itemize}
Just to distinguish, we will name the bubble (B) in above picture as {\it removable bubble} for the reason becoming clear in section \ref{section:localmoves}. 
Here is an example of triangle move and bubble reduction:
$$\includegraphics[scale=0.6]{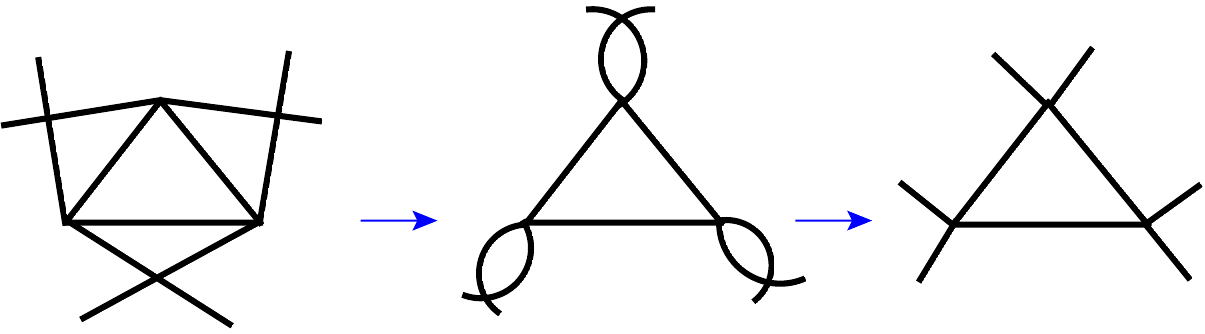}$$
where we have applied a triangle move and three bubble reductions in that order. It was proposed in~\cite{HW} that the connection between cells of $OG_{k+}$ and the medial graph is the following:
\begin{equation}
\text{Cells of $OG_{k+}$ is in one-to-one correspondence with distinct reduced medial graphs.}. \nonumber  
\end{equation}
In the following we will prove the above statement by embedding the medial graphs of $OG_{k+}$ into bipartite graphs, $Gr(k,2k)_{+}$. Since from the work of~\cite{PostGrass, TLam}, we know that the stratification of $Gr(k,n)_{+}$ is in one-to-one correspondence with the bipartite graphs, then the above statement follows. This will also allow us to derive canonical positive coordinates for $OG_{k+}$ which is equivalent to the coordinates obtained from the iterative gluing procedure discussed in~\cite{HW}.

\subsection{The embedding of OG$_{k+}$ in $Gr(k,2k)_{+}$}
A cell of $Gr(k,2k)_{+}$ can be represented by a bipartite network. Since OG$_{k}$ can be considered as a reduction of $Gr(k,2k)$, it is natural to try to transform the medial graph defined in last subsection into a bipartite network. The crucial thing is to solve the orthogonal constraints on the coordinates once a bipartite network is found.

\subsubsection{Review of bipartite network}
The various properties of OG$_{k}$ can be derived by considering a strata preserving embedding of OG$_{k}$ within $Gr(k,2k)$. The positroid stratification of $Gr(k,n)$ can be one-to-one identified with bipartite graphs. These are graphs that consist of trivalent vertices, with each vertex associated with two possible colors, black and white, and any each edge is connected to two vertices of opposite color. As an example one has:
$$\includegraphics[scale=0.5]{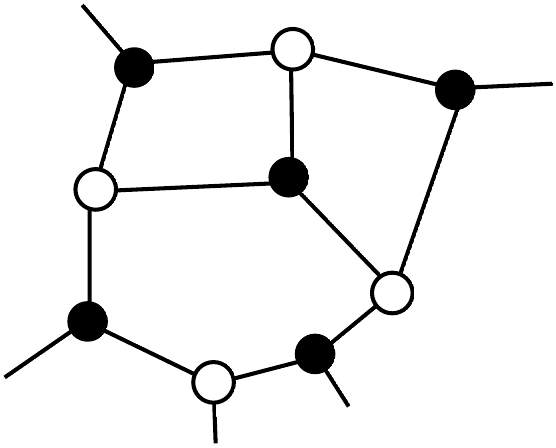}$$
For a given bipartite graph, one can associate each face with a variable $f_i$ satisfying $\prod_i f_i=1$. Due to this constraint, the dimension of the graph is $(n_f-1)$, where $n_f$ is the number of faces. 

One can extract an explicit representation of the $C$-matrix by providing a perfect orientation to the graph. 
A perfect orientation is given by assigning an edge with an arrow, and for each white vertex, one must have one incoming, and two outgoings, whereas for a black vertex one has two incomings and one outgoing:
$$\includegraphics[scale=0.7]{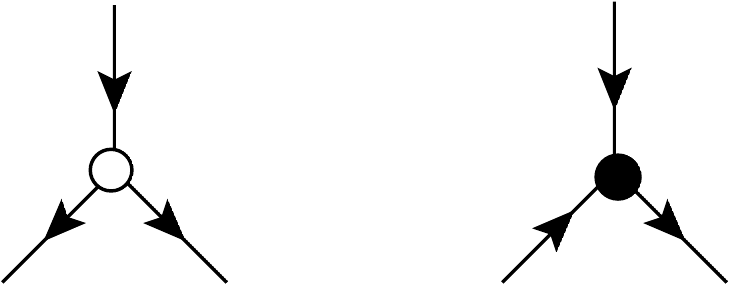}$$
For example, for the four-dimensional top cell in $Gr(2,4)_{+}$, one has:
\eq
\includegraphics[scale=0.5]{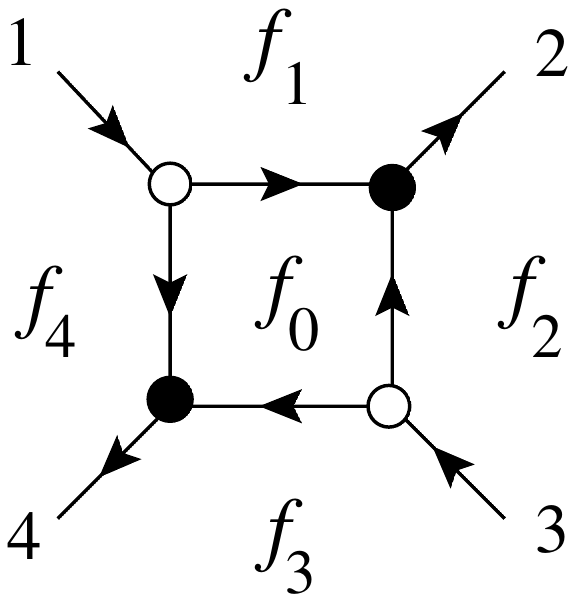}
\label{FaceParemeterization}
\eqe
A given oriented decoration correspond to a particular gauge choice for the $C$-matrix, with the incoming arrows of the external legs (the source set) indicating the columns that are set to unity. The boundary measurements are then given by the products of face variables on the right hand side of the path that connects the source to the sink (the outgoing external legs). If there are closed paths, then one obtains a geometric series weighted by $(-1)^{w}$, where $w$ is the number of times one encircles the closed circle. In the example above, we have:
\eq
M_{12}=1/f_1, \; M_{14}=f_4,\quad M_{32}=f_2,\quad M_{34}=1/f_3
\eqe
Using the boundary measurement, we can define the $C$-matrix. Denote the source set of the network as $I=(i_1,i_2,\ldots, i_k)$, then $C$ is defined as
\begin{itemize}
\item The sub matrix $C_I$ is an identity matrix 
\item The remaining element $C_{rj}=(-1)^sM_{i_{r},j}$, where $s$ is the number of elements of $I$ strictly between $i_r$ and $j$.
\end{itemize}
So for the network in figure \ref{FaceParemeterization}, the source set is $(1,3)$, and the C matrix is 
\begin{equation}
C=\begin{pmatrix} 
1& 1/f_1 &0&-f_4\\
0 & f_2 &1&1/f_3
\end{pmatrix} \, .
\label{Gr24Face}
\end{equation}
Using the boundary measurement and the definition of $C$ matrix, one find a parameterization of cells of $Gr(k,n)_{+}$. 
\begin{center}
\begin{figure}[htbp]
\small
\centering
\includegraphics[width=7cm]{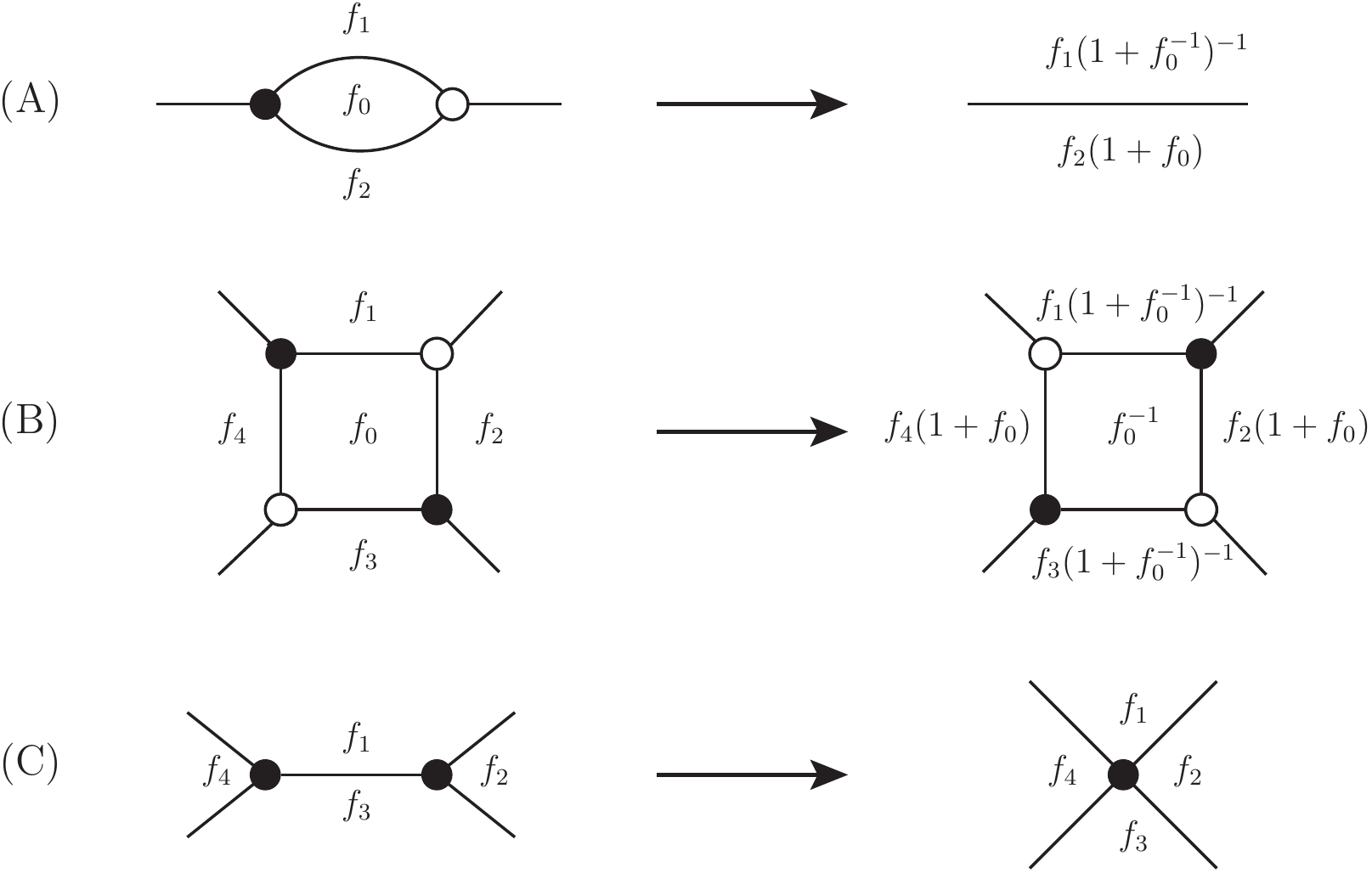}
\caption{}
\label{bipartitemove}
\end{figure}
\end{center}
The equivalence moves for the bipartite network are shown in figure \ref{bipartitemove}, where we have also indicated how the face variables transform under the moves (it is actually the cluster transformation on the $A$-variables). Note that the transformation rules are ``subtraction-free", i.e. there are no minus signs in the transformation rules. An important consequence is that once $f_i$ are restricted to be positive, this property will be preserved. Again a reduced graph is defined as:
\begin{itemize}
\item A bipartite graph is called reduced network if no bubble is formed after doing any sequence of square moves.
\item Two irreducible bipartite graphs  are equivalent and representing the same cell if they can be related by square moves.
\end{itemize}
It is proven in \cite{PostGrass} that there is a one-to-one correspondence between the positive cells and bipartite network up to equivalence. Using the parameterization from reduced bipartite network, one can define a simple volume form on the positive cells: 
\begin{equation}
{d f_1\over f_1}\wedge {d f_2\over f_2} \wedge \ldots \wedge{d f_n\over f_n}=d\ln f_1\wedge d\ln f_2 \wedge \ldots \wedge d \ln f_n,
\end{equation}
here $f_i$'s are the face variables in the corresponding bipartite network. The remarkable thing about this volume form is that it is invariant under the cluster transformation, up to an overall factor $(-1)^n$.

\subsubsection{From medial graph to bipartite network}
\begin{center}
\begin{figure}[htbp]
\small
\centering
\includegraphics[width=6cm]{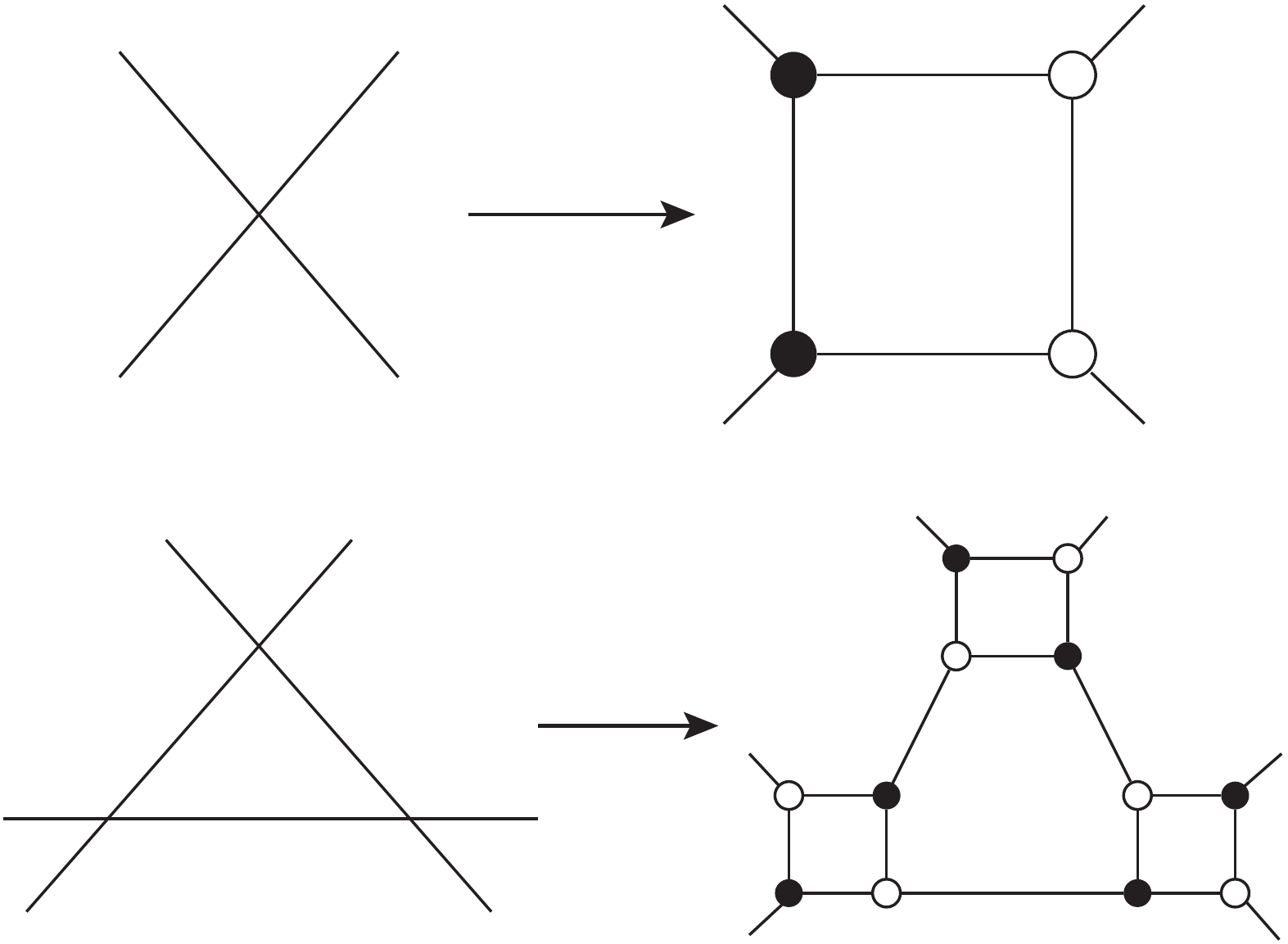}
\caption{}
\label{blowup}
\end{figure}
\end{center}

Now we are ready to consider the ``image" or our medial graph in a bipartite network. The procedure is straightforward: one simply replaces a internal four vertex with a square, see figure \ref{blowup}. Notice that we can choose any of two 
orderings of black-white vertices, since these two orderings are related by the square move and therefore would give the same point in Grassmannian.  
The canonical choice is found by doing the blow-up such that the corresponding network is bipartite. It will be advantageous to consider the blow up such the source legs connected to white vertices. 
The next question is how to find a $k$-dimensional parameterization of the face variables, such that the resulting $C$-matrix of $Gr(k,2k)$ is in fact orthogonal. Let us  begin with simplest case, OG$_{2+}$. Comparing eq.(\ref{AlterGauge}) and eq.(\ref{Gr24Face}), one immediately see that orthogonal constraint can be solved by setting
\eq
f_1=\frac{1}{c},\;f_{4}=s,\;f_2=s,\;f_3=\frac{1}{c}\,,
\eqe 
or diagrammatically:
\eq
\includegraphics[scale=0.4]{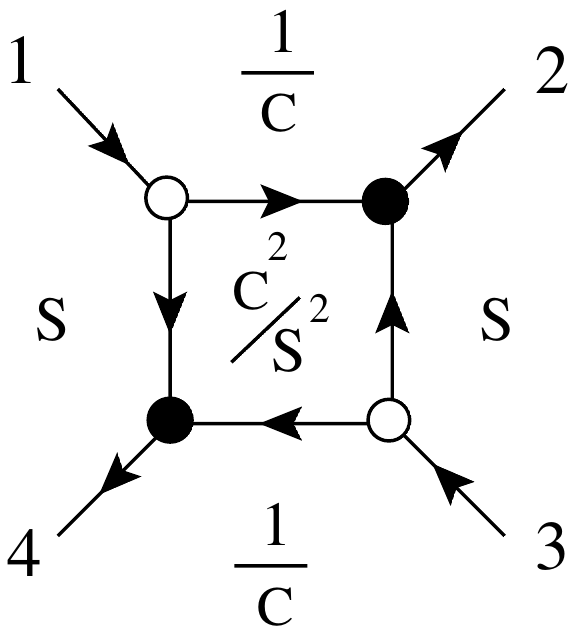}\,.
\eqe
If one chooses another perfect orientation such that the source set is $\mathsf{I}=(1,2)$, one can see that eq.(\ref{CanonicalGauge}) is reproduced using the same face variables. 
As we reviewed earlier, the boundary measurement is unchanged by doing cluster transformation, and we have a different parameterization if we have a different canonical 
blow up for $OG_{2+}$:
\eq
\includegraphics[scale=0.3]{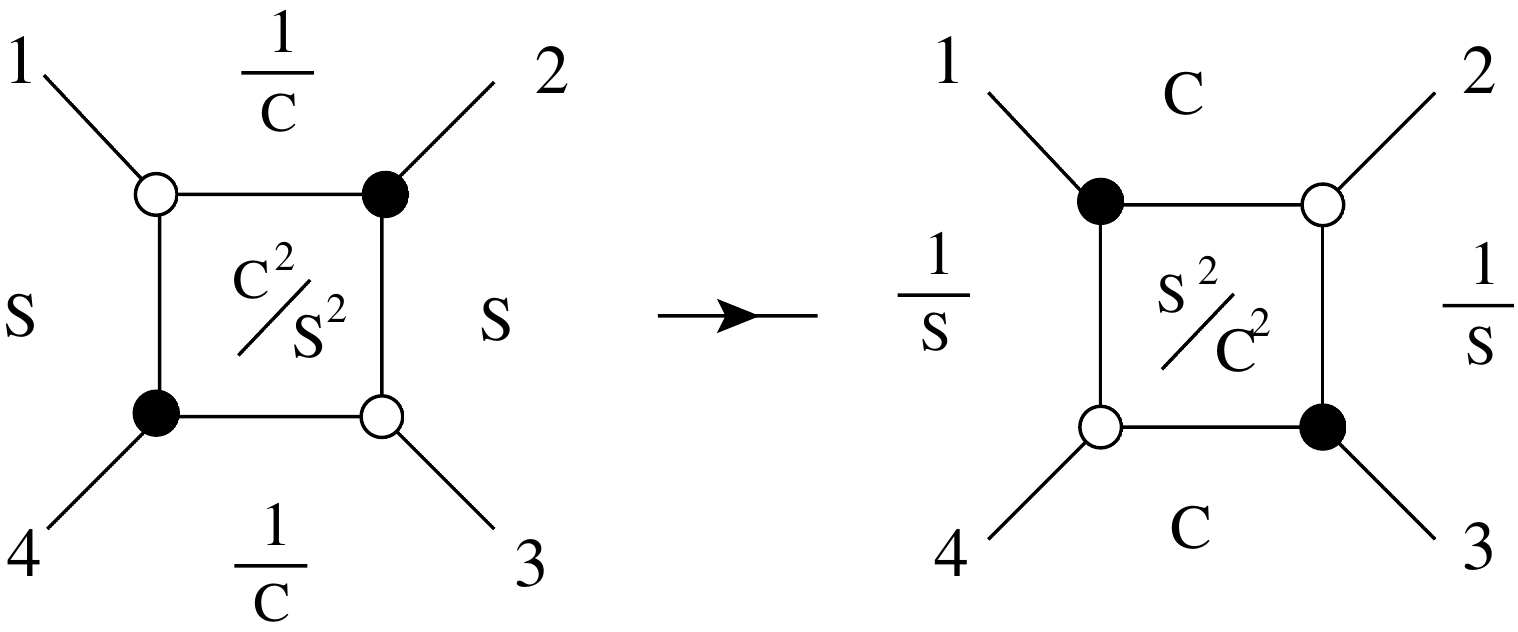}\,.
\eqe
Note that the cluster transformation is nothing but a change of variable from $\theta\rightarrow \theta+\pi/2$, in $OG_{k}$.

The canonical coordinates in the bipartite network for a general $OG_{k}$ image is given by:
\begin{itemize}
\item The variable for the $k$ new faces associated with the blow up of each four-point vertex is simply $f_v={c_v^2\over s_v^2}$.
\item For the remaining faces that were present in the medial graph, identify all vertices that inclose the face and take a clockwise orientation on each face. The contribution from each vertex is $1/c$ if in the bipartite network, one first encounters the black
vertex under this orientation, otherwise the contribution is $s$. In other words, the face variables are either $f=s_1 s_2\ldots s_i$ or $f={1\over c_1 c_2\ldots c_i}$ depending on the orientation. 
\end{itemize} 
As an example consider the following embedding for $OG_3$
\eq
\includegraphics[scale=0.45]{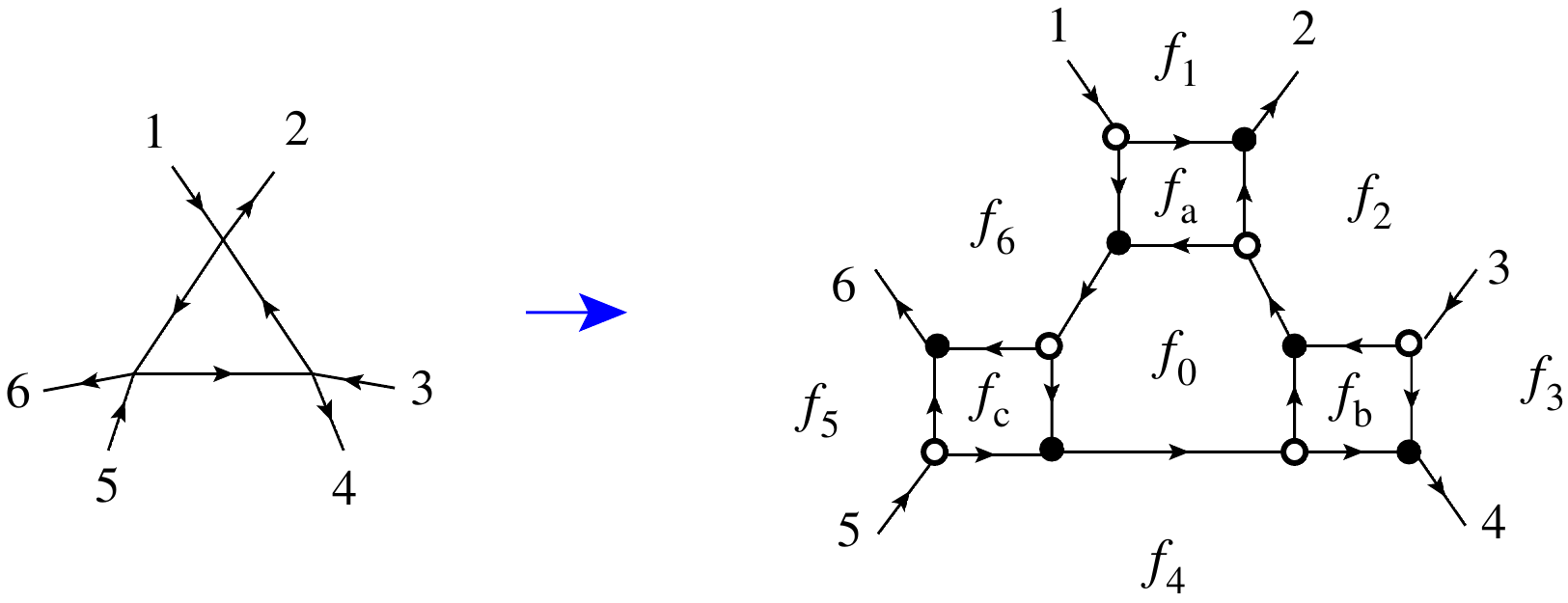}\,.
\eqe
The face variables are given by:
\eqa
\nonumber &&(f_a,f_b,f_c)=(c^2_1/s^2_1, c^2_2/s^2_2, c^2_3/s^2_3),\; f_0=\frac{1}{c_1c_2c_3}\\
&&f_1=\frac{1}{c_1},\;f_2=s_1s_2,\;f_3=\frac{1}{c_2},\;f_4=s_2s_3,\;f_5=\frac{1}{c_3},\;f_6=s_1s_3
\eqae
The boundary measurement is given by:
\eqa\label{Gr36M}
\nonumber &&M_{16}=f_6/(1+f_0^{-1}),\;M_{12}=f_1^{-1}+\frac{1}{f_1f_a(1+f_0)},\;M_{14}=f_4f_5f_6f_c/(1+f_0^{-1})\\
\nonumber &&M_{32}=f_2/(1+f_0^{-1}),\;M_{34}=f_3^{-1}+\frac{1}{f_3f_b(1+f_0)},\;M_{36}=f_1f_2f_6f_a/(1+f_0^{-1})\\
\nonumber&&M_{54}=f_4/(1+f_0^{-1}),\;M_{56}=f_5^{-1}+\frac{1}{f_5f_c(1+f_0)},\;M_{52}=f_3f_4f_2f_b/(1+f_0^{-1})\,.\\
\eqae
One can straightforwardly verify that the $C$-matrix obtained from the boundary measurements are indeed orthogonal. Note that this prameterization is also subtraction free, ensuring that the image of OG$_{k+}$ is indeed in $Gr(k,2k)_{+}$.

It is important to show that the equivalence moves of medial graph can be reproduced using the equivalence moves of bipartite graph,
\eq
\includegraphics[scale=0.6]{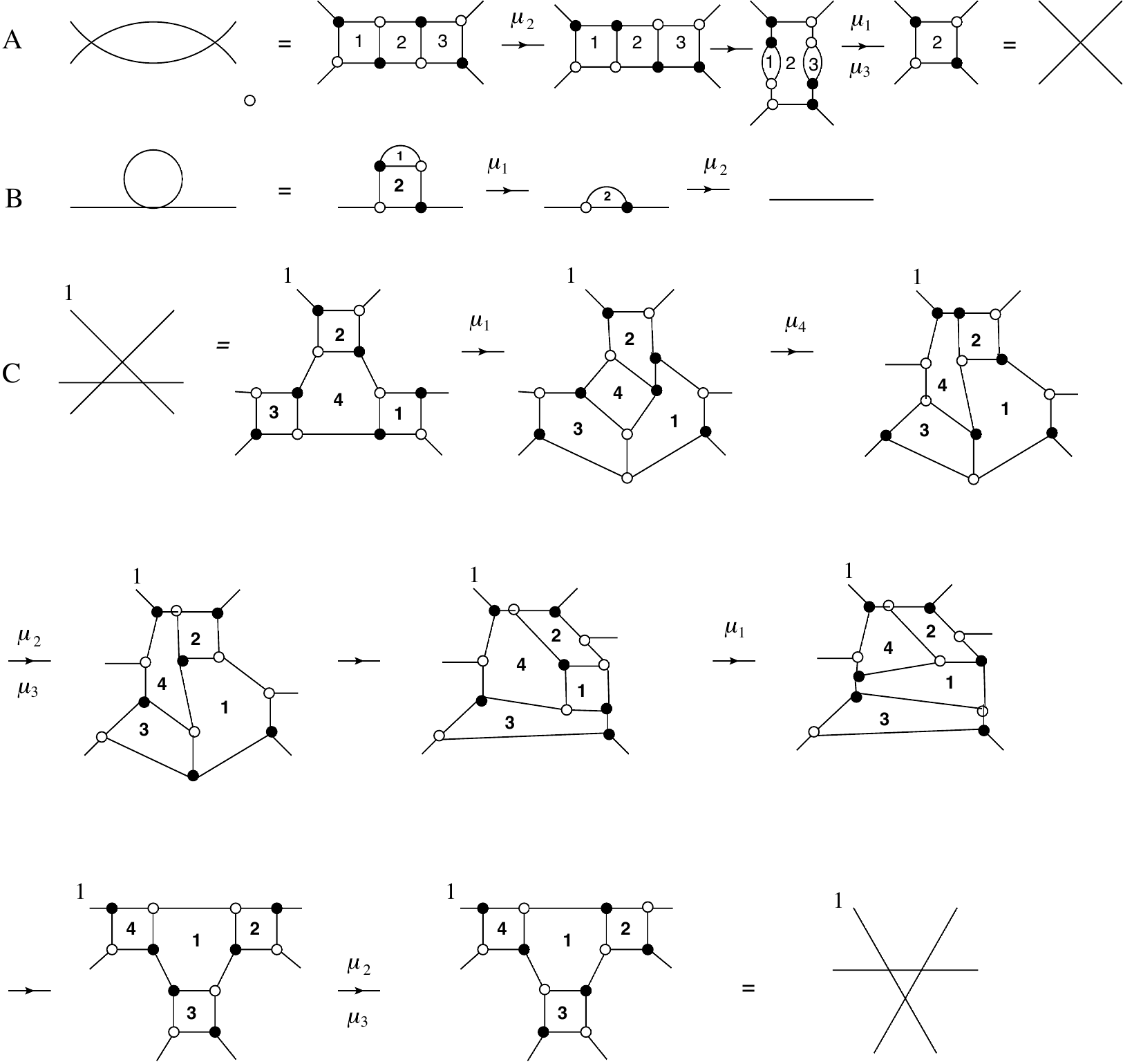}\,.
\label{moverelation}
\eqe
In the above, we have marked the faces where we perform a square move or a bubble reduction. Since all equivalence and reduction moves for the medial graph are faithfully represented by their images in the bipartite network, this confirms the statement that $OG_{k+}$ is a stratification preserving the reduction of $Gr(k,2k)_{+}$. However, it is obvious that not all cells in $Gr(k,2k)_{+}$ contains the image of $OG_{k+}$. Using the ``left-right permutation paths" of~\cite{NimaBigBook}, one can easily identify the cells that do contain such images. These are the ones whose left-right path connects \textit{both ways} between pairs of external vertices, i.e. if there is a left-right path that begins with $i$ and ends in $j$, then the path beginning with $j$ will end in $i$.

\subsection{Back to medial graph}
\subsubsection{Coordinates and boundary measurement}
As discussed in~\cite{NimaBigBook} the parameterization of cells in the orthogonal Grassmannian can also be read off by looking at the medial graph alone. To define the boundary measurement, we need to choose an orientation of the four-point vertex. 
Using the blow up picture of a four-point vertex, we can see that there are two kinds of orientation: a) alternating orientation, b) adjacent orientation, as shown below
\eq
\includegraphics[width=6cm]{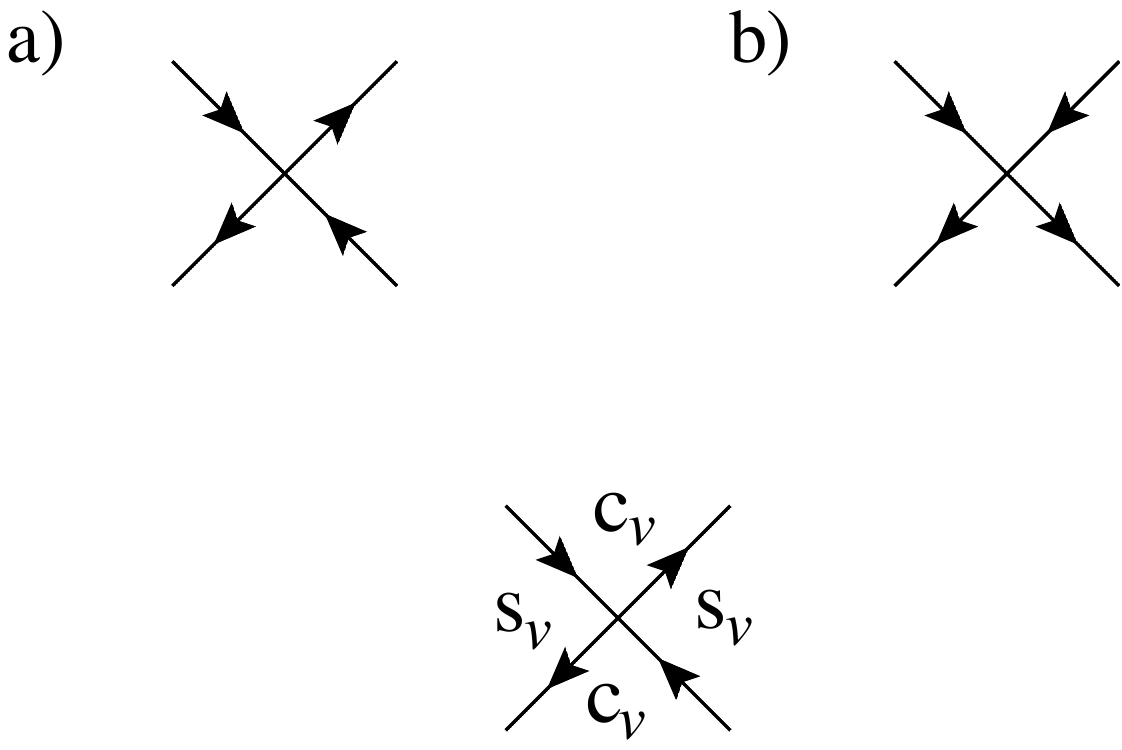}
\label{vertexorientation}
\eqe
Here (a) correspond to the cyclic gauge, eq.(\ref{AlterGauge}), and (b) correspond to eq.(\ref{CanonicalGauge}) which we will denote as the canonical gauge. It is easy to see, for a given medial graph, it is always possible to only use orientation a) for each vertex. On the other hand, such statement is not true for orientation b).  For example, for the following diagram, we see that we must use at least one cyclic gauge vertex (denoted by the black dot):
$$\includegraphics[scale=0.8]{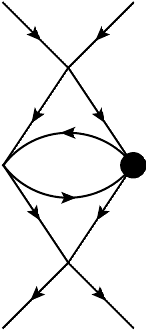}$$
For simplicity, we choose the alternative gauge for each vertex. As discussed in~\cite{HW}, the cyclic gauge is advantageous in that the corresponding parameterization covers all co-dimension boundaries of a given cell. With this choice a given graph can have two orientations related by flipping all arrows in the graph. After choosing the orientation, the graph has an 
interesting feature: every simple face is oriented\footnote{We call a face simple if it does not include other faces.}. Two faces have the same orientation if they share a single vertex, and two faces have the opposite orientation if they share an edge.

We now associate an angle variable $\theta_v$ and functions $c_v=\cos \theta_v$ and $s_v=\sin \theta_v$ with each angle of vertex
according to the rule shown in figure \ref{vertexorientation}.  The definition of the boundary measurement is similar to the bipartite network case we reviewed earlier: for a oriented path passing through a vertex, we assign function $c_v$ or $s_v$ depending on how the path turns around this vertex,  see figure \ref{vertexorientation}.
Using the above rule, we can define a measurement between a sink and source node using the path between them:
\begin{equation}
M_{ij}=\sum _{P:i\rightarrow j} (-1)^{w(p)} \prod_{v\in P} (c_v~\text{or}~s_v)
\end{equation}
where $w(p)$ is the winding number of the path. It is easy to see that if there is a loop in a path, the local contribution of this loop would be
\begin{align}
& {1\over 1+c_1c_2c_3\ldots}~~~~\text{counterclockwise}  \nonumber\\
&  {1\over 1+s_1s_2s_3\ldots}~~~~\text{clockwise}
\end{align}
Notice that the boundary measurement defined above is actually equivalent to the one defined using the embedding to $Gr(k,2k)_{+}$. 
\subsubsection{The invariant volume form}
Recall that for $Gr(k,n)_{+}$, there is an associated volume form that takes a canonical $d\log$ form in terms of face variables, and is invariant under equivalence moves. In this section we would like to consider a similar volume form for $OG_{k+}$. We define a canonical volume form to be that of having only logarithmic singularity on the boundaries of the positive region ($\theta=0$ and $\theta=\pi/2$), and is preserved under equivalence moves. In~\cite{HW}, a canonical volume form for $OG_{2+}$ was given as 
\eq
d\log (s^2/c^2)=d\log (f_v),
\eqe 
where in the second equality, we have emphasized the fact that the argument of the $d\log$ is simply the new face variable when $OG_{2+}$ is embedded in $Gr(2,4)_{+}$. So the measure behaves as $dx/x$ on the boundary of the positive region. For more complicated graphs, it was understood that the canonical volume form is given by:
\eq
\mathcal{J} \times \prod_{i\in n_v}d\log f_i \,,
\eqe
where $\mathcal{J}$ is an additional Jacobian factor that appears when there are closed faces in the graph. Note that this is unlike the bipartite network, where the canonical volume form is a product of $d\log$ that by itself is invariant. We referred to $\mathcal{J}$ as the Jacobian factor associated with the medial graph, that is because it has its origin as the Jacobian factor that arises from the gluing of on-shell diagrams and solving a set of linear (super) momentum-conservation constraints, see~\cite{HW} for more details.\footnote{In fact in order to correctly produce the extra $\mathcal{J}$ factor, one must have $\mathcal{N}=6$ SUSY.}  Here we give a set of simple rules to produce the correct Jacobian for any medial graph.

The Jacobian for a given medial graph can be decomposed into several contributions from distinct closed loops. Note that when two closed-loops share an edge, they are necessary of opposite orientation, where as if two share 2 vertices, they are of the same orientation and can combine to define a bigger loop. We list the distinct contributions as follows: 
\begin{itemize}
\item $\mathcal{J}_1 $: This contains contributions from all single closed loops $J_i$, products of disjoint pairs of closed loops $J_i J_j$, disjoint triples of closed loops $J_i J_j J_k$ and so on. We will denote this part as $\mathcal{J}_1$, and it takes the following form:
\eq
\mathcal{J}_1 = \sum_{\rm single } J_i + \sum_{\rm disjoint \, pairs} J_i J_j
+ \sum_{\rm disjoint \, triples } J_i J_j J_k + \ldots \, . 
\eqe
For our graph, every face forms a oriented closed loop, according to the choice of fundamental vertex (\ref{AlterGauge}), if the loop is closed in anti-clock wise, it is a product of $c$'s, otherwise it is a product of $s$'s. 

\item $\mathcal{J}_2$: This takes into account the contributions for closed loops sharing a single vertex. This contribution is given by products of $c$'s or $s$'s of the corresponding loops, except that of the shared vertex. 

\item $\mathcal{J}_3$: It is for closed loops sharing two vertices without sharing an edge. This is given by the sum of products of $c$'s (or $s$'s) of all vertices involved plus the product of all vertices \textit{except} the two shared vertices. 

\item $\mathcal{J}_{13}$ and $\mathcal{J}_{23}$: Since in the case of $\mathcal{J}_3$, the closed loops form a bigger loop, we need to go back and consider case 1 and 2 from the viewpoint of this bigger loop. They are denoted as $\mathcal{J}_{13}$ (and $\mathcal{J}_{23}$) and given by combining rules $\mathcal{J}_3$ with that of $\mathcal{J}_1 $(and $\mathcal{J}_2$). 

\item $\mathcal{J}$: The final Jacobian $\mathcal{J}$ of a medial graph is then given by $1$ plus all above contributions, 
\eq
\mathcal{J} = 1 + \mathcal{J}_1 + \mathcal{J}_2 + \mathcal{J}_3 + \mathcal{J}_{13} + \mathcal{J}_{23} \, .
\eqe

\item Finally, removable bubbles in the diagram can be fully decoupled from the diagram. One only needs to compute the Jacobian for the rest diagram after decoupling according to the rules described above, and in the end multiply with the Jacobian of each removable bubble.

\end{itemize} 
Let us now consider a few concrete examples to illustrate the rules. For instance, consider following two diagrams,
\eq \label{Jacobi1}
\includegraphics[scale=0.4]{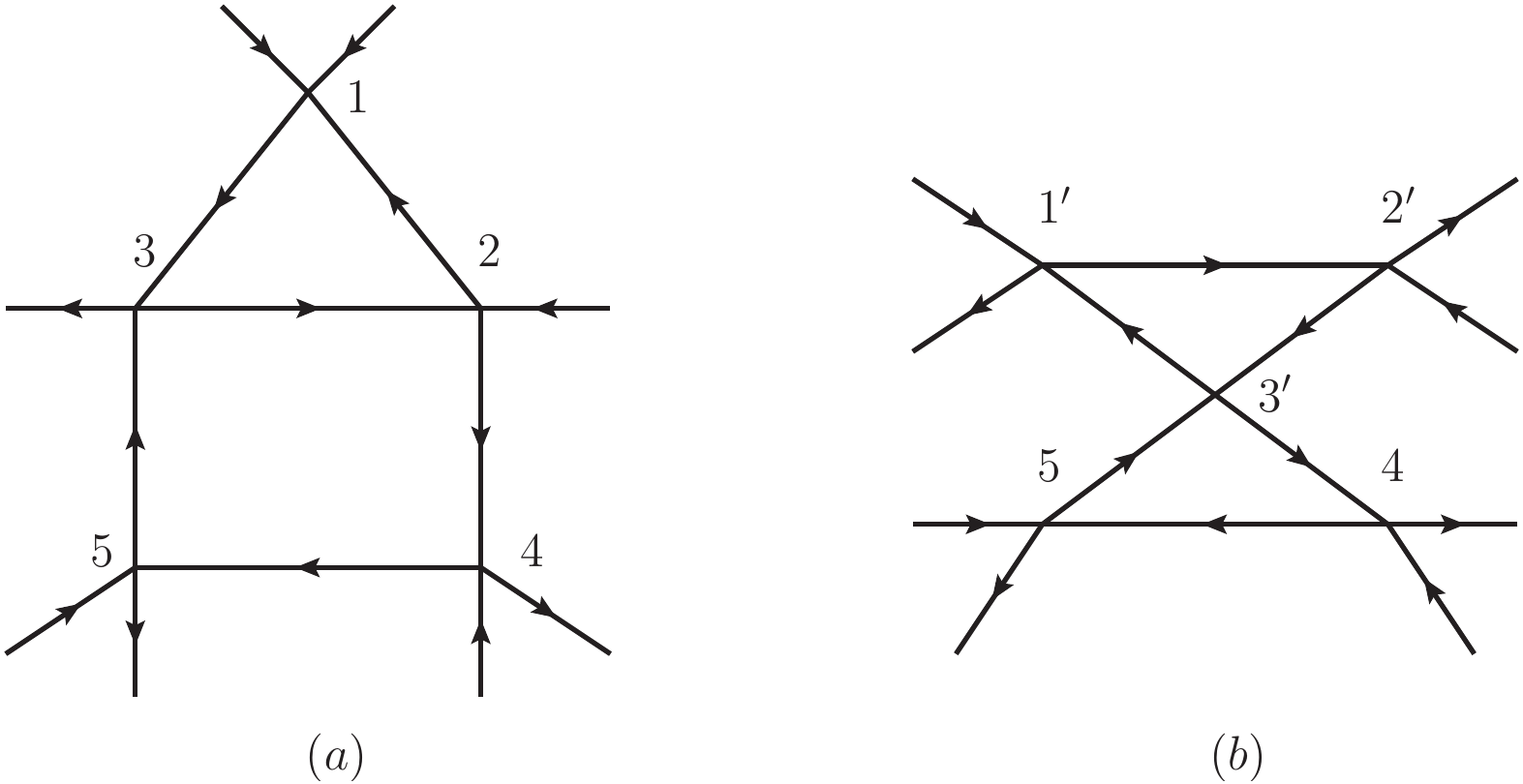} \, .
\eqe
Note these two diagrams are related to each other by a triangle move. The Jacobian of the diagram $(a)$ in above picture only gets contribution from $\mathcal{J}_1$, which is given by 
\eq
\mathcal{J}_{a} = 1 + \mathcal{J}_{1 \, a} = 1 + (c_1 c_2 c_3 + s_2 s_3 s_4 s_5) \, .
\eqe
While the diagram $(b)$ gets contributions from $\mathcal{J}_1$ as well as $\mathcal{J}_2$, the result is given by
\eq
\mathcal{J}_{b} = 1 + \mathcal{J}_{1 \, b}+ 
\mathcal{J}_{2 \, b} = 1 + (s'_{1} s'_{2} s'_3 + s'_3 s_4 s_5) +( s'_1 s'_2 s_4 s_5 ) \, .
\eqe
As we will discuss in next section, the volume forms of diagrams related by triangle moves should be invariant, namely 
\eq \label{relation1}
{\mathcal{J}_{a}} \prod^5_{i=1} d \log ({\rm tan}_i) 
= {\mathcal{J}_{b}} \prod^5_{i=1} d \log ({\rm tan'}_i)  \, 
\eqe
where ${\rm tan}_i = s_i/c_i$. Here ${\rm tan'}_i = {\rm tan}_i$ for $i=4, 5$ whereas other ${\rm tan'}_i$ are related to ${\rm tan}_i$ according to the triangle move transformation given later in eq.(\ref{triangletransform}). From eq.(\ref{identi}) and above relation (\ref{relation1}), we find $\mathcal{J}_{a}$ and $\mathcal{J}_{b}$ should satisfy following relation, 
\eq
{\mathcal{J}_{a} \over 1+ c_1 c_2 c_3}  = {\mathcal{J}_{b} \over 1+ s'_1 s'_2 s'_3} \, .
\eqe
It is straightforward to verify the above equivalence.

Now we move on to something more complicated. Consider the following two diagrams that are related by two triangle moves: 
\eq \label{Jacobi}
\includegraphics[scale=0.4]{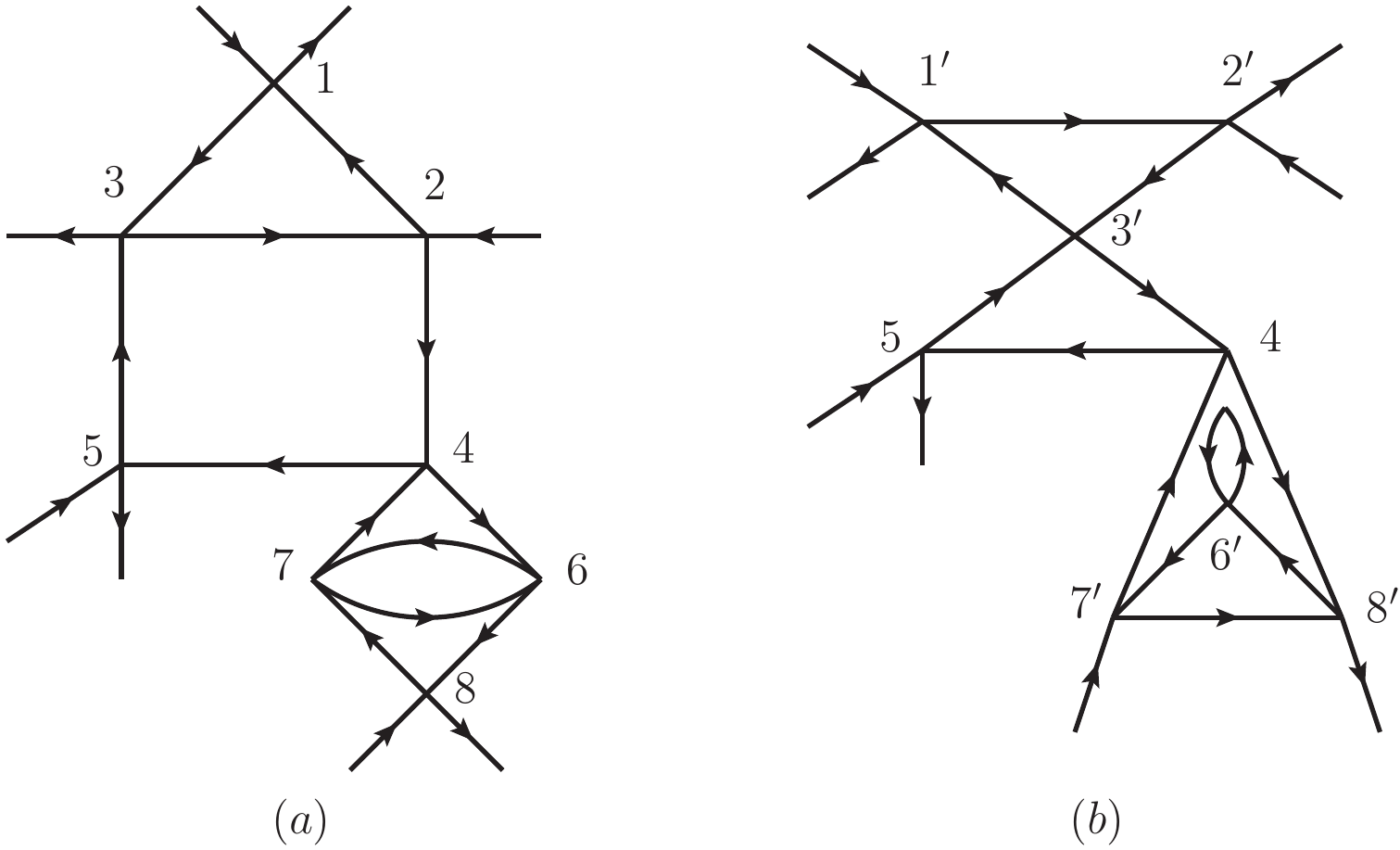} \, .
\eqe
According to the rules, it is straightforward to read off all the contributions from the diagram. For the diagram $(a)$, we have
\eqa
{\mathcal{J}_a}_1 &=& c_1 c_2 c_3 +  s_2 s_3 s_4 s_5+ s_4 s_6 s_7 + c_6 c_7 + s_6 s_7 s_8 \cr
&+& c_1 c_2 c_3 (s_4 s_6 s_7 +  c_6 c_7 + s_6 s_7 s_8 )
 + s_2 s_3 s_4 s_5 (c_6 c_7 + s_6 s_7 s_8) \, , \cr
 {\mathcal{J}_a}_2 &=& s_2 s_3 s_5 s_6 s_7  \, , \cr
 {\mathcal{J}_a}_3 &=& s_4 s_8 + s_4 c_6 c_7 s_8 \, .
\eqae
Finally there are contributions from ${\mathcal{J}}_{13}$ and ${\mathcal{J}}_{23}$, which give  
\eqa
{\mathcal{J}_a}_{13} &=& c_1 c_2 c_3(s_4 s_8 + s_4 c_6 c_7 s_8) \, , \cr
{\mathcal{J}_a}_{23} &=& s_2 s_3 s_5(s_8 + c_6 c_7 s_8) \, .
\eqae
So the Jacobian of the diagram $(a)$ is then
\eq
\mathcal{J}_a = 1 + {\mathcal{J}_a}_1 + {\mathcal{J}_a}_2 +{\mathcal{J}_a}_3 + {\mathcal{J}_a}_{13} + {\mathcal{J}_a}_{23} \, . 
\eqe
Similarly, we can read off the Jacobian for the diagram $(b)$ in eq.(\ref{Jacobi}), which contains a removable bubble,
\eq
\mathcal{J}_b =(1 + c'_{6})( 1 + {\mathcal{J}_b}_1 + {\mathcal{J}_b}_2 )\, ,
\eqe
where ${\mathcal{J}}_3$ vanishes in this case, and ${\mathcal{J}_b}_1, {\mathcal{J}_b}_2$ are given by
\eqa
{\mathcal{J}_b}_1 &=& s'_{1} s'_{2}s'_{3} + s'_{3}s_4 s_5 + s_4 s'_{7} s'_{8} + c'_{7}c'_{8}+ s'_{3}s_4 s_5 c'_{7}c'_{8} \cr
&+& s'_{1} s'_{2}s'_{3} ( s_4 s'_{7} s'_{8} + c'_{7}c'_{8} ) + s'_{1}s'_{2}s_4 s_5 c_7 c_8  \cr
{\mathcal{J}_b}_2 &=& s'_{1} s'_{2} s_4 s_5 + s'_{3}s_5 s'_{7} s'_{8} + s'_{1} s'_{2} s_5 s'_{7} s'_{8} \, .
\eqae
Again the differential forms of these two diagrams in eq.(\ref{Jacobi}) should be invariant under the moves, 
\eq \label{relation}
\mathcal{J}_a  \prod^8_{i=1} d \log {\rm tan}_i 
= \mathcal{J}_b  \prod^8_{i=1} d \log {\rm tan'}_i  \, ,
\eqe
which leads to a very non-trivial relation between $\mathcal{J}_a$ and $\mathcal{J}_b$, 
\eq
{\mathcal{J}_a \over (1 + c_1 c_2 c_3)(1+ s_6 s_7 s_7) } = { \mathcal{J}_b \over (1 + s'_{1} s'_{2}s'_{3} )(1+ c'_{6} c'_{7}c'_{8}) } \, , 
\eqe
where $s'_4 =s_4, s'_5 = s_5$ and other $c'_i$ and $s'_i$ are related to $c_i$ and $s_i$ according to eq.(\ref{triangletransform}). We have checked numerically that the above relation indeed holds. 

Before closing this section, let us comment on the fact that the Jacobian as well as the boundary measurement of a big graph may be obtained by considering its subgraphs inside the big graph as effective vertices\footnote{In some sense it is the inverse procedure of embedding a medial graph into a bipartite network, where one expands a four-point vertex into a box diagram.}. This will also be proved to very useful in next section. For instance for the graph on the right in (\ref{Jacobi1}), one may consider, for instance, the triangle on the top effectively as a six-point vertex shown as following
\eq \label{effectivevertex}
\includegraphics[scale=0.5]{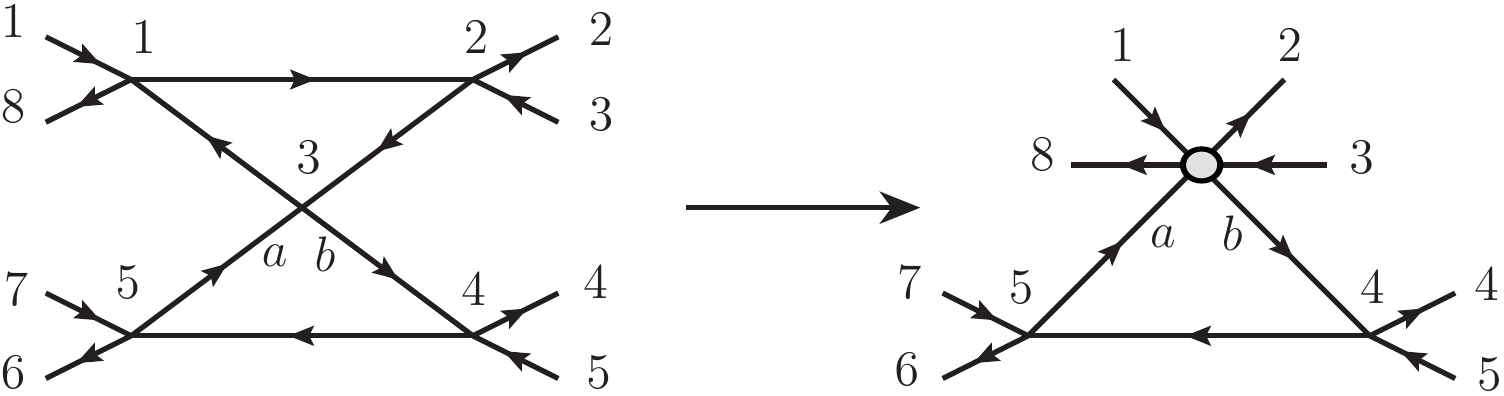} \, .
\eqe
We will call the graph with effective vertex as an {\it effective graph}. The content of the effective vertex is its own boundary measurements. When constructing the boundary measurement of the entire graph, if the oriented path passes through the effective vertex, instead of the usual $s_i$ or $c_i$, one picks up the boundary measurement of that effective vertex. For the case in eq.(\ref{effectivevertex}), denote the measurements as $m_{ij}$, we have for example, 
\eq \label{mab}
m_{ab}= {s_3 + s_1 s_2 \over 1+ s_1 s_2 s_3} \, .
\eqe
To calculate the boundary measurements and Jacobian of the full graph, one can use exactly the same rules described in previous sections. The only change is that when encounter the effective vertex, one needs to use its measurements. So the Jacobian of the full graph is simply the Jacobian of the effective graph multiplied with the Jacobian coming from the effective vertex itself. For the example we are considering, we have
\eq
\mathcal{J} = (1 + m_{ab} s_4  s_5)(1+ s_1 s_2 s_3)   \, ,
\eqe
where $(1 + m_{ab} s_4  s_5)$ is the Jacobian of the effective graph, and $(1+ s_1 s_2 s_3)$ is the Jacobian of the effective vertex itself. Substitute the result of $m_{ab}$, eq.(\ref{mab}), we find the result we obtained earlier. 

\subsubsection{Coordinates change under local moves} \label{section:localmoves}
Here we demonstrate that the Jacobian factors discussed previously is necessary for the volume form to be preserved under the triangle equivalence move. Furthermore, in the bubble reduction, the Jacobian factor is necessary such that after the reduction, one obtains a simple $d\log$ form. 
Let's start with the simplest case, the reduction of a removable bubble 
\eq
\includegraphics[scale=0.65]{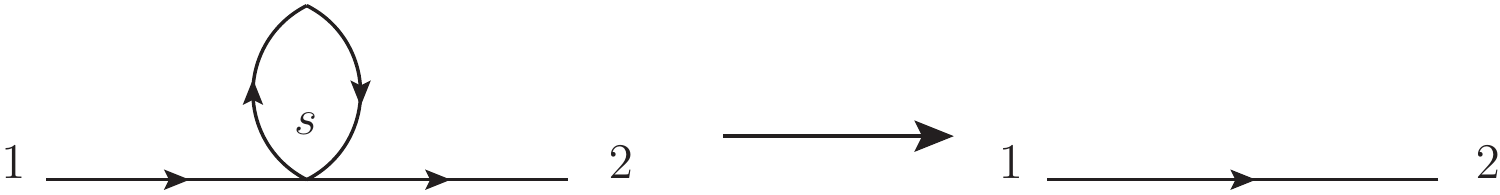} \, .
\label{removablebubble}
\eqe 
The measurement of the removable bubble is trivial, $M_{12}=1$, and the volume form is simply 
\eq
d \log(1/s -1) \, .
\eqe
Since the boundary measurement of the removable bubble is trivial, which means that one can always decouple it from the rest of diagram, that's the reason we call such bubble as removable. This fact was used to obtain the Jacobian factor for a given graph in previous section. We now move on to consider the other type of bubble, and the triangle equivalence move, as shown following
\eq
\includegraphics[scale=0.5]{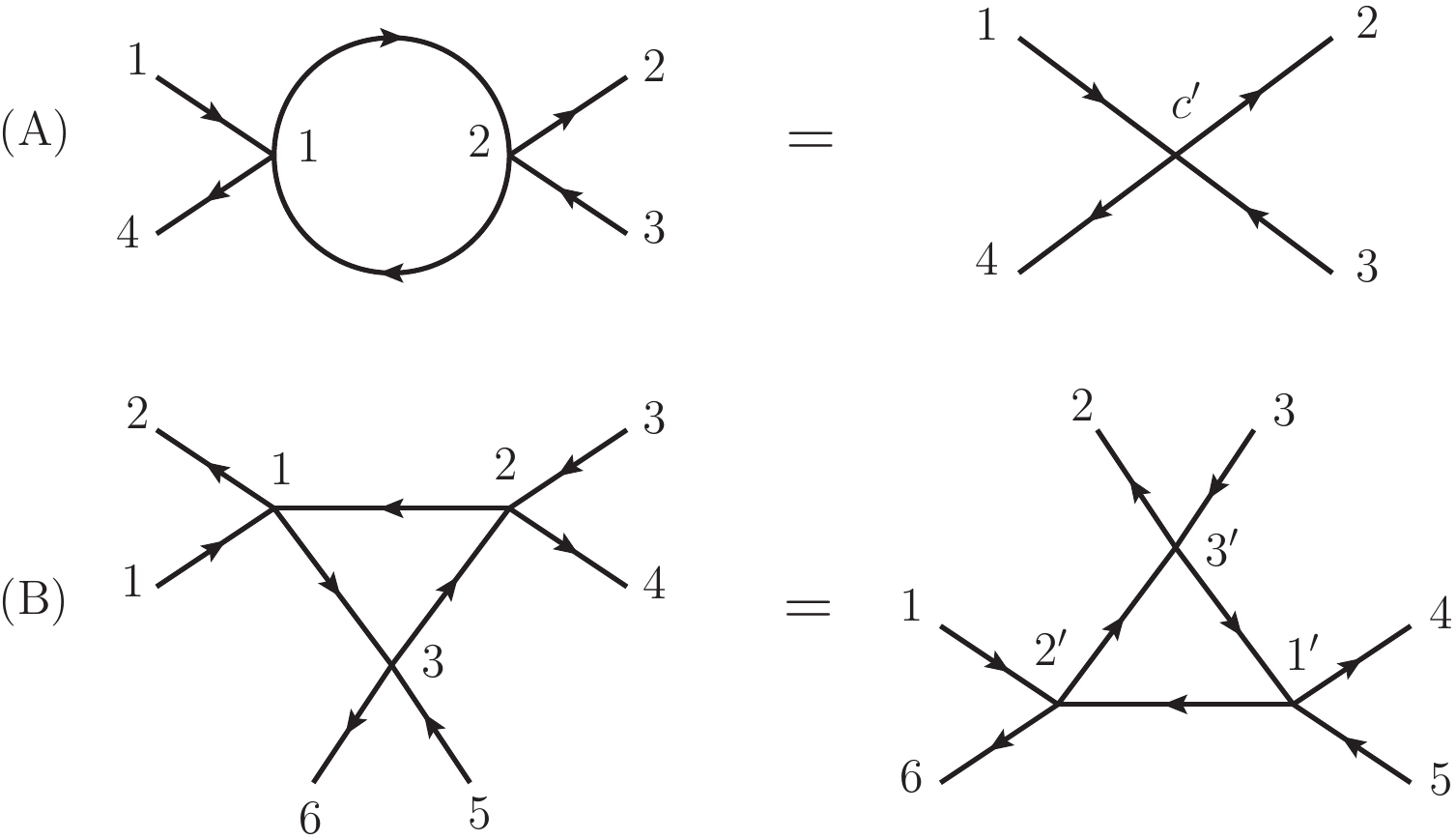}
\label{coordinatechange}
\eqe
The boundary measurement  for two graphs shown in figure. \ref{coordinatechange}A is 
\begin{align} \label{bubblemeasurement}
{\rm LHS}\; \ref{coordinatechange}A:\quad & M_{12}=M_{34}={c_1c_2\over 1+s_1 s_2},~~~~~~~~~~~~~~M_{14}=M_{32}={s_1+s_2\over 1+s_1 s_2} \nonumber \\
{\rm RHS}\; \ref{coordinatechange}A:\quad& M_{12}=M_{34}=c^{\prime},~~~~~~~~~~~~~~~~~~~~~~~ M_{14}=M_{32}=s' 
\end{align}
To make the boundary measurement to be the same, the coordinates are then related by
\begin{equation} \label{bubbleeffective}
c'={c_1 c_2 \over 1+s_1 s_2},~~~~~~~~s^{\prime}={s_1+s_2 \over 1+s_1 s_2}.
\end{equation}
We now calculate how the integration measure is changed. The initial integration measure is 
\begin{equation}
{ (1+s_1 s_2)} {d\log \tan_1}\wedge {d\log \tan_2} \, .
\label{measure}
\end{equation}
In the primed coordinates, we can see that the new integration measure is given as\footnote{The differential form we write down is slightly different, but equivalent, to the one obtained in~\cite{HW}.}
\begin{equation}
d\log \tan' \wedge d\log p=d\log(\tan') \wedge d \log({s_2 c_1\over s_1 c_2}) \, .
\end{equation}
Thus we see that equipped with the Jacobian factor in eq.(\ref{measure}), the bubble reduction directly gives a $d\log$-form where one separates out the degree of freedom that does not appear in the $C$-matrix.

We have checked the local change of volume form, now we will prove that the above calculation is also 
valid for an arbitrary graph. Consider following general graph with a bubble inside
\eq
\includegraphics[scale=0.6]{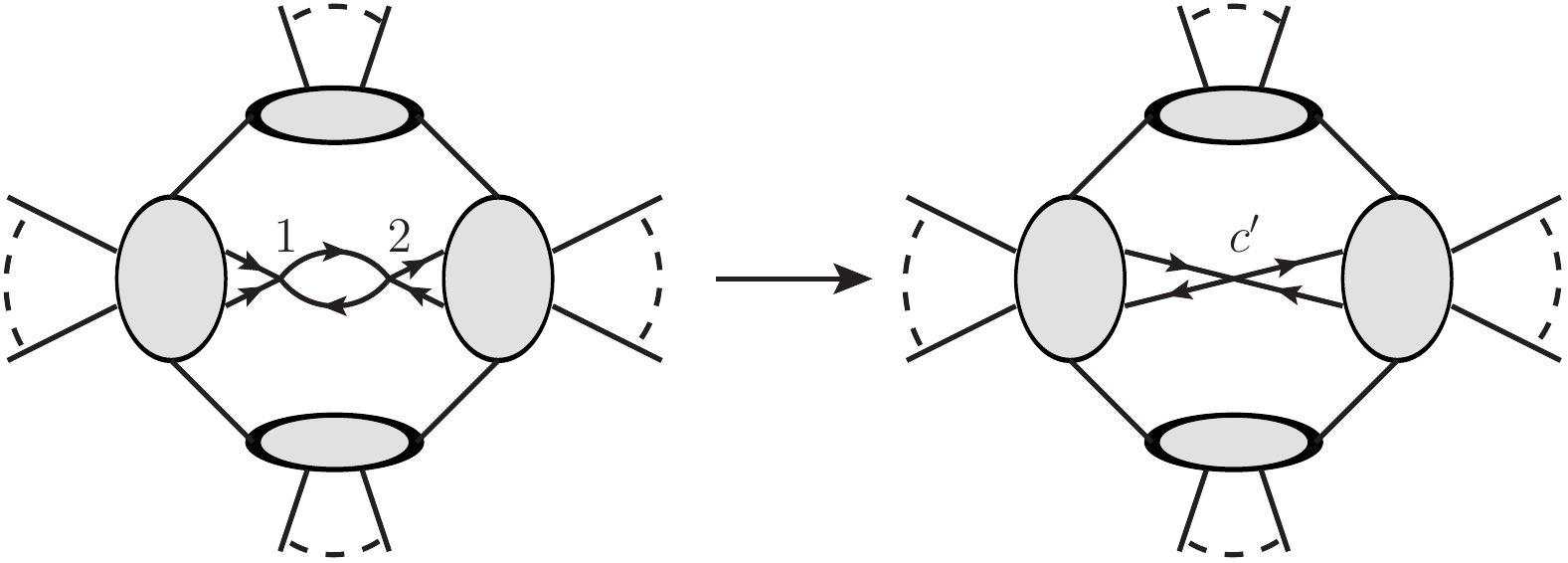}
\label{bubblegeneral}
\eqe
As we discussed at the end of previous section, one can view the bubble in the above graph effectively as a fundamental four-point vertex as shown in (\ref{bubblegeneral}) by the graph on the right. The Jacobian of the graph is given by 
\eq
\mathcal{J} = \mathcal{J}_{\rm bubble} \times \mathcal{J}_{\rm eff} = (1 + s_1 s_2) \mathcal{J}_{\rm eff} \, .
\eqe
Thus the volume form can be rewritten as
\eqa 
\mathcal{J} \prod_{i} {d\log \tan_i}
&=&(1 + s_1 s_2) {d\log \tan_1}\wedge {d\log \tan_2} \, \mathcal{J}_{\rm eff} \prod_{i>2} {d\log \tan_i} \cr
 & =& d\log({s_2 c_1\over s_1 c_2})  \times \mathcal{J}_{\rm eff} d\log \tan' \prod_{i>2} {d\log \tan_i}   \, ,
\eqae
with transformation given by (\ref{bubbleeffective}). This is nothing but the statement of bubble reduction for a general graph, with $d\log({s_2 c_1\over s_1 c_2})$ decoupled from the rest of graph. 

Let us move on to consider the transformation rule for the vertex coordinates under the triangle move. For the left graph in figure. \ref{coordinatechange}B, the boundary measurements are given as
\begin{align} \label{triangle1}
& M_{12}={c_1+c_2c_3\over 1+c_1c_2c_3},~~M_{34}={c_2+c_1c_3\over 1+c_1c_2c_3},~~M_{56}={c_3+c_1c_2\over 1+c_1c_2c_3} \nonumber\\
& M_{14}={s_1c_3 s_2\over 1+c_1c_2c_3},~~M_{36}={s_3c_1s_2\over 1+c_1c_2c_3},~~ M_{52}={s_3c_2s_1\over 1+c_1c_2c_3}\nonumber\\
& M_{16}={s_1s_3 \over 1+c_1c_2c_3},~~M_{32}={s_2s_1\over 1+c_1c_2c_3},~~M_{54}={s_2s_3\over 1+c_1c_2c_3}
\end{align}
After the triangle move, we have
\begin{align} \label{triangle2}
& M_{12}={c'_2 c'_3\over 1+s'_1 s'_2 s'_3},~~M_{34}={c'_3 c'_1\over 1+s'_1 s'_2 s'_3},~~ 
M_{56}={c'_1 c'_2\over 1+s'_1 s'_2 s'_3}\nonumber\\
& M_{14}={c'_2 s'_3 c'_1\over 1+s'_1 s'_2 s'_3},~~M_{36}={c'_3 s'_1 c'_2\over 1+s'_1 s'_2s'_3} ,~~M_{52}={c'_1s'_2 c'_3\over 1+s'_1s'_2s'_3}\nonumber\\
& M_{16}={s'_2+s'_3s'_1 \over 1+s'_2s'_3s'_1},~~M_{32}={s'_3+s'_2s'_1 \over 1+s'_1s'_2s'_3},~~M_{54}={s'_1+s'_2s'_3 \over 1+s'_1s'_2s'_3}
\end{align}
By requiring the measurements to be the same, we find
\begin{align} \label{triangletransform}
&s'_1={s_3c_1 s_2\over c_1+c_2c_3} ,~~ c_1={c'_3s'_1 c'_2\over s'_1+s'_2s'_3 }\nonumber\\
& s'_2={s_3c_2 s_1\over c_2+c_1c_3},~~~ c_2={c'_1s'_2 c'_3\over s'_2+s'_3s'_1 }\nonumber\\
&s'_3={s_1c_3 s_2\over c_3+c_1c_2},~~~ c_3={c'_2s'_3 c'_1\over s'_3+s'_1s'_2 }  \, .
\end{align}
It is straightforward to show that the volume form is also invariant under the above coordinate transformation:
\begin{equation}
({ 1+c_1 c_2 c_3}){d\log\tan_1}\wedge d\log\tan_2\wedge d\log\tan_3=
({1+s'_1 s'_2 s'_3}) {d\log\tan'_1}\wedge d\log\tan'_2\wedge d\log\tan'_3
\label{identi}
\end{equation}
Indeed as we have emphasized the Jacobian factor is crucial to ensure the invariance of the volume form.
Again, we have only checked the invariance of local piece of integration measure, but it is enough for us. To see this, let's consider the following general graphs related to each other by a triangle move, 
\eq
\includegraphics[scale=0.5]{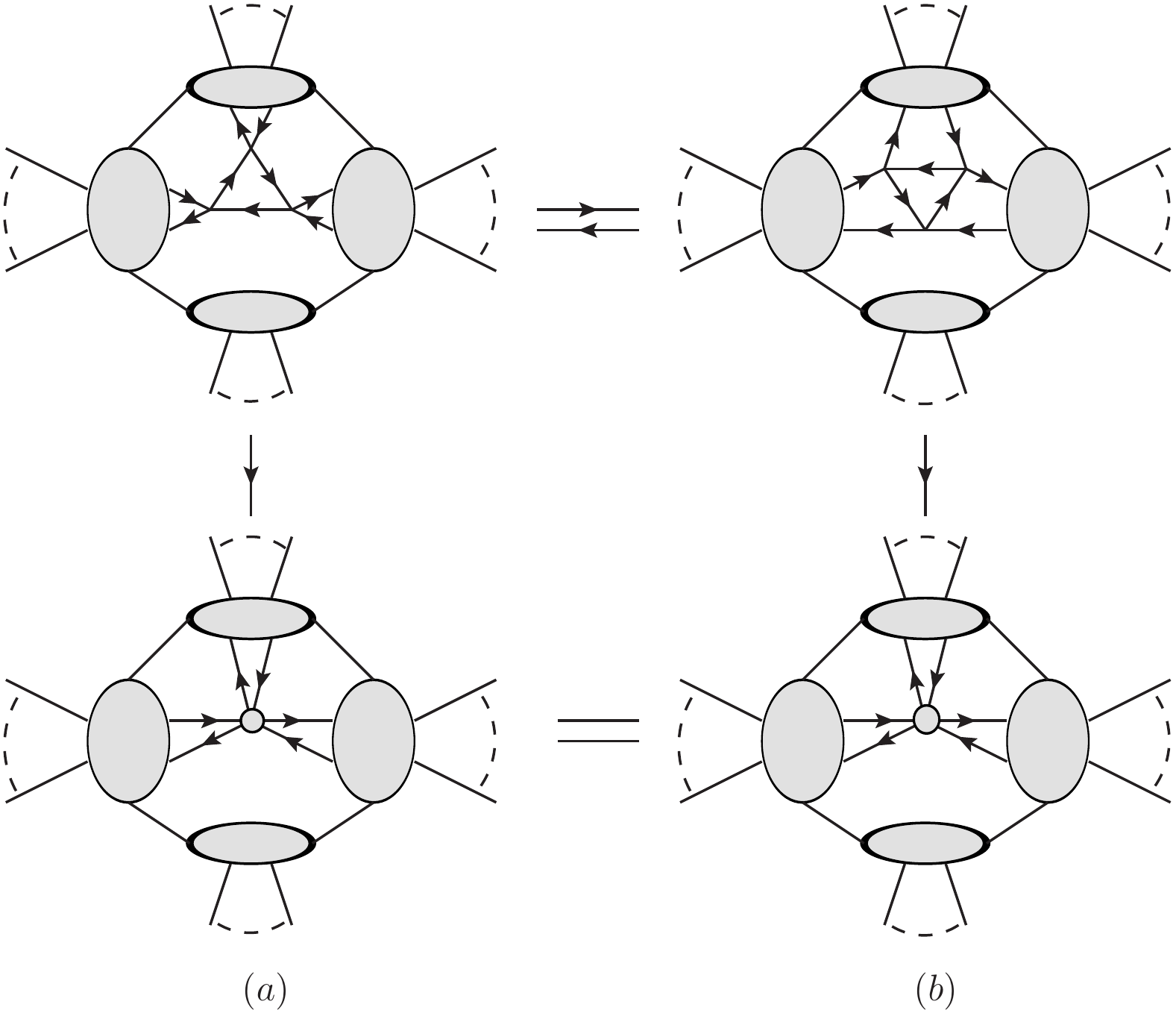}
\label{trianglegeneral}
\eqe
As shown in the above picture, two graphs on the top are related by a triangle move, where two diagrams on the bottom are the corresponding graphs by considering the triangles as effective vertices. According to the rules, the Jacobi of two graphs are respectively given by 
\eq
\mathcal{J}_a &=& {\mathcal{J}_a}_{\rm triangle} \times {\mathcal{J}_a}_{\rm effective} \, , 
\quad
\mathcal{J}_b &=& {\mathcal{J}_b}_{\rm triangle} \times {\mathcal{J}_b}_{\rm effective} \, ,
\eqe
where we denote ${\mathcal{J}_i}_{\rm triangle}$ as the Jacobian of the triangle in graph $(i)$, whereas ${\mathcal{J}_i}_{\rm effective}$ is the Jacobian of the effective graph $(i)$, for $i=a, b$. Since the only difference between effective graph $(a)$ and effective graph $(b)$ is that they have different six-point effect vertices. However we have proved that measurements of two six-point effective vertices are actually the same, which then leads to 
\eq
{\mathcal{J}_a}_{\rm effective} = {\mathcal{J}_b}_{\rm effective}  \,. 
\eqe
Using this result and the identity (\ref{identi}), we thus deduce 
\eq
\mathcal{J}_a \prod_{i} d\log\tan_{a \, i} = \mathcal{J}_b \prod_{i} d\log\tan_{b \, i} \, .
\eqe
Namely the volume form of a general graph is invariant under the triangle move, as we also have shown explicitly by examples in previous section.


\subsection{Global aspects of medial graph}
\subsubsection{Global characterization of medial graph: decorated permutation}
In this section, we would like to consider following questions:
\begin{itemize}
\item When a medial graph reduced?
\item How to tell two that two graphs are equivalent? 
\end{itemize}
Note that since all equivalence and reduction moves are preserved by the embedding of $OG_{k+}$ in $Gr(k,2k)_{+}$, the reducibility of a given medial can be rephrased as a question of reducibility of its image in the bipartite network. In other words the criteria of reducibility is exactly the same as that of the bipartite network.  

Reducibility of bipartite network can be deduced by considering the permutation associated with the  ``left-right path"~\cite{PostGrass, NimaBigBook}:  as one enters the graph take a left turn when one hits a white vertex and a right turn when hits a black vertex. Starting with the external vertex $i$, through the above rule one reaches another external vertex $\sigma(i)$. This defines a set of permutation $i\rightarrow \sigma(i)$ for all external vertices. Equipped with this definition, a reducible bipartite network can be identified as that where closed paths are formed, or two sets of permutation paths crosses each other along more than one common edge in the form of:
\eq
\includegraphics[scale=0.45]{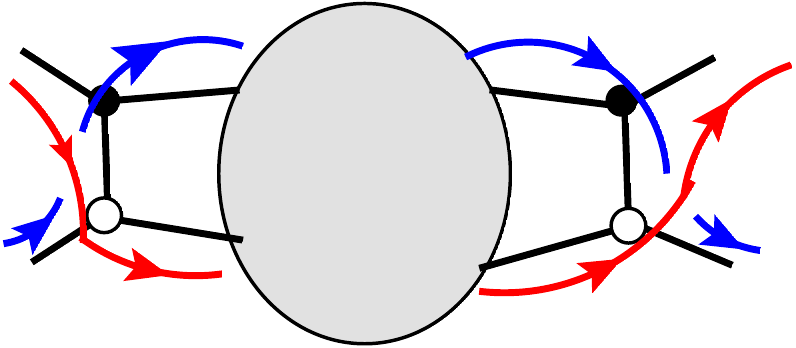}\,.
\eqe

As discussed previously, the bipartite network that contains the image of $OG_{k+}$, is that whose permutation paths all have the property that if $\sigma(i)=j$, then $\sigma(j)=i$. From the above analysis one can also deduce the corresponding rule for reducible medial graphs:
\begin{itemize}
\item A medial graph is reduced if there is no self intersecting permutation paths. 
\end{itemize}
As an example consider the following two diagrams: 
\eq
\includegraphics[scale=0.4]{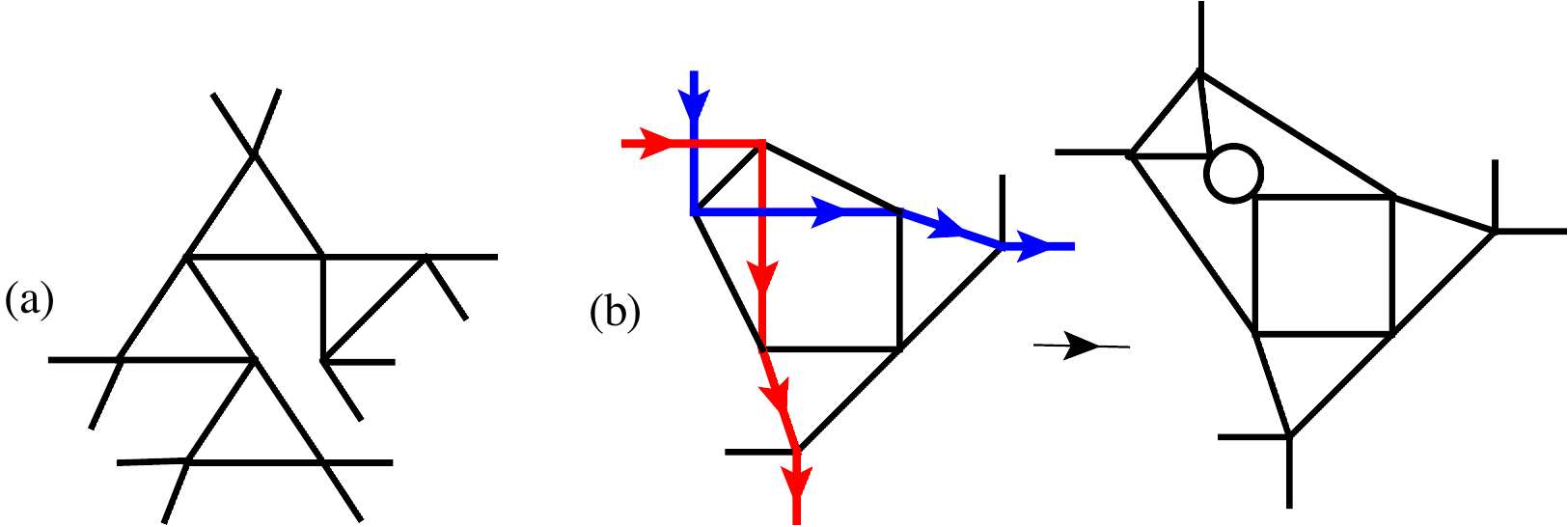}\,.
\eqe
As one can see while (a) is reduced, diagram (b) is reducible via triangle move. Once we have a set of reduced graphs, we can further classify them by the following rule: 
\begin{itemize}
\item Two reduced graphss are equivalent if and only if they define the same decorated permutation. 
\end{itemize}
All inequivalent medial graphs now correspond to distinct cells in $OG_{k+}$.

\subsubsection{Going to the boundary: removable vertices and Eulerian poset}
For a given cell in $OG_{k+}$ we would like to consider its co-dimension one boundaries. These are associated with one extra linear dependency condition among consecutive columns in the $C$-matrix. The boundaries can be read off from the medial graph via the opening of the four-point vertices~\cite{NimaBigBook}:
\eq
\includegraphics[scale=0.4]{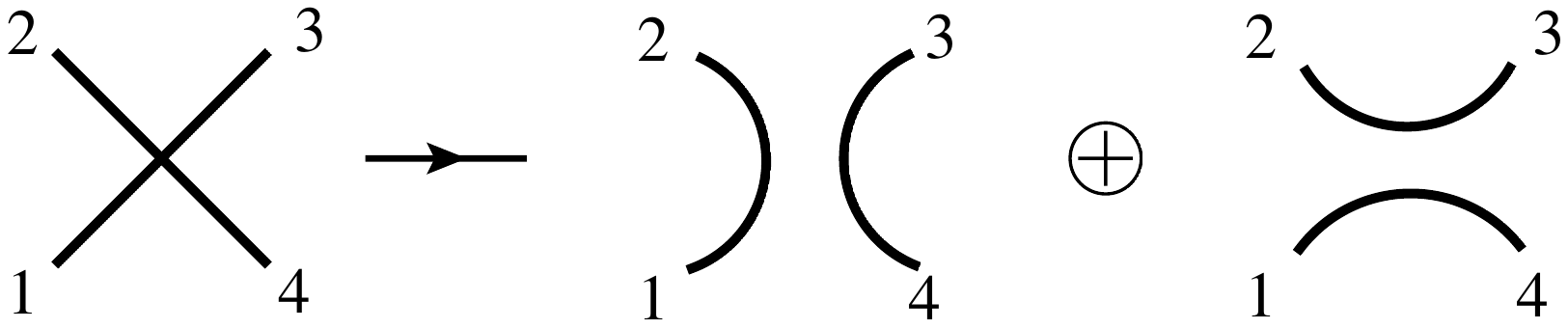}
\eqe
If the resulting graph is also reduced they are cells which represent the co-dimension one boundaries. Through this procedure, we obtain a  poset, i.e. we have $A>B$ if one can obtain the reduced graph corresponding to B from opening vertices in $A$. 

Remarkably the positroid stratification of the $OG_{k+}$ forms an Euler poset, i.e. the number of even-dimensional positroids are one more than the number of odd-dimensional ones. For example consider $OG_{2+}$, which as a one-dimensional top cell with 2 zero-dimensional boundaries, giving $2-1=1$. Let us consider $OG_{3+}$, starting with the 3-dimensional top-cell, we have:
\eq
-1+3-6+5=1
\eqe
Since each stratification can be represented by an on-shell diagram, we can illustrate the enumeration of the elements in each dimension as follows:
\begin{equation}
\includegraphics[scale=0.7]{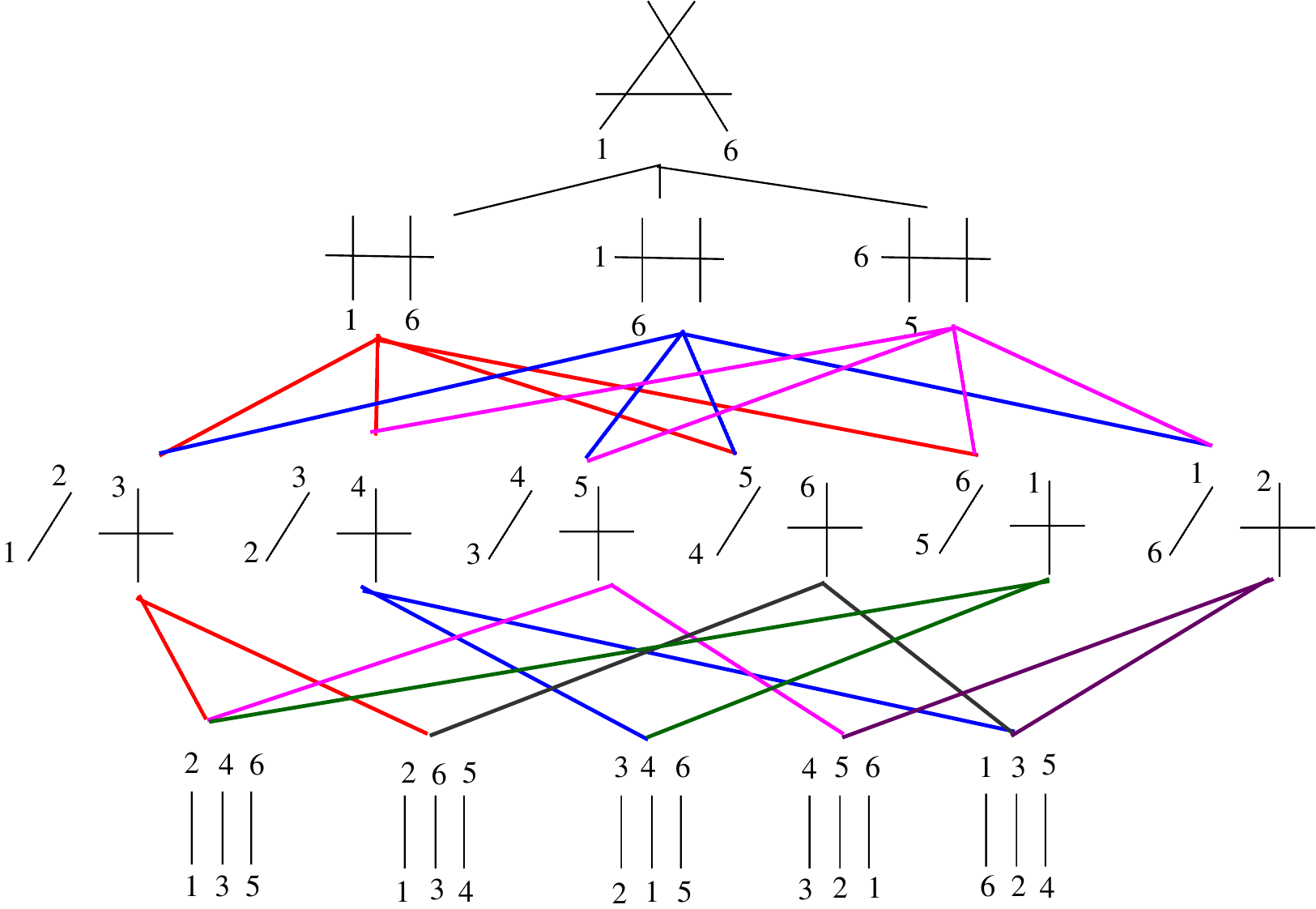}\,.
\end{equation}
Note that as the face-lattice of a convex polytope forms a Eulerian poset, the above observation suggests that the cell stratification of $OG_{k+}$ is combinatorially a polytope. Recently a new way of enumeration distinct cells for $OG_{k+}$ was developed in~\cite{SangminNew}, and a closed form of a generating function for the enumeration starting from a top cell was found. Use this result, the authors have explicitly shown that the graded poset of all top cells in $OG_{k+}$ are indeed Eulerian.

However, this is not the end of the story. In fact if we begin in the middle of some stratification, one still obtains an Eulerian poset. For example
\eq
\vcenter{\hbox{\includegraphics[scale=0.4]{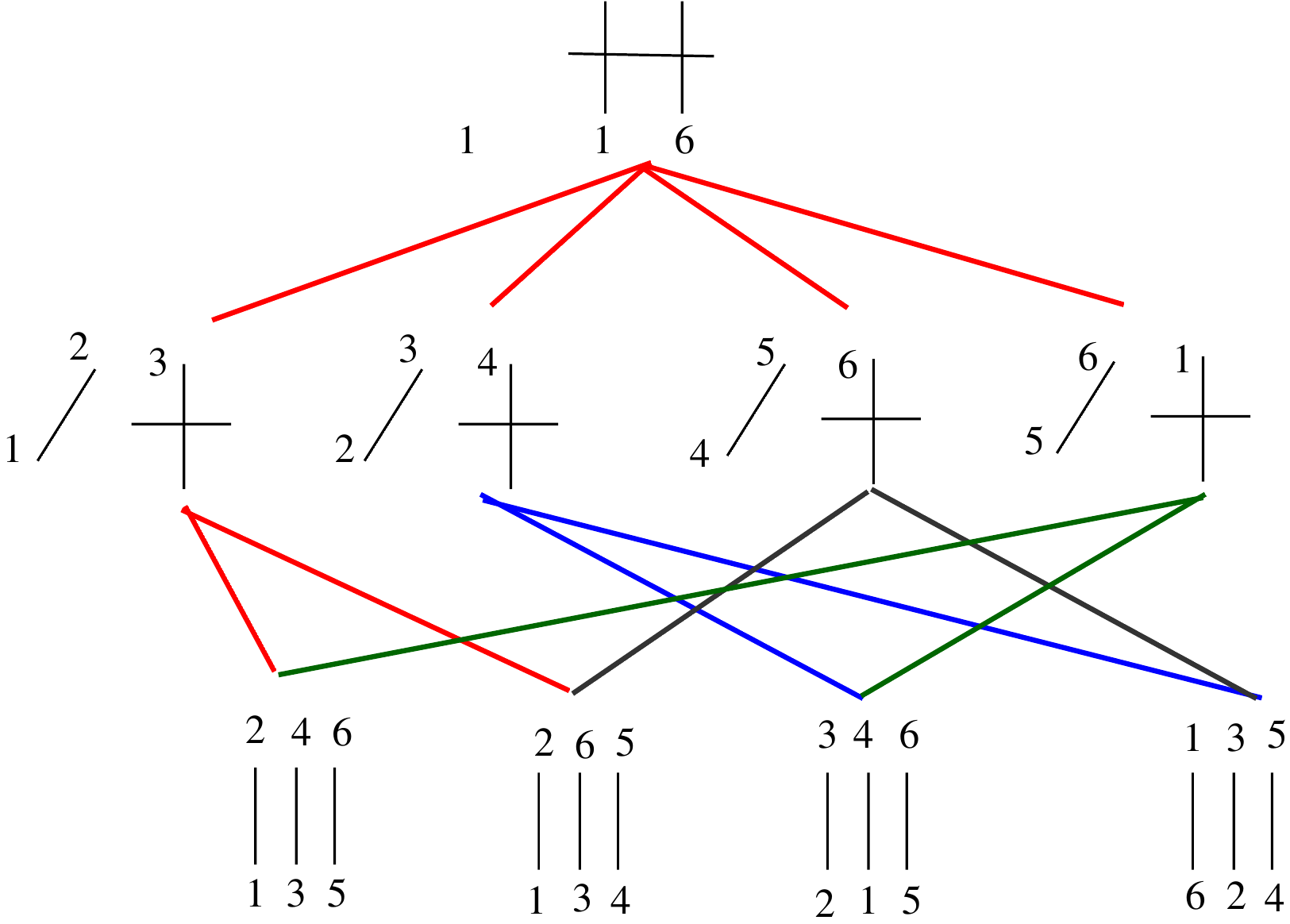}}}\quad\rightarrow\quad 1-4+4=1
\eqe
As a further non-trivial example, let us consider the stratification that corresponds to the permutation path $(16)\,(25)\,(38)\,(47)$, namely $1 \rightarrow 6$ and $6 \rightarrow 1$, etc. We list the elements of each stratification by their permutation paths
\eq
\begin{tabular}{c|c|c}
\hline
   dimensions & cell  & multiplicity \\
   4 &  (16)(25)(38)(47)  &1 \\\hline 
  3 &  (23)(47)(58)(16),\,(45)(16)(27)(38),\,(18)(25)(36)(47),\,(67)(38)(14)(25) &\;   \\
  \;&(28)(16)(47)(35),\,(24)(38)(16)(57),\,(17)(38)(25)(46),\,(13)(25)(47)(68)&8\\\hline 
   2 &   (23)(48)(57)(16),\,(23)(68)(47)(15),\,(23)(17)(58)(46),\,(45)(26)(38)(17)  &\;  \\
   \; &  (45)(13)(27)(68),\,(45)(16)(28)(37),\,(18)(26)(47)(35),\,(18)(57)(36)(24) &\;   \\
  \;&(18)(46)(25)(37),\,(67)(25)(48)(13),\,(67)(35)(14)(28),\,(67)(24)(38)(15) &\;\\
  \;&(12)(35)(47)(68),\,(28)(17)(35)(46),\,(34)(28)(16)(57),\,(13)(24)(57)(68) &\;\\
    \;&(56)(24)(38)(17),\,(78)(46)(25)(13)&18\\\hline
1 &   (35)(18)(46)(57),\,(23)(14)(57)(68),\,(23)(56)(17)(48),\,(23)(67)(48)(15)  &\;  \\
   \; & (45)(23)(17)(68),\,(45)(36)(28)(17),\,(45)(78)(26)(13),\,(45)(18)(26)(37)  &\;   \\
  \;&(18)(67)(24)(35),\,(18)(27)(35)(46),\,(18)(34)(28)(57),\,(67)(58)(13)(24) &\;\\
  \;&(67)(45)(28)(13),\,(67)(12)(35)(48),\,(23)(78)(46)(15),\,(45)(12)(37)(68) &\;\\
  \;&(18)(56)(24)(37),\,(12)(78)(35)(46),\,(12)(34)(68)(57),\,(34)(56)(17)(28) &\;\\
    \;&(34)(67)(15)(28),\,(56)(78)(13)(24)&22\\\hline
    0 &   (23)(18)(45)(67),\,(23)(18)(56)(47),\,(23)(14)(56)(78),\,(23)(14)(67)(58)  &\;  \\
   \; &  (45)(23)(78)(16),\,(45)(36)(78)(12),\,(45)(36)(18)(27),\,(18)(67)(34)(25) &\;   \\
  \;&(18)(27)(34)(56),\,(67)(58)(12)(34),\,(67)(45)(12)(38),\,(12)(34)(56)(78) &12\\
 \end{tabular}
\eqe
Thus one indeed finds $1-8+18-22+12=1$. For all the examples we have checked this is always the case, it leads us to conjecture that all the cells in $OG_{k+}$ should be Eulerian. It would be of great interest to generalize the result of~\cite{SangminNew} to find the generating function starting from a general middle cell.

\subsection{Integrability in $(2+1)$ dimensions} \label{integrability}
Finally, we would like to answer the question we raised earlier: what is so special about the parametrization of the top cell in $Gr(3,6)$ that is given by the embedding of $OG_{3}$? To answer this question, let's consider the $(2+1)$-dimensional analog of the factorizability condition of scattering amplitudes for particles in $(1+1)$-dimensions: the Yang-Baxter equation. A suitable generalization was proposed long ago by Zamolodchikov~\cite{ZZ1,ZZ2}, who considered the scattering of infinite straight strings in $(2+1)$-dimensions. Instead of the fundamental $2\rightarrow2$ S-matrix in $(1+1)$-dimensions, the first nontrivial amplitude is the $3\rightarrow3$ process for three strings $(a,b,c)$ illustrated in the following diagrams:
\eq
\label{3StringScattering}
\includegraphics[scale=0.8]{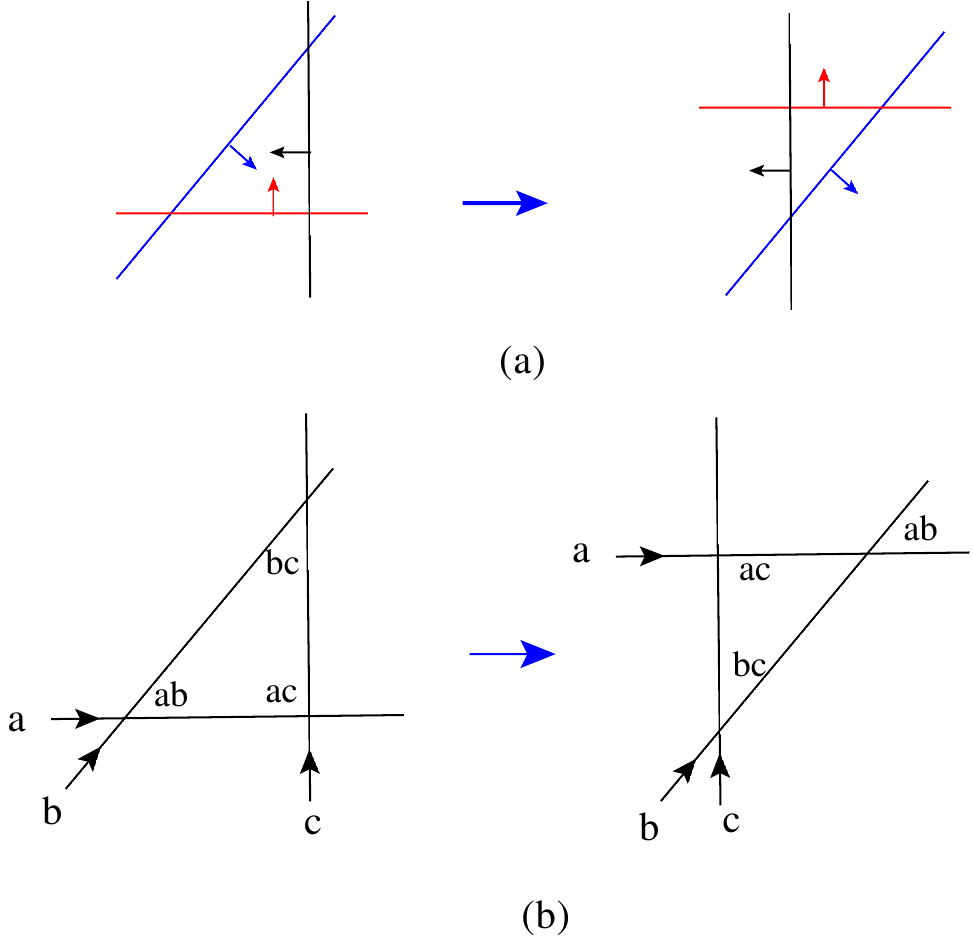}
\eqe
Diagram $(a)$ denotes the scattering process. The configuration of the three strings can be characterized by three angles $(\theta_{ab}, \theta_{ac}, \theta_{bc})$. The angles are defined by considering the strings as intersection of three infinite planes, and the angles are associated with the inner product of the normal vectors $n_i$, of each plane:
\eq
n_a\cdot n_{b}=-\cos\theta_{ab},\quad n_a\cdot n_{c}=-\cos\theta_{ac},\quad n_{c}\cdot n_{b}=-\cos\theta_{cb}\,.
\eqe
The scattering amplitude is then a map between the initial $| \theta'_{ab},\theta'_{ac},\theta'_{bc}\rangle$ and final state $| \theta_{ab},\theta_{ac},\theta_{bc}\rangle$:
\eq
| \theta'_{ab},\theta'_{ac},\theta'_{bc}\rangle=R_{abc}(\theta_{ab},\theta_{ac},\theta_{bc})|\theta_{ab},\theta_{ac},\theta_{bc}\rangle\,.
\eqe 
A diagrammatic representation of the scattering process is to assign a positive direction to each of the string as shown in diagram $(b)$ of eq.(\ref{3StringScattering}). Denoting each vertex by the two strings that intersect, before scattering we move from vertex $(ab)$ to $(ac)$ to $(bc)$ along the positive direction of each string, while after the scattering the same path moves along the negative direction of each string.  

Factorizability of the three-dimensional scattering process is then encoded in the following \textit{Tetrahedron equation}\footnote{We thank Thomas Lam for pointing out the possible connection between ABJM on-shell diagram and eq.(\ref{Tetra}) }
\eq
R_{acb}R_{abd}R_{acd}R_{bcd}=R_{bcd}R_{acd}R_{abd}R_{acb} \, .
\label{Tetra}
\eqe
This equation represent the equivalence of two distinct sequences of 4 (three-string)scattering process that results in the same final string configuration. Diagramatically the scattering process is represented as:
\eq
\includegraphics[scale=0.7]{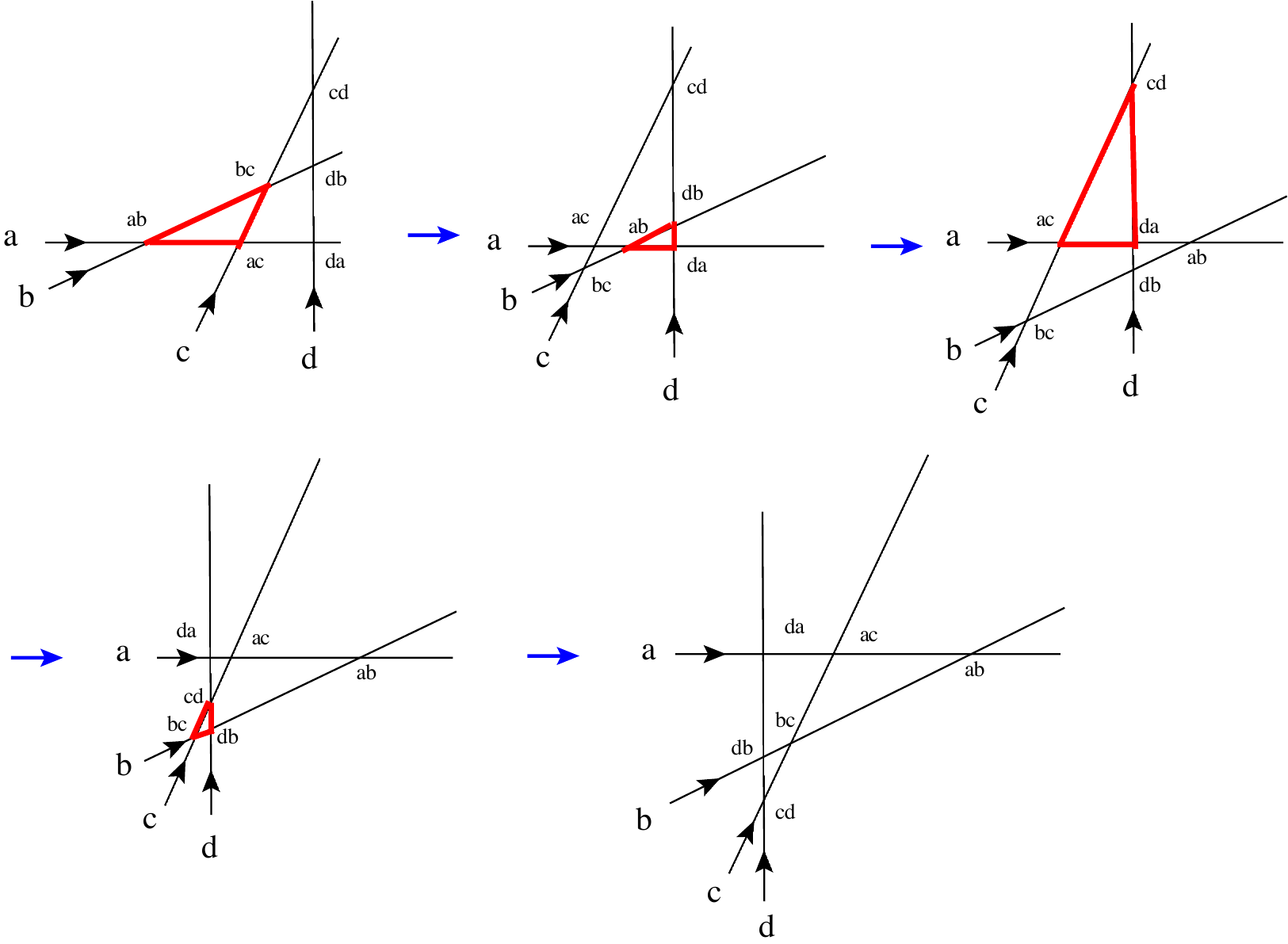}\,,
\eqe
and 
\eq
\includegraphics[scale=0.7]{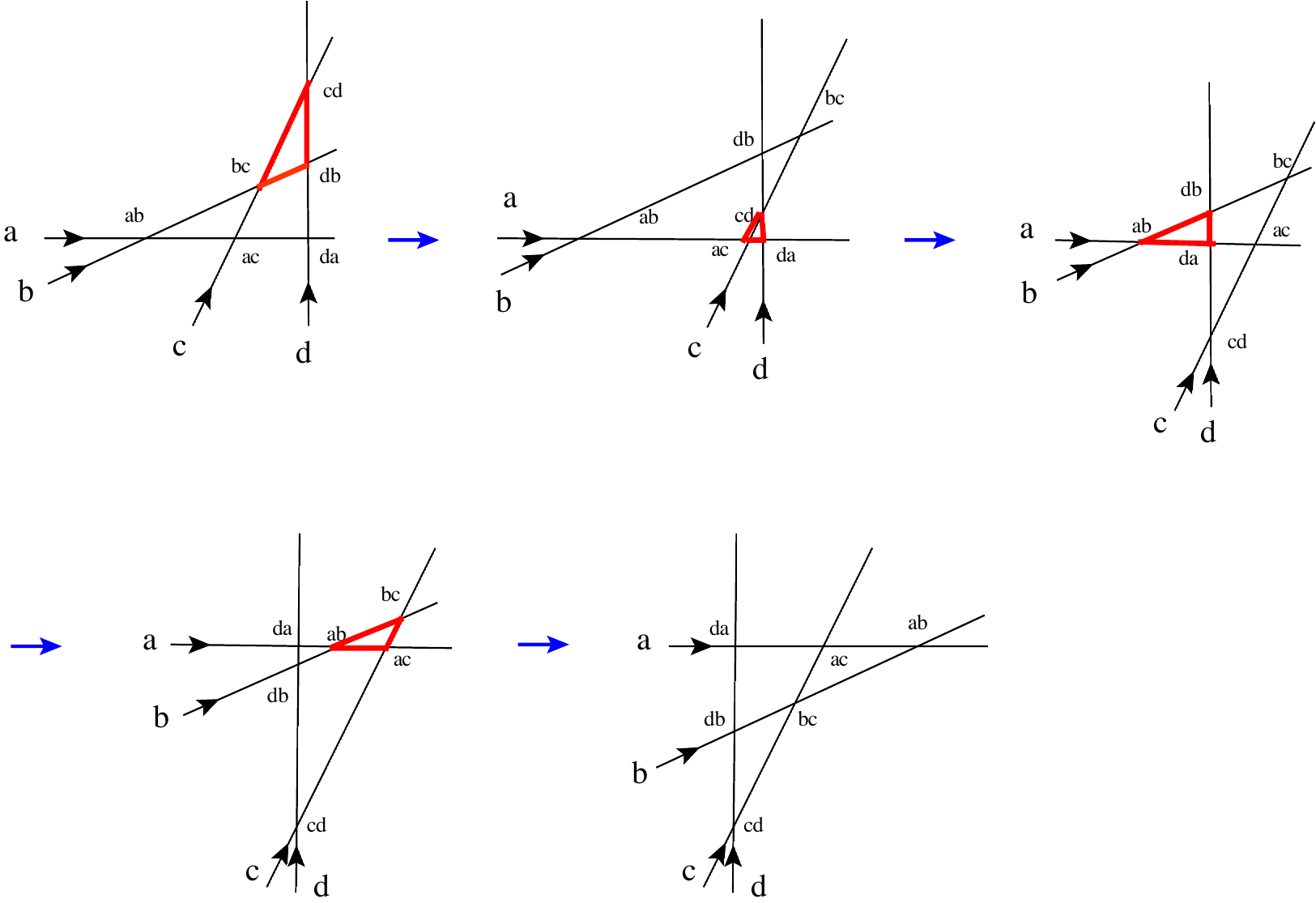}\,.
\eqe

A well-known solution to eq.(\ref{Tetra}) is the functional map between two equivalent networks in electric network~\cite{CIM, dVGV}. Two electric devices are the same between the ``star" and ``triangle" configuration (the Y-$\Delta$ transformation):
\eq
\includegraphics[scale=0.6]{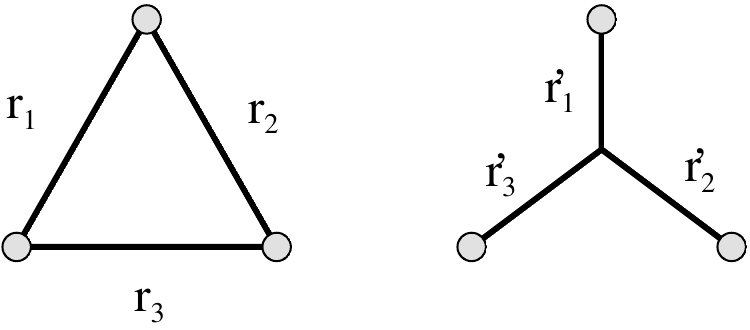}
\eqe
if their resistance satisfies:
\eq
r'_1r'_2+r'_3r'_2+r'_1r'_3=\frac{r_1r_2r_3}{r_1+r_2+r_3}\,.
\eqe
A nontrivial solution is given as: 
\begin{equation}
\label{StarTriMap}
R: \quad (r_1,r_2,r_3)\quad \rightarrow \quad \left({r_3 r_2 \over r_1+r_2+r_3},{r_1 r_3 \over r_1+r_2+r_3},{r_3 r_2 \over r_1+r_2+r_3}\right)\, .
\end{equation}
The transformation can be shown to satisfy the standard tetrahedron equation, eq.(\ref{Tetra}), by making following identification, 
\eqa
\theta_1 &=& -{1 \over r_1}\, , \quad \theta_2 = r_2 \, , \quad  \theta_3  =  -{1 \over r_3} \, ,   \cr 
\theta'_1 &=& r'_1 \, , \quad  \theta'_2 = -{ 1 \over r'_2} \, ,  \quad \theta'_3 = r'_3 \, .
\eqae
It then leads to a map from $\theta_i$ to $\theta'_i$ given by
\begin{equation}
\label{StarTriMap2}
R: \quad (\theta_1,\theta_2,\theta_3)\,\,\, \rightarrow \,\,\, \left({\theta_1 \theta_2 \over \theta_1+\theta_3 - \theta_1 \theta_2 \theta_3}, \theta_1+\theta_3 - \theta_1 \theta_2 \theta_3,{\theta_2 \theta_3 \over \theta_1+\theta_3 - \theta_1 \theta_2 \theta_3}\right)\, .
\end{equation}
It is straightforward to check the map given above indeed satisfies eq.(\ref{Tetra}).

A general class of solutions to eq.(\ref{Tetra}) was found by Kashaev, Korepanov and Sergeev~\cite{TetraSol} by considering the map that arises from the solution to the following equation: 
\eqa\label{BlockId}
\nonumber\left(\begin{array}{ccc} A_{1} & B_{1} & 0 \\C_{1} & D_{1} & 0 \\0 & 0 & 1\end{array}\right)\left(\begin{array}{ccc} A_{2} & 0 & B_{2} \\0 & 1 & 0 \\C_{2} & 0 & D_{2}\end{array}\right)\left(\begin{array}{ccc}1 & 0 & 0 \\0 & A_{3} & B_{3} \\0 & C_{3} & D_{3}\end{array}\right)=\left(\begin{array}{ccc}1 & 0 & 0 \\0 & A'_{3} & B'_{3} \\0 & C'_{3} & D'_{3}\end{array}\right)\left(\begin{array}{ccc} A'_{2} & 0 & B'_{2} \\0 & 1 & 0 \\C'_{2} & 0 & D'_{2}\end{array}\right)\left(\begin{array}{ccc} A'_{1} & B'_{1} & 0 \\C'_{1} & D'_{1} & 0 \\0 & 0 & 1\end{array}\right)\\
\eqae
where each element in the block $2\times2$ matrix is a function of one parameter, i.e. $A_i(\theta_i)$, $B_i(\theta_i)$, $C_i(\theta_i)$ and $D_i(\theta_i)$. It was shown in~\cite{TetraSol}, and references there in that given a solution to eq.(\ref{BlockId}), the corresponding map $R:\; \theta_i\rightarrow \theta'_i$ will then satisfy the tetrahedron identity in eq.(\ref{Tetra}). It is known that using the canonical gauge in eq.(\ref{CanonicalGauge}), the triangle diagram for the non-trivial part of $OG_{3+}$ ($OG_{3+}$ without the $3\times3$ the unity matrix) can be written as products of $3\times3$ matrices in the form eq.(\ref{BlockId})~\cite{HW},\footnote{If the bilinear $\eta^{ij}$ takes all $+$ signature, then the $3\times3$ matrices are simply rotation matrices, corresponding to the Euler decomposition of the three-dimensional rotation group SO(3).} with,
\eq \label{ABCD}
\left(\begin{array}{cc}A_i(\theta_i) & B_i(\theta_i) \\C_i(\theta_i) & D_i(\theta_i)\end{array}\right)=\left(\begin{array}{cc}-\cot(\theta_i) & -\csc(\theta_i) \\ \csc(\theta_i) & \cot(\theta_i) \end{array}\right)\,.
\eqe 
In other words, the coordinate transformation associated with the triangle move satisfies the tetrahedron identity in eq.(\ref{Tetra})! The corresponding medial graphs related by the triangle move are shown below
\eq
\vcenter{\hbox{\includegraphics[scale=0.6]{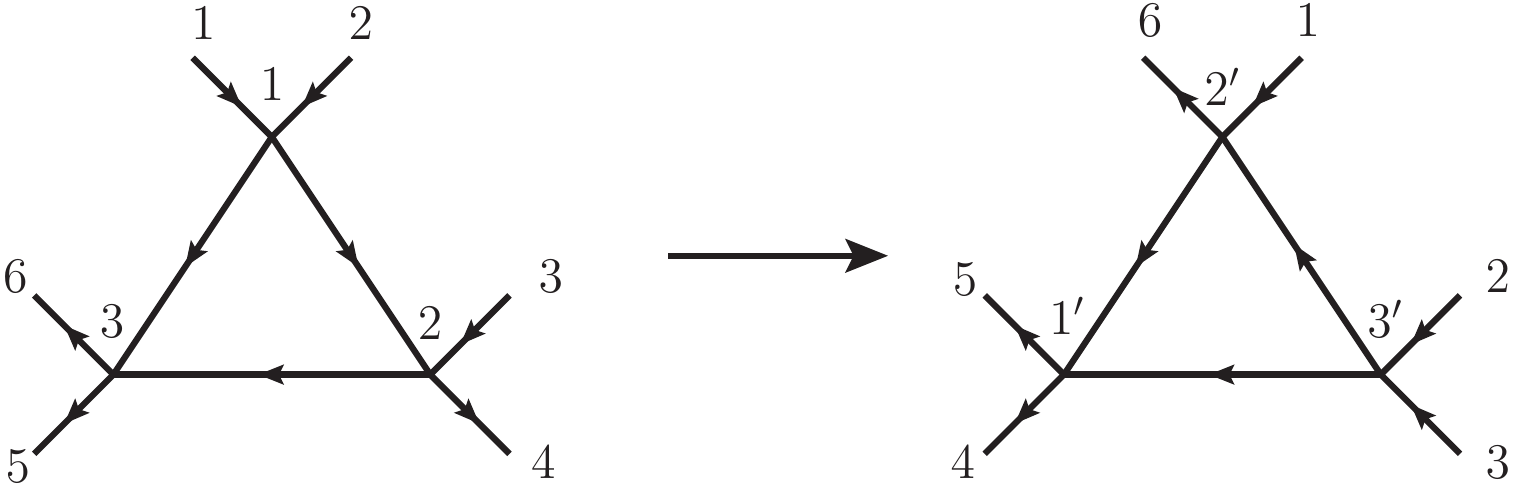}}}\, .
\eqe
Solve eq.(\ref{BlockId}) with $A_i(\theta_i), B_i(\theta_i), C_i(\theta_i)$ and $ D_i(\theta_i)$ given in eq.(\ref{ABCD}), we obtain the triangle-move transformation of adjacent gauge
\eq \label{triangletransform2}
R: \quad (c_1,s_2,c_3) \rightarrow 
\left(\frac{c_1 c_2 s_3 }{c_1 + s_2c_3}, 
-\frac{s_1 s_2 s_3}{s_2 + c_1 c_3}, 
\frac{s_1 c_2 c_3}{c_3 + c_1s_2}\right) \, .
\eqe 
Indeed the above map induces a map of $\theta_i \rightarrow \theta'_i$, 
\eq 
(\theta_1, \theta_2, \theta_3) \rightarrow 
\left( -{\rm ArcCos} \left(\frac{c_1 c_2 s_3 }{c_1 + s_2c_3} \right),
-{\rm ArcSin} \left(\frac{s_1 s_2 s_3}{s_2 + c_1 c_3} \right),
-{\rm ArcCos} \left( \frac{s_1 c_2 c_3}{c_3 + c_1s_2} \right)    \right)
\eqe
which can be checked to satisfy the tetrahedron equation in the positive region, namely for $\theta_i \in (0, \pi/2) $. As for the cyclic gauge~eq.(\ref{AlterGauge}), its corresponding transformation, eq.(\ref{triangletransform}), can be made to be the same as eq.(\ref{triangletransform2}) by some trivial identifications,
\eq 
\{s'_2 \rightarrow -s'_2 \, ,c_2 \leftrightarrow s_2\, ,  s'_1 \leftrightarrow c'_1 \, ,   s'_3 \leftrightarrow c'_3 \} \, ,
\eqe
and with the rest untouched. So with the above identifications the triangle-move transformation of the cyclic gauge satisfies the tetrahedron equation as well, although, interestingly, the Grassmannian of this gauge cannot be decomposed into a factorization form as eq.(\ref{BlockId}). 

Thus we see that while a generic one-dimensional subspace of $G_{2,4+}$ allows us to define a map $R_{123}(\theta_1,\theta_2,\theta_3)$ through the triangle equivalence. What is special about the subspace defined by $OG_{2+}$ is that the corresponding map also defines a S-matrix in $(2+1)$-dimensional describing the scattering of straight strings that is integrable! Note that the diagrams having this interpretation correspond to ones where no two lines intersect twice which implies all such diagrams are reduced. Since these diagrams correspond to three-string scatterings, there must be a total of $k(k-1)(k-2)/6$ number of triangles within a such diagram.

\section{Scattering amplitudes of ABJM theory} \label{section:Scattering}
The combinatorial aspects of the network of positive orthogonal Grassmannian is closely related to the on-shell diagram (leading singularity) of ABJM theory. In fact, the network using four-point vertex is exactly the on-shell diagram of ABJM theory. Indeed the boundary measurement associated with each four-vertex trivalent graph can be obtained via:
\eq
\left(\prod_{j\in I_{dia}}\int d^{2|3}\Lambda_j\right) \left(\prod_{i\in v_{dia}}\int d\log\tan_i \delta^{2|3}\left(C(\alpha_i)\cdot\Lambda\right)\right)
\eqe
where one integrates the over all $\Lambda_j$ associated with internal edges $I_{dia}$, with the integrand being constructed by the product of fundamental $OG_2$ integrals for each vertex. Note that while the boundary measurements are the same irrespective to the degree of delta function one has. However, in order for the integration measure to remain invariant under the equivalence moves, it is necessary for the degree of fermonic delta functions to be \textit{greater} than the bosonic delta function by 1. This fixes the degree 3 for the fermionic delta function. Thus the combinatorics of the positive orthogonal Grassmannian is naturally associated with a three-dimensional theory with $\mathcal{N}=6$ supersymmetry.  An immediate consequence of this identification is:
\begin{itemize}
\item As the boundary of the positive-cells both for $Gr(k,2k)$ and $OG_k$ are configurations where adjacent columns become linearly dependent, any on-shell diagram which is not a top-cell can be interpreted as the residue of an integral over the top-cell, localized on the zeroes of consecutive minors. The residue of an integral over $OG_k$ localized on consecutive minors are known to be the leading singularities of ABJM theory~\cite{LeeOG}.
\item After localizing the internal spinor integrals $\int \lambda_I$, the total number of bosonic delta functions for a given graph is $2k-3$. Thus the leading singularities are one to one correspondent to the positive cells of $OG_{k+}$ which are $2k-3$-dimensional. 
 
\end{itemize}

However, it was long-known that the leading singularities for ABJM theory come in pairs~\cite{Gang}, due to the fact that the propagator can be expressed as a quadratic equation in the BCFW deformation parameter.\footnote{Note that while the leading singularities of $\mathcal{N}=4$ SYM also comes in pair, they are trivially related by parity conjugation} This appears to contradict the discussion so far that each leading singularity is associated with a cell in $OG_{k+}$. This contradiction disappears once one realizes that there are two disconnected spaces of $OG_k$: $OG_{k+}$ and $OG_{k-}$. Recall that $OG_{k+}$ is defined by ${M_{\mathsf{ I}}/M_{\bar{\mathsf{ I}}}}=1$. Not surprisingly, $OG_{k-}$ can be defined by ${M_{\mathsf{ I}}/M_{\bar{\mathsf{ I}}}}=-1$. In the following we will use the tree-level recursion to demonstrate how the amplitude is given by the sum of cells in OG$_{k+}$ and OG$_{k-}$.

\subsection{ABJM amplitudes as sum of OG$_{k+}$ and OG$_{k-}$}
Here, we give a brief review to how the ABJM amplitudes can be given by the on-shell diagrams. The four-point amplitude of ABJM is given by the an integral over OG$_2$. The two distinct amplitudes are given as:
\eqa
\nonumber \mathcal{A}_4(\bar{1}2\bar{3}4)=\int \frac{d^{2\times4}C}{(12)(23)}\delta^{3}(C\cdot C^T)\delta^{2\times 2|3}(C\cdot\Lambda)\\
\mathcal{A}_4(1\bar{2}3\bar{4})=\int \frac{d^{2\times4}C}{(23)(34)}\delta^{3}(C\cdot C^T)\delta^{2\times 2|3}(C\cdot\Lambda)\,.
\label{4ptDef}
\eqae
As discussed in \cite{HW}, the above integral can be rewritten as a \textit{sum or difference the two different branches} in OG$_2$. To see this, recall that due to the orthogonal constraint we have:
\eq
OG_{2+}:\quad (12)=(34), \quad OG_{2-}:\quad (12)=-(34) \,.
\eqe
Thus on the positive branch, $\mathcal{A}_4(\bar{1}2\bar{3}4)=\mathcal{A}_4(1\bar{2}3\bar{4})$ where as on the negative branch  $\mathcal{A}_4(\bar{1}2\bar{3}4)=-\mathcal{A}_4(1\bar{2}3\bar{4})$. Thus if $\mathcal{A}_4(\bar{1}2\bar{3}4)$ is given by the sum of the two branches, $\mathcal{A}_4(1\bar{2}3\bar{4})$ would be given by the difference. Starting from 
\begin{equation}
C=\begin{pmatrix} 
1& a &0&b\\
0 &c &1&d\,.
\end{pmatrix}
\end{equation}
explicitly solving the orthogonal constraints in eq.(\ref{4ptDef}), we find that\footnote{In solving the delta functions, we treat $\delta(ax)=\oint dx\frac{1}{ax}=\frac{1}{a}\oint dx\frac{1}{x}$, to avoid any absolute values on the Jacobian factors. } 
\eqa
\nonumber \mathcal{A}_4(\bar{1}2\bar{3}4)&=&\sum_{\alpha=\pm} \int \frac{d\theta}{2cs}~~ \delta^{2\times 2|3}(C\cdot\Lambda)\\
\mathcal{A}_4(1\bar{2}3\bar{4})&=&\sum_{\alpha=\pm} \alpha\int \frac{d\theta}{2cs}~~ \delta^{2\times 2|3}(C\cdot\Lambda)\,,
\label{4ptCyclic}
\eqae
where now the $OG_{2}$ matrix is given by 
\begin{equation}
C=\begin{pmatrix} 
1&  c &0& -\alpha s\\
0 & s &1&\alpha c\,
\end{pmatrix}\,.
\label{AdjacentC}
\end{equation}
The form of the four-point amplitude in eq.(\ref{4ptCyclic}) can also be found in \cite{Beisert1Loop}. The parameter $\alpha$ takes value in $+1,-1$ and determines the branch of $OG_{2}$. Thus we see that while $\mathcal{A}_4(\bar{1}2\bar{3}4)$ is given by the sum of the two branches, and $\mathcal{A}_4(1\bar{2}3\bar{4})$ is given by the difference.

As discussed in~\cite{HW}, the two branches in the orthogonal Grassmannian is intricately related to a unique feature of three-dimensional kinematics. At four-points, through momentum conservation we can identify 
\eq
\langle 12\rangle=\pm \langle 34\rangle\,.
\label{KinConfig}
\eqe
Note that $+$ and $-$ are inequivalent kinematic configurations. The relative signs for the remaining configurations are given by the following identity as a result of momentum conservation:
\eq
\frac{\langle 12\rangle}{\langle 34\rangle}=\frac{\langle 14\rangle}{\langle 23\rangle}=\frac{\langle 42\rangle}{\langle 13\rangle}\,.
\eqe
In fact, the two distinct kinematic configurations in eq.(\ref{KinConfig}) can be mapped to the two branches in OG$_2$. Indeed from eq.(\ref{AdjacentC}), we can see that the bosonic delta functions $\delta^2(C\cdot\lambda)$ implies that:
\eq
\left\{\begin{array}{c}\langle 41\rangle+ c\langle 42\rangle=0  \\ \langle 23\rangle+\alpha c\langle 24\rangle=0 \end{array}\right.\Rightarrow \quad \frac{\langle14\rangle}{\langle23\rangle}=1/\alpha
\eqe
Thus the $+$ and $-$ of OG$_2$ precisely corresponds to the  $+$ and $-$ of eq.(\ref{KinConfig}). We can denote the two branches graphically as:
\eq
\includegraphics[scale=0.5]{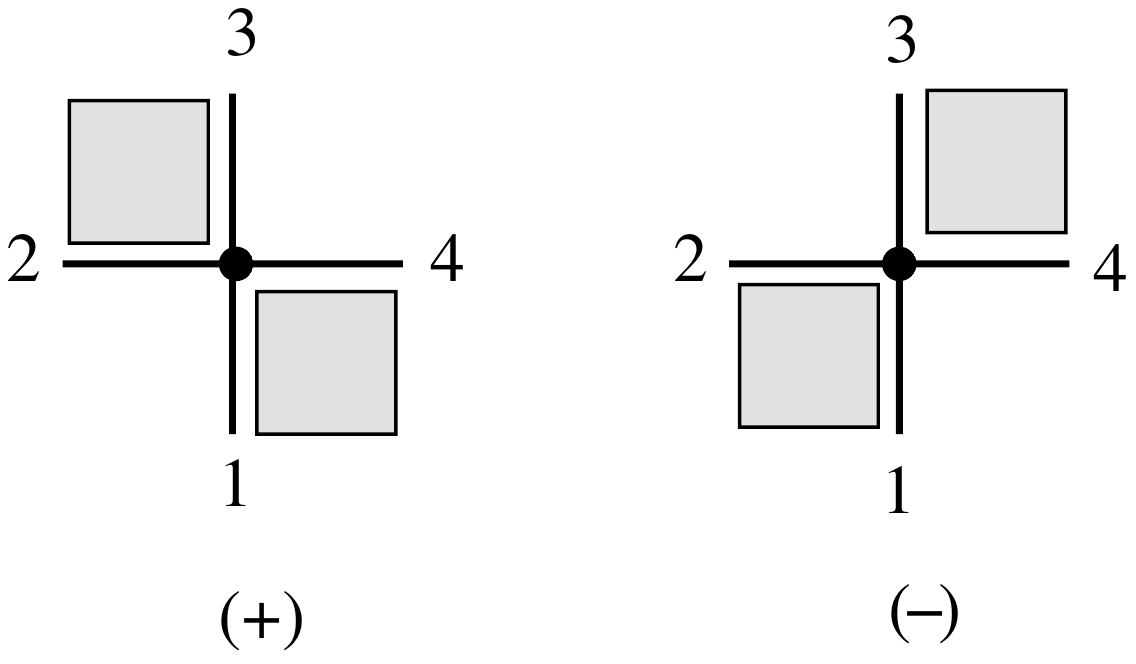}\,,
\label{4ptBranchFig}
\eqe
where the shaded region indicates the kinematic invariant that is constructed clockwise out of the two legs that bound the region are equivalent up to a $-$ sign, where as the unshaded region indicates the opposite, i.e
\eq
(+): \quad \langle 12\rangle= \langle 34\rangle, \;\;\langle 23\rangle= -\langle 41\rangle,\quad (-): \quad \langle 12\rangle= -\langle 34\rangle, \;\;\langle 23\rangle= \langle 41\rangle\,.
\eqe

The BCFW recursion construction of higher-point tree-level amplitudes in ABJM theory~\cite{Gang}, can be mapped into the gluing of the fundamental $OG_2$'s into $OG_k$~\cite{NimaBigBook,HW}. This is schematically represented as 
$$\includegraphics[scale=0.6]{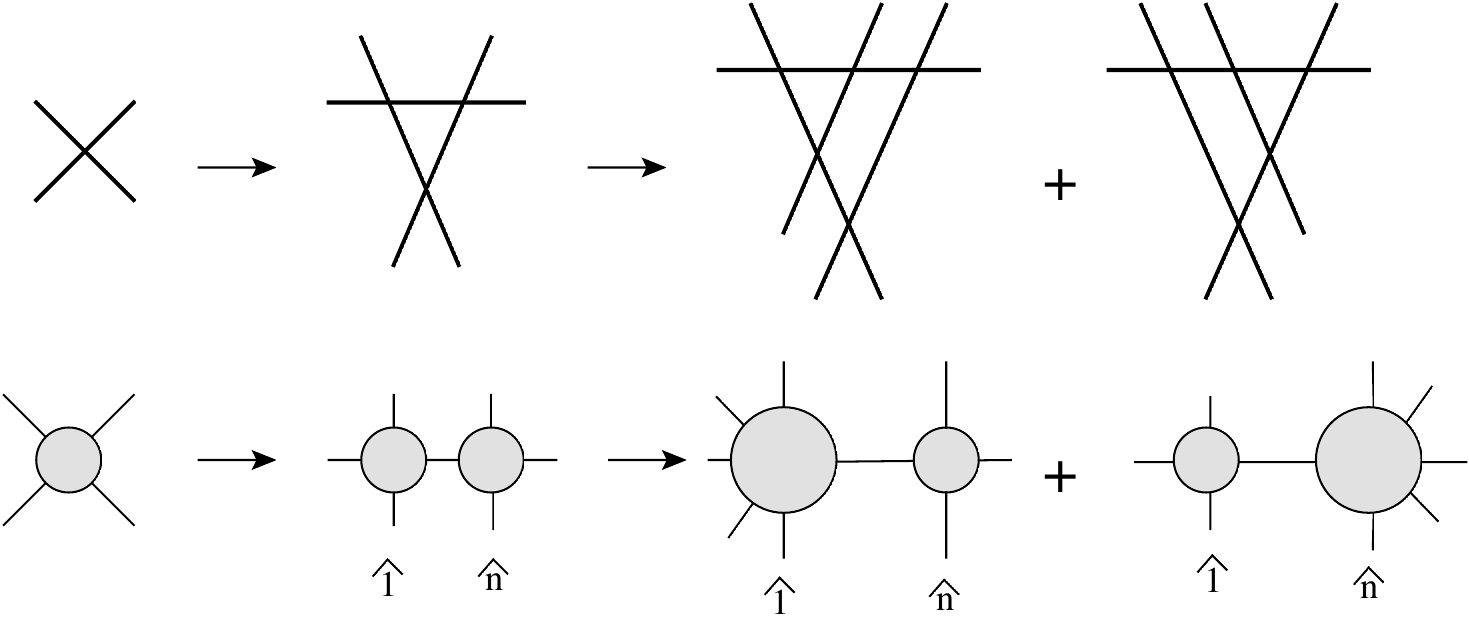}\,.$$
In other words, each term in the BCFW expansion can be represented as a sum of on-shell diagrams. Note that since each term in the BCFW expansion is a rational function, this implies that the on-shell diagrams must be of dimension $2k-3$, such that the bosonic delta functions associated with each diagram, i.e. $\delta^{2k}(C\cdot \lambda)$, completely localizes these degree of freedom. 

However, this cannot be the whole story since we have not indicated which branch each the individual vertex should take value in. Naively, one would just sum over all possible local branches resulting in a sum of $2^{n_v}$ number of terms, where $n_v$ is the number of vertices in a given on-shell diagram. On the other hand, by now we are familiar with the fact that distinct configurations in OG$_{k+}$ can be faithfully represented by on-shell diagrams that are inequivalent under various equivalent moves. Thus it is impossible for a given on-shell diagram to represent $2^{n_v}$ number of distinct terms. In fact, for any on-shell diagram, one can only have two distinct configurations, OG$_{k+}$ and OG$_{k-}$. Thus half of the $2^{n_v}$ terms that correspond to the same branch must be equivalent to each other via redefinition. Indeed this is the case. Thus summing over all $2^{n_v}$ terms is equivalent to summing over arbitrary two representative of each branch. 

Just as the four-point amplitude, the above discussion is also reflected in the branches of kinematics. It is instructive to consider an example beyond four-point, let us merge two vertices and consider all four possible combination of the local OG$_2$ branches:
$$\includegraphics[scale=0.5]{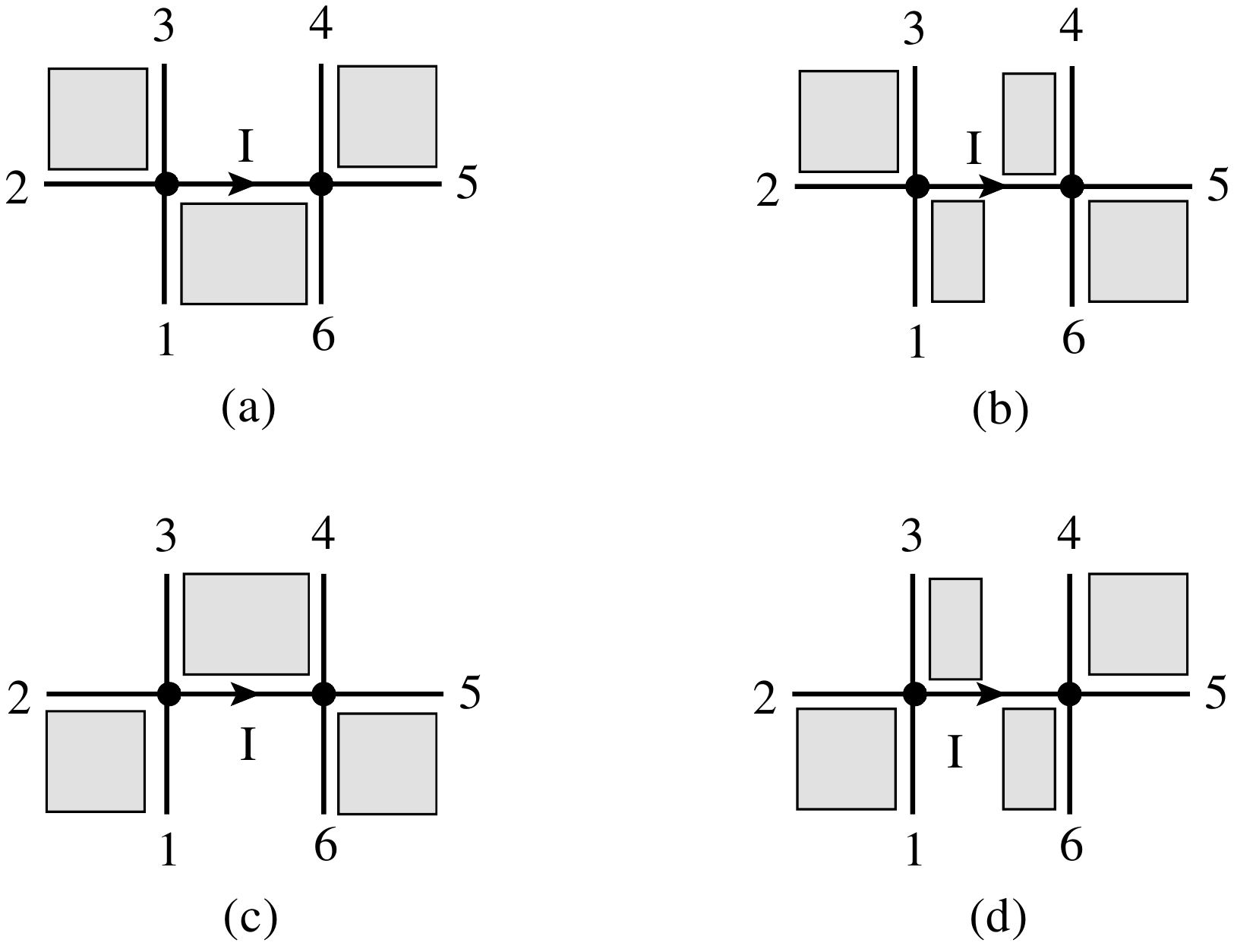}\,,$$
Configuration $(a)$ and $(c)$ correspond to two-dimensional cells in OG$_{3+}$, where as $(b)$ and $(d)$ are cells in OG$_{3-}$. Instead of showing the equivalence of the two pairs, it is more instructive to show that each pair correspond to the same branch for the external data. Indeed using the prescription in fig.(\ref{4ptBranchFig}) one can straightforwardly deduce that the four configurations imply:
\eq
(a),(c):\quad i\langle 3|p_5+p_6|4\rangle=\langle 12\rangle\langle 56\rangle,\quad (b),(d):\quad i\langle 3|p_5+p_6|4\rangle=-\langle 12\rangle\langle 56\rangle
\eqe
The right hand side are precisely the two distinct six-point kinematic configuration in the factorization limit as can be seen using the following identity:
\eq
\langle i|p_j+p_k| l\rangle^2+(p_j+p_k+p_l+p_i)^2(p_j+p_k)^2=(p_i+p_j+p_k)^2(p_j+p_k+p_l)^2\,.
\eqe
Next, we consider the diagram with closed loop:
$$\includegraphics[scale=0.6]{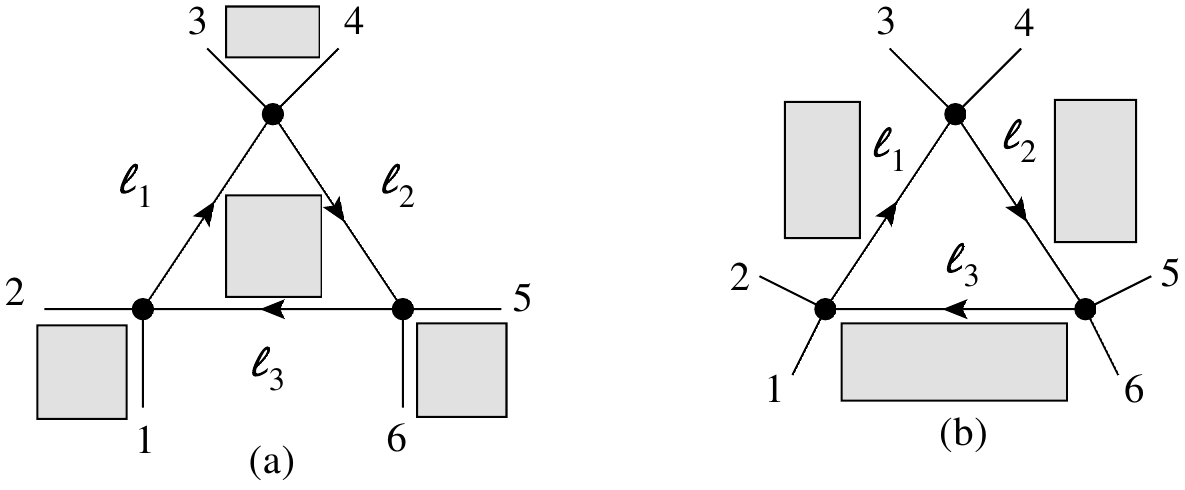}$$
It is straightforward to show that the two configurations imply:
\eq
(a): \langle 12\rangle \langle 34\rangle \langle 56\rangle = i \langle \ell_3\ell_1\rangle \langle \ell_1\ell_2\rangle \langle \ell_2\ell_3\rangle,\quad\quad (b): \langle 12\rangle \langle 34\rangle \langle 56\rangle = -i \langle \ell_3\ell_1\rangle \langle \ell_1\ell_2\rangle \langle \ell_2\ell_3\rangle
\eqe
where we've used $\lambda_{-I}=i\lambda_{I}$. Again one can straightforwardly check that all other configurations simply correspond to one of these branches. Not surprisingly configurations of OG$_{3-}$ correspond to $(a)$, while those of OG$_{3+}$ correspond to $(b)$.

Finally, a question one might pose is whether or not summing over the local branches of a given vertex is valid if the vertex is identified with a BCFW bridge. Let us consider the attachment of a BCFW bridge is to connect two tree-amplitudes:  
\eq
\includegraphics[scale=0.4]{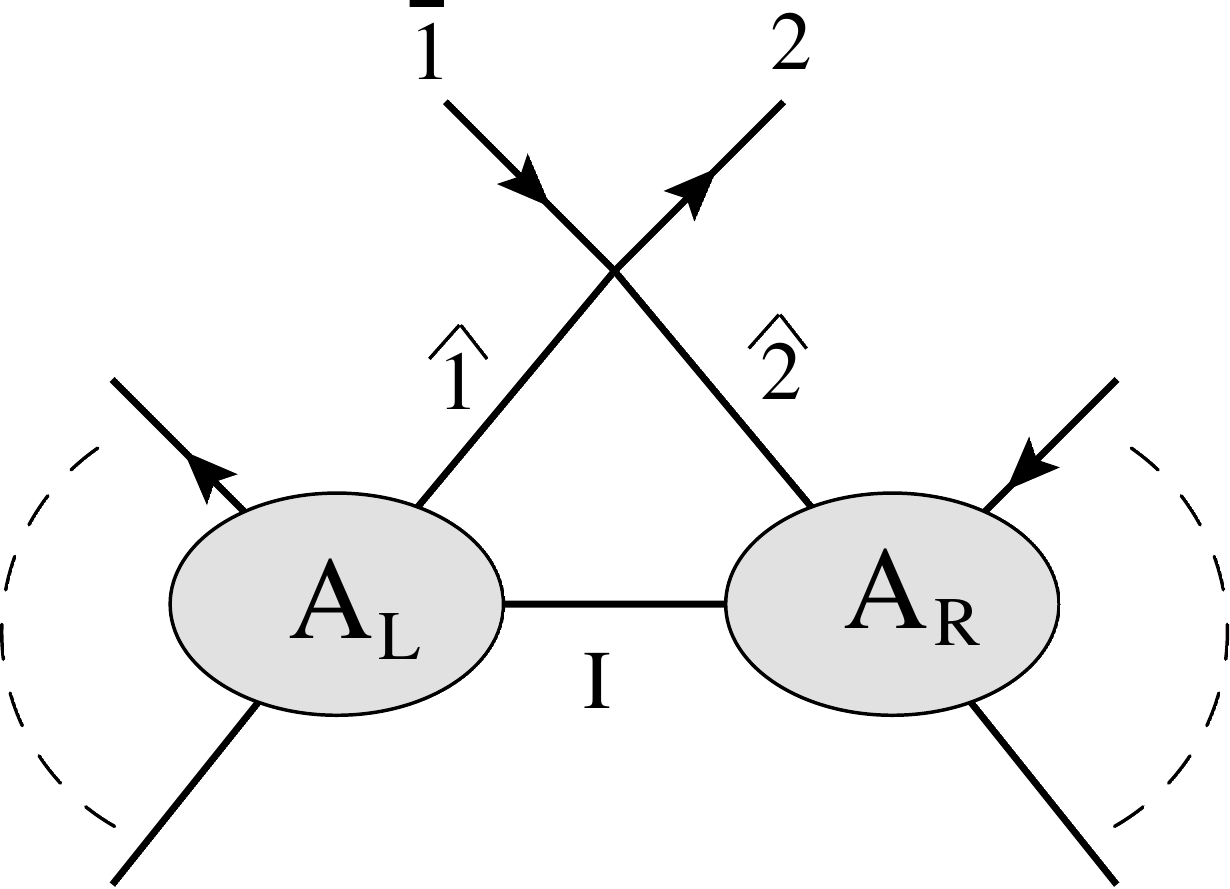}\,,
\label{twobranchBCFW}
\eqe
where the blobs labelled by A$_L$ and A$_R$ are collections of on-shell diagrams whose sums give the tree-amplitudes. The hatted lines indicate that the are expressed in terms of external spinors. It is important to keep in mind that legs $\hat{1}$ sits on a barred leg in A$_L$, and $\hat{2}$ sits on an un-barred leg in A$_R$. An immediate consequence of this is that if $\lambda_{\hat{1}}\rightarrow -\lambda_{\hat{1}}$, A$_L\rightarrow -$A$_L$, whereas if $\lambda_{\hat{2}}\rightarrow -\lambda_{\hat{2}}$, A$_R\rightarrow $A$_R$. Now if we sum over the two branches of the upper top vertex, this diagram is given as:
\eqa
dia\,(\ref{twobranchBCFW}) &=& {1 \over 2} \sum_{\alpha}\int d^{2|3}\Lambda_{\hat{1}}d^{2|3}\Lambda_{\hat{2}}d^{2|3}\Lambda_{\rm I} \int {d\theta \over cs} \delta^{4|3}(C(\alpha)\cdot\Lambda)A_LA_R\\ 
\eqae
The integral over $\Lambda_{\hat{1}}$ and $\Lambda_{\hat{2}}$ is to be localized by the delta function in the integrand. This localizes 
\eq
\Lambda_{\hat{1}}=\frac{1}{\alpha s}(\Lambda_1+c\Lambda_2),\quad \Lambda_{\hat{2}}=-\frac{1}{s}(\Lambda_2+c\Lambda_1)\,.
\eqe
This generates a Jacobian factor $\alpha s$. Now, we see that summing over the two branches, again labeled by $\alpha$, amount to summing over $\Lambda_{\hat{1}}$ and $-\Lambda_{\hat{1}}$. As discussed previously, under the flip of the sign in front of $\Lambda_{\hat{1}}$, the only $\Lambda_{\hat{1}}$ dependent term in the integrand A$_L$ also flips a sign which precisely compensates for the extra sign that arise from the Jacobian factor $\alpha s$. Thus the integrand actually takes the same form on the two branches. This proves that in the BCFW recursion, the vertex that is identified as the BCFW bridge can be safely averaged over the two branches of the original the vertex.

Thus in conclusion, the tree-level amplitude of ABJM can be schematically written as:
\eq
A_{n}(\bar{1}2\cdots n)=\sum_{dia}\bigg[\prod_{j\in I_{dia}}\int d^{2|3}\Lambda_j\bigg] \bigg[\prod_{i\in v_{dia}}\left(\sum_{\alpha_i}\frac{1}{2}\int d\log\tan_i \right)\delta^{2k|3k}\left(C(\alpha_i)\cdot\Lambda\right)\bigg]\,.
\eqe

\section{Loop recursion relation} \label{section:loop}
In this section we study the recursion formula for loop-level amplitudes of ABJM theory, utilizing on-shell diagrams. The recursion relation can diagrammatically represented as:
\eq \label{looprecursion}
\mathcal{A}_n^{\ell}=\quad\sum_{\ell_{1}+\ell_{2}=\ell}\sum_{i=4}^{n-2}\;\;\vcenter{\hbox{\includegraphics[scale=0.4]{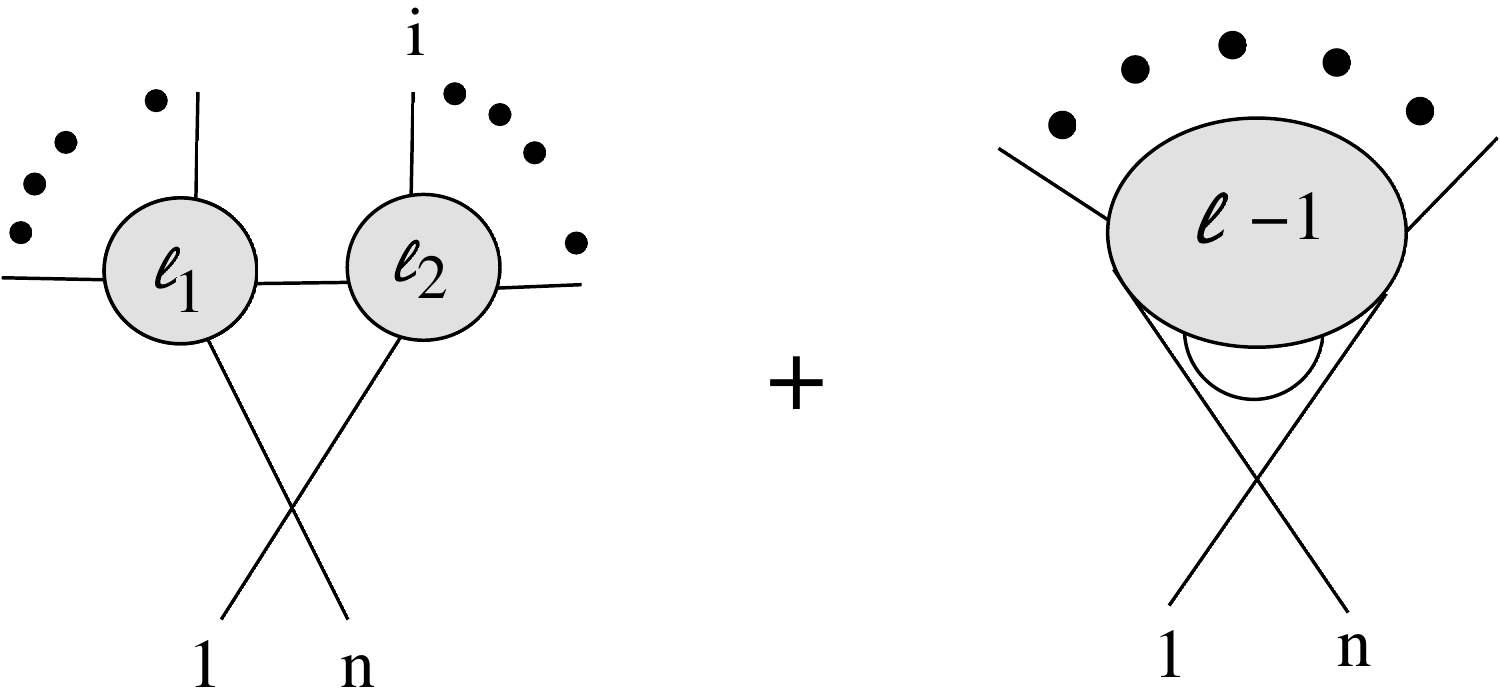}}}
\eqe
The validity for loop-level BCFW recursion can be established by demonstrating that all physical singularities are present in the form, and any spurious singularities cancels in pairs. This was done for $\mathcal{N}$=4 SYM~\cite{NimaBigBook}. In the following we will study the singularity structures of the solution to the recursion relation for ABJM theory. We will find:
\begin{itemize}
\item All spurious singularities cancel out in pairs.
  \item At loop level, half of the physical forward-limit singularities are visible in the solution via opening of BCFW bridges of external vertices. The other half originate from internal vertices along external edges. For the recursion with BCFW shifts on legs $(\bar{1}, 2)$, all forward-limit singularities associated with the channel between $(\overline{2i+1},2i+2)$ are of the former type, where as singularities for $(\overline{2i},2i+1)$ are of the latter type.
    \item Just as the tree-level solution discussed in~\cite{HW}, the solution to the loop-level recursion exhibits manifest two-cite cyclic symmetry. 
\end{itemize}
These properties allow us to conclude the solutions to the loop-recursion indeed gives the correct loop amplitudes of ABJM theory. Note that the last two properties are special for ABJM amplitudes, which are distinct from that of $\mathcal{N}=4$ SYM. The reason is due to the simplicity of the on-shell diagrams of $OG_k$. Unlike $Gr_{k,n}$ where the equivalence move include merging and expanding, the only equivalence move allowed for $OG_k$ is the triangle move. Thus if all physical singularity is present by opening of external vertices, as there are no equivalence move that can move an internal vertex to external, one can inductively show that the solution must exhibit certain cyclic symmetry. Let us consider the tree-level solution, where the only physical poles are factorization channels. For example the channel (23456|781) factorization pole of eight-point tree amplitude can be seen by either opening up (12) or (67). Thus the minimum solution to the  requirement that
\begin{itemize}
  \item All factorization poles are present 
  \item All physical poles are accessible via opening external vertices
\end{itemize}
is to have a $i\rightarrow i+2$ symmetry, indeed a representation with manifest two-site cyclic symmetry was found in \cite{HW}. For loop level, one now needs to include forward-limit singularities. If the forward-limit singularities are only accessible through opening of external BCFW vertices, which would require full cyclic symmetry instead of two-site one. However, this would double up the number of factorization singularities. To be consistent, and as we will show explicitly, at loop level half of the forward-limit singularities must appear as opening of internal vertices to maintain $i\rightarrow i+2$ symmetry.

\subsection{Four-point one-loop amplitude}
Let us start with the simplest one-loop amplitude, four-point scattering. Since there is no three-point amplitude, the only the contribution to the loop recursion relation is from the forward limit of the six-point tree-amplitude. Thus the result of the loop-recursion is given by:
\eq \label{4ptRecur}
\quad \includegraphics[scale=0.5]{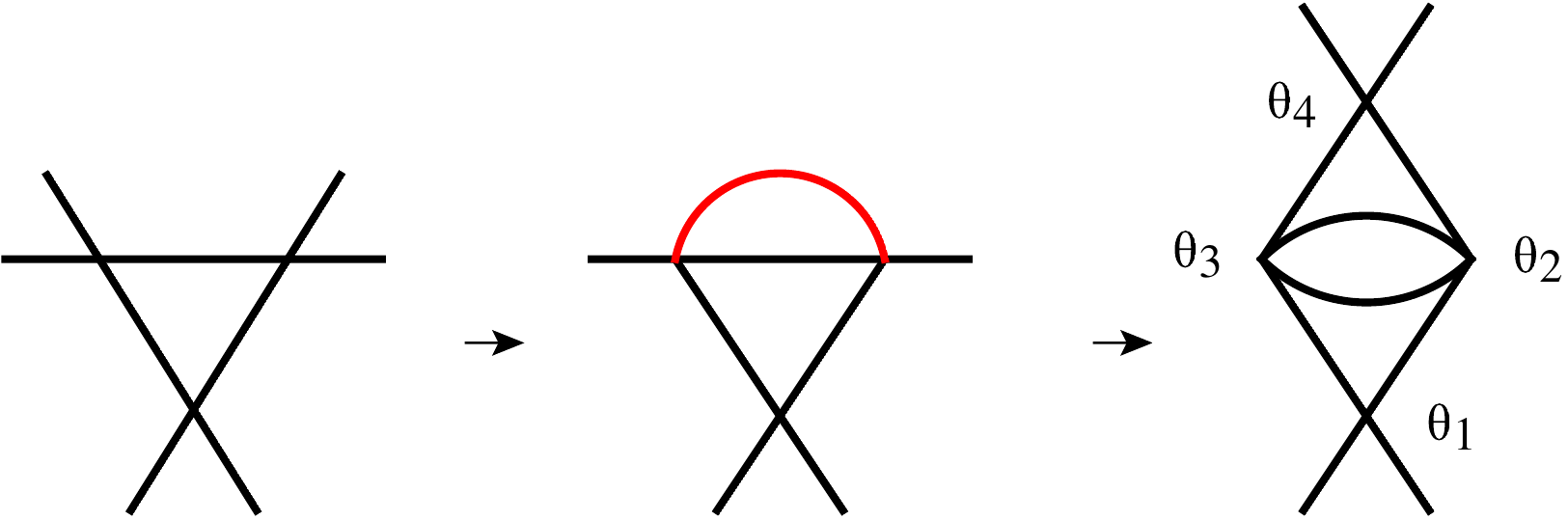}\,,
\eqe
where we have illustrated the procedure of obtaining the forward limit of the six-point amplitude, and then attaching a BCFW bridge to obtain the one-loop amplitude. 
\subsubsection{Forward limit}
Before obtaining the one-loop amplitude, as an exercise let us first show that forward-limit on-shell diagram reproduces the correct single cut from the known integrand. Consider the following forward-limit diagram: 
\eq \label{4ptforward}
F^{(4,1)}_4=\vcenter{\hbox{\includegraphics[scale=0.4]{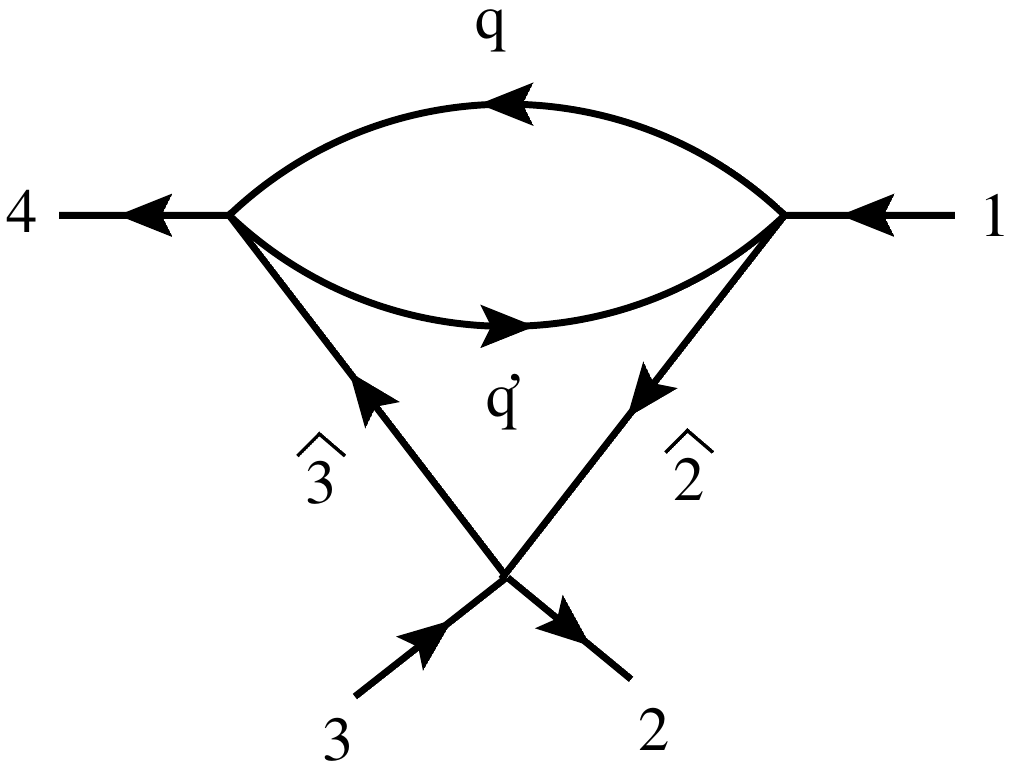}}}\,,
\eqe 
where the arrows indicate the gauge choice, and all external momenta are flowing outwards and $q,q'$ are flowing rightward in the diagram. This diagram can be regarded as attaching a BCFW bridge to a bubble diagram, where the latter can be conveniently expressed as:\footnote{The final line can be derived by a change of variable $\lambda_{q'}=a\lambda_3+b\lambda_4$, and noting that:
$$\delta^3 ( P_L)=\delta(\langle qq'\rangle\mp \langle 34\rangle)\delta(\langle q'3\rangle\pm \langle 4q\rangle)\delta(\langle q3\rangle\pm \langle q'4\rangle)$$
where the $\pm$ indicates the two branches of the four-point kinematics. For either branch, one can use two delta functions to localize $a,b$, leaving behind $\delta\left((q+p_3+p_4)^2\right)$.} 
\eqa
\vcenter{\hbox{\includegraphics[scale=0.5]{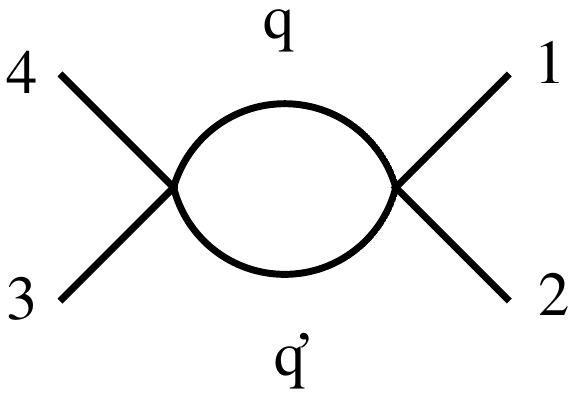}}} &= &
-\delta^3 ( P_{full})\delta^6 ( Q_{full}) \int d^2 \lambda_q d^2 \lambda_{q'} { \langle 34  \rangle\over 
\langle 1 q  \rangle   \langle q 4  \rangle }  \delta^3 ( P_L) \cr
\nonumber& = &
-\delta^3 ( P_{full})\delta^6 ( Q_{full})\int d^2\lambda_q  { \langle 34  \rangle\over 
\langle 1 q  \rangle   \langle q 4  \rangle }\delta\left((q+p_3+p_4)^2\right)\equiv B_4\,.\\
\eqae
We can now obtain the forward limit diagram by applying a BCFW vertex to the bubble diagram. The explicit form is now given as
\eqa
\nonumber F^{(4,1)}_4&=&\sum_{\alpha=\pm}\frac{1}{2}\int {d\log\tan\theta}\; \delta^{2|3}(C\cdot\Lambda)\;B_4\, \\
=&&-\delta^3 ( P_{full})\delta^6 ( Q_{full})\sum_{\alpha=\pm}\frac{1}{2}\int {d\theta\over s}\;\int d^2 \lambda_q \; \delta((q + p_{\hat{3}} + p_4)^2) 
{ \langle \hat{3} 4 \rangle \over \langle 1 q\rangle  \langle q 4\rangle } 
\eqae
where the shifted $\lambda_{\hat{3}}$ is defined as
\eq
\lambda_{\hat{3}} = -\sec(\theta) \lambda_3 + \alpha \tan(\theta) \lambda_2 \, . 
\eqe
Note that since the integrand is little-group even on leg 2, we can simply combine the two branches. As usual we treat the delta-function, $\delta((q + p_{\hat{3}} + p_4)^2) $, as a contour integral. The only other place where there is $\theta$ dependence is in the integration measure, thus we can deform the counter and instead localize on the pole $s=0$, i.e. $\theta=0,\pi$. Since the integrand contain odd number of $\lambda_{\hat{3}}$, the two residue combine to give: 
\eq \label{41singlecut}
F^{(4,1)}_4=\delta^3 ( P_{full})\delta^6 ( Q_{full})\int d^2 \lambda_q 
{ \langle 3 4 \rangle \over \langle 1 q\rangle  \langle q 4\rangle (q + p_{3} + p_4)^2 } \, ,
\eqe
which agrees precisely the single-cut of one-loop four-point amplitude \cite{BH}:
\eqa\label{BigHeadResult}
\nonumber\mathcal{A}_4^{\rm tree}\int d^3 \ell  { -\langle 3 4 \rangle^2  \langle  4 |\ell|1 \rangle \langle 14 \rangle +  \ell^2\langle 12\rangle\langle 24\rangle\langle 41\rangle \over  \ell^2 (\ell - p_1 )^2 (\ell + p_4)^2 (\ell - p_1 - p_2)^2 }\;\;\underrightarrow{\; \ell^2=0\; }\int d^2 \lambda_q 
{\delta^3 ( P_{full})\delta^6 ( Q_{full}) \langle 3 4 \rangle \over \langle 1 \ell\rangle  \langle \ell 4\rangle (\ell + p_{3} + p_4)^2 } \, .\\
\eqae

It is instructive to consider the forward limit entirely in terms of on-shell variables. We begin with the explicit integrand associated with the pre-reduced on-shell diagram:
\eq \label{41singlecut1}
J_{\rm F} \, {d\tan_1 } \wedge {d\tan_2} \wedge {d\tan_3} \, 
\eqe
where the labelling of the vertices are consistent with fig.(\ref{4ptRecur}), and the Jacobian factor $J_{\rm F}$ may be determined according to the rules given in section \ref{section:localmoves} 
\eq
J_{\rm F} = 1 + c_2 c_3 + s_1 s_2 s_3 \, .
\eqe
The Grassmannian $C_{\rm F}$ in $\delta^2( C_{\rm F} \cdot \lambda )$ is given as,  
\eqa
 C_{\rm F}=\left(\begin{array}{cccc}1 & \frac{c_4 c_1}{1 + s_1 s_4} & 0 & -\frac{s_1 + s_4 }{1 + s_1 s_4} 
 \\0 & \frac{s_1 + s_4 }{1 + s_1 s_4}  & 1 & 
\frac{c_4 c_1}{1 + s_1 s_4} \end{array}\right)\, ,
\eqae
with $s_4$(and $c_4$) given by:
\eq \label{transformationfw0}
s_4 = {s_2 s_3 \over 1+ c_2 c_3 } \, , \quad  c_4 = {c_2 + c_3 \over 1+ c_2 c_3 } \,.
\eqe 
Starting from this integrand, we can explicitly preform two-step bubble reductions, under which the diagram is reduced to a tree-level four-point amplitude, 
\eq 
\vcenter{\hbox{\includegraphics[scale=0.6]{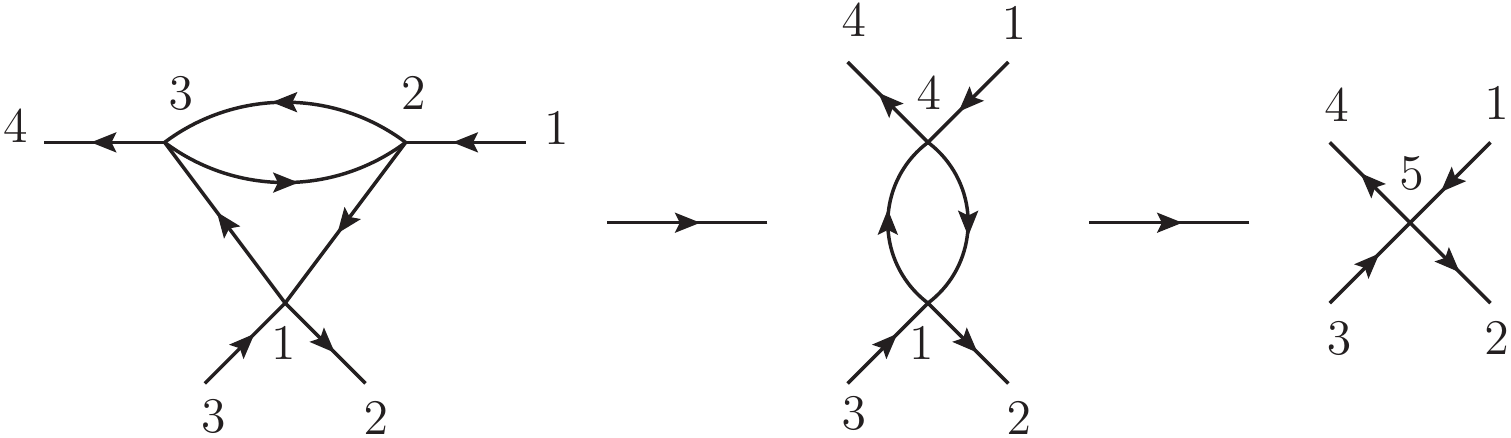}}}\,. 
\eqe 
The result of this reduction, one finds that $F^{(4,1)}_4$ can be written in a nice $d\log$ form, 
\eq\label{41singlecut2}
F^{(4,1)}_4 = A^{\rm tree}_4 \int d\log({\tan_2 \over \tan_3} ) \wedge d\log( {\tan_4 \over \tan_1}) \, , 
\eqe
with $s_4$($c_4$) defined in eq.(\ref{transformationfw0}) and $s_5$($c_5$) given by, 
\eqa \label{transformationfw}
s_5 &=& {s_4 + s_1 \over 1+ s_4 s_1 } \, , \quad  c_5 = {c_4 c_1 \over 1+ s_4 s_1 } \, .
\eqae
For simplicity we only focus on the positive branch, and $s_5$($c_5$) are fixed by external kinematics,
\eq
 s_5 = - i {\langle  12\rangle \over  \langle  42\rangle } = -i {\langle 34 \rangle \over  \langle  42\rangle }\, , \quad c_5 =  i {\langle 41 \rangle \over  \langle  42\rangle } = -i {\langle 23 \rangle \over  \langle  42\rangle } \, .
\eqe
The equivalence between eq.(\ref{41singlecut}) and eq.(\ref{41singlecut2}) can be verified by exchanging  $\lambda_{q}$ in eq.(\ref{41singlecut}), in terms of on-shell variables,
\eq\label{Qdef}
\lambda_{q} = {1 \over s_3}(\lambda_4 + c_3 \lambda_{\hat{3}}  ) 
= 
{1 \over s_3}(\lambda_4 - {c_3 \over c_1} \lambda_3 + {c_3 s_1 \over c_1} \lambda_2  ) \, .
\eqe
Thus in summary, we see that the forward limit on-shell diagram not only reproduces the correct single cut, but it reveals the simplicity of the result: it is nothing but a product of two $d\log$'s. Note that this structure only reveals itself after one performs bubble reduction, and the $d\log$ form crucially depends on the presence of the Jacobian factor. 

\subsubsection{One-loop amplitude}

We will now show that the full one-loop amplitude is also given by three $d\log$'s.  Again, the one-loop on-shell diagram is given as:
\eq 
\label{Reduc1}
\includegraphics[scale=0.3]{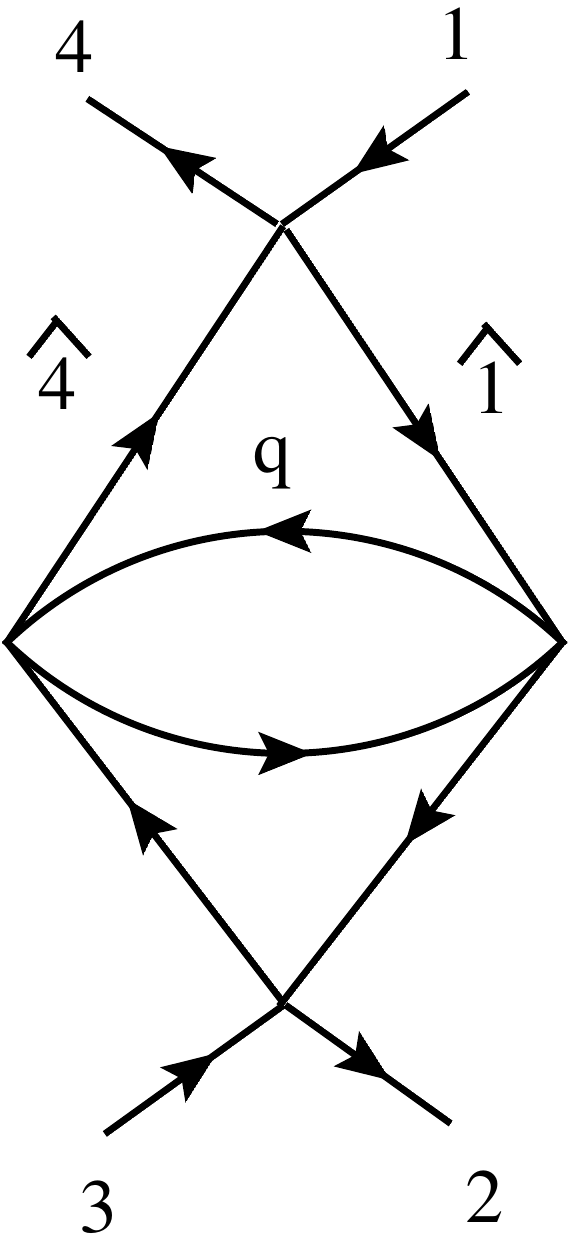} \,.
\eqe
The associated integrand is:
\eq \label{looponshell}
J_{\rm loop} \;\;{ d\log\tan_1 } \wedge { d\log\tan_2 } \wedge { d\log\tan_3 } \wedge { d\log\tan_4 } \, ,
\eqe
with the Jacobian
\eq
J_{\rm loop}
= 1 + s_1 s_2 s_3 + s_2 s_3 s_4 + c_2 c_3 + s_1 s_4 + c_2 c_3 s_1 s_4 \, .
\eqe
Just as with the case of forward-limit diagram, one can also perform bubble reductions on four-point one-loop diagram, leading to:
\eq \label{lowerorderpole}
\includegraphics[scale=0.6]{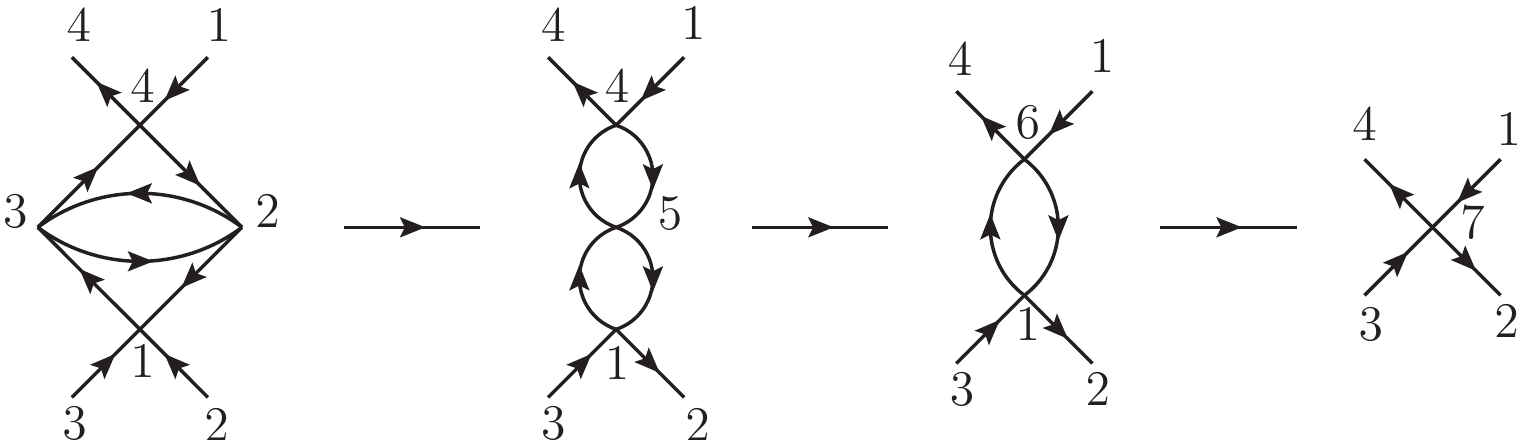}\, .
\eqe
Remarkably, after the reduction the final form reveals itself to be a wedge product of $d\log$ forms: 
\eq\label{One-Loop4ptIntegrand}
A^{\rm loop}_4 = \int d\log (\tan_7)\prod_{a=1,2}\delta^{2|3}(C_{a}\cdot\Lambda) \int d\log( {\tan_2\over \tan_3}) \wedge d\log( {\tan_4\over \tan_5} ) 
\wedge d\log({\tan_6\over \tan_1} )\,, 
\eqe
with following identification:
\eqa  \label{transformationloop}
s_5 &=& {s_2 s_3 \over 1+ c_2 c_3 } \, , \quad  c_5 = {c_2 + c_3 \over 1+ c_2 c_3 } \cr
s_6 &=& {s_4 + s_5 \over 1+ s_4 s_5 } \, , \quad  c_6 = {c_4 c_5 \over 1+ s_4 s_5 } \cr
s_7 &=& {s_6 + s_1 \over 1+ s_6 s_1 } \, , \quad  c_7 = {c_6 c_1 \over 1+ s_6 s_1 } \, .
\eqae
The parameterization of the Grassmannian is given as:
\eqa
 C_{\rm loop}=\left(\begin{array}{cccc}1 & c_7 & 0 & -s_7 
 \\0 & s_7 & 1 & c_7 \end{array}\right)\, .
\eqae
Since only $s_7$ ($c_7$) appears in the Grassmannian, the prefactor in eq.(\ref{One-Loop4ptIntegrand}) is simply the tree-amplitude as expected from the reduction diagram eq.(\ref{lowerorderpole}). Thus we see that the four-point amplitude given by a product of the tree-amplitude and three $d\log$s.

For the purpose of studying the singularities associated with the on-shell diagram, it is simpler to work with eq.(\ref{looponshell}). It is easy to see that at the singularity $s_4 = 0$, the differential form reduces to the one of forward-limit diagram eq.(\ref{41singlecut1}) with corresponding Jacobian and Grassmannian. Similarly $s_1=0$ corresponds to the forward-limit of opening up BCFW vertex with legs $2$ and $3$, as one would expect. What is more interesting is the singularities at $s_2=0$, as well as $s_3=0$. For instance at $s_2=0$, the differential form (\ref{looponshell}) reduces to 
\eq
J_{\rm s_2=0} \;\;{ d\log\tan_1 }  \wedge { d\log\tan_3 } \wedge { d\log\tan_4 }  \, ,
\eqe
with the following Jacobian and Grassmannian 
\eqa \label{s2=0}
J_{\rm s_2=0}
&=& (1 + c_3) (1 + s_1 s_4)\, , \\
 C_{\rm s_2=0}&=&\left(\begin{array}{cccc}1 & \frac{c_4 c_1}{1 + s_1 s_4} & 0 & -\frac{s_1 + s_4 }{1 + s_1 s_4} 
 \\0 & \frac{s_1 + s_4 }{1 + s_1 s_4}  & 1 & 
\frac{c_4 c_1}{1 + s_1 s_4} \end{array}\right)\, .
\eqae
Diagrammatically this singularity can be represented as\footnote{Here we have flipped the bubble inside out, which effectively exchanges $s_3$ and $c_3$ in the expression of eq.(\ref{s2=0}).}
\eq \label{4ptSing}
\includegraphics[scale=0.4]{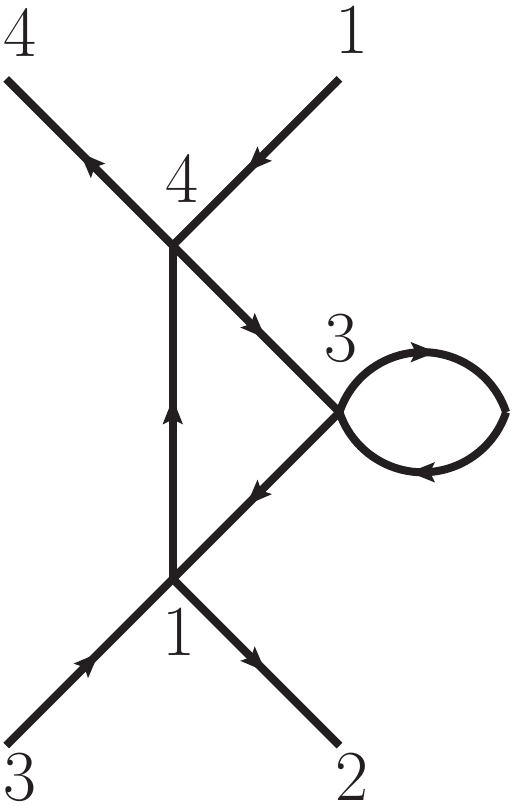}\,.
\eqe
This is nothing but the forward-limit singularity of one-loop four-point amplitude by opening up the BCFW vertex with legs $1$ and $2$, which can be shown explicitly by performing a triangle move on the diagram \ref{4ptforward}. The same analysis can be applied to the singularity at $s_3 =0$. We can summarize our finding in the following diagram,
\eq \label{4ptSingSum}
\includegraphics[scale=0.5]{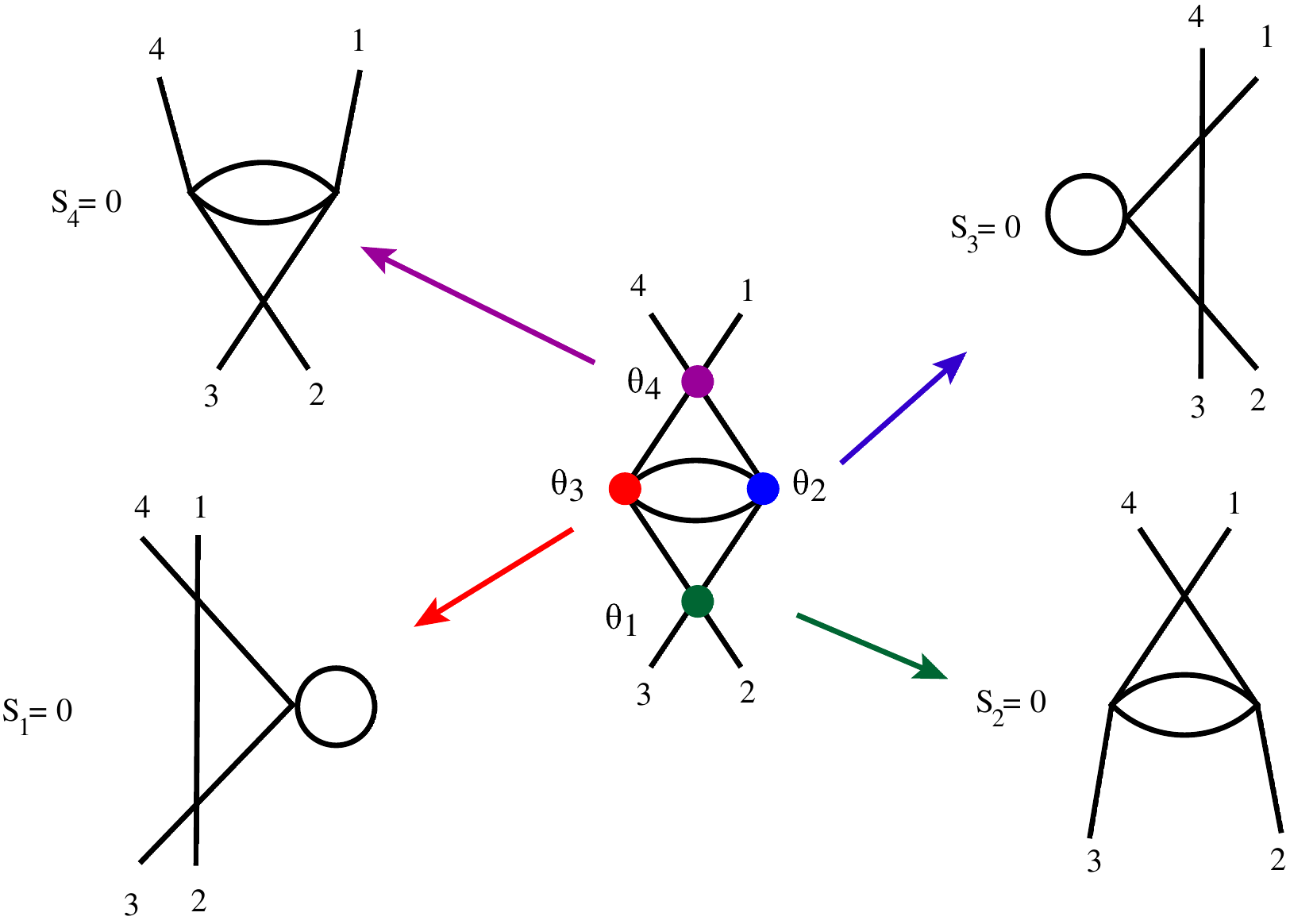}\,.
\eqe
So all four physical forward-limit singularities are presented, two of them appear trivially whereas the other two appear in a rather non-trivial way by opening up the internal vertices. The fact that half of forward-limit singularities are obtained by opening up internal vertices, turns out to be a general feature of loop amplitudes in ABJM theory.

Note the singularities coming from opening internal vertex corresponds to collinear limit between $q$ and BCFW shifted momentum $p_{\hat{1}}$ (and $p_{\hat{4}}$). To see that the collinear limit $p_{\hat{1}}||q$ indeed arises from the singularity denoted by the vertex $\theta_1$, notice that the residue contains a tadpole attached to the line between $p_{\hat{1}}$ and $p_{\hat{2}}$. The constraint implied by such configuration can be found by explicitly identify the two legs on a four vertex:
\eq
\begin{array}{c}\delta^{2|3}(\Lambda_{\hat{1}}+c\Lambda_{\hat{2}}-s\Lambda_q) \\ \delta^{2|3}(\Lambda_{q'}+s\Lambda_{\hat{2}}+c\Lambda_q)\end{array}\quad
\vcenter{\hbox{\includegraphics[scale=0.35]{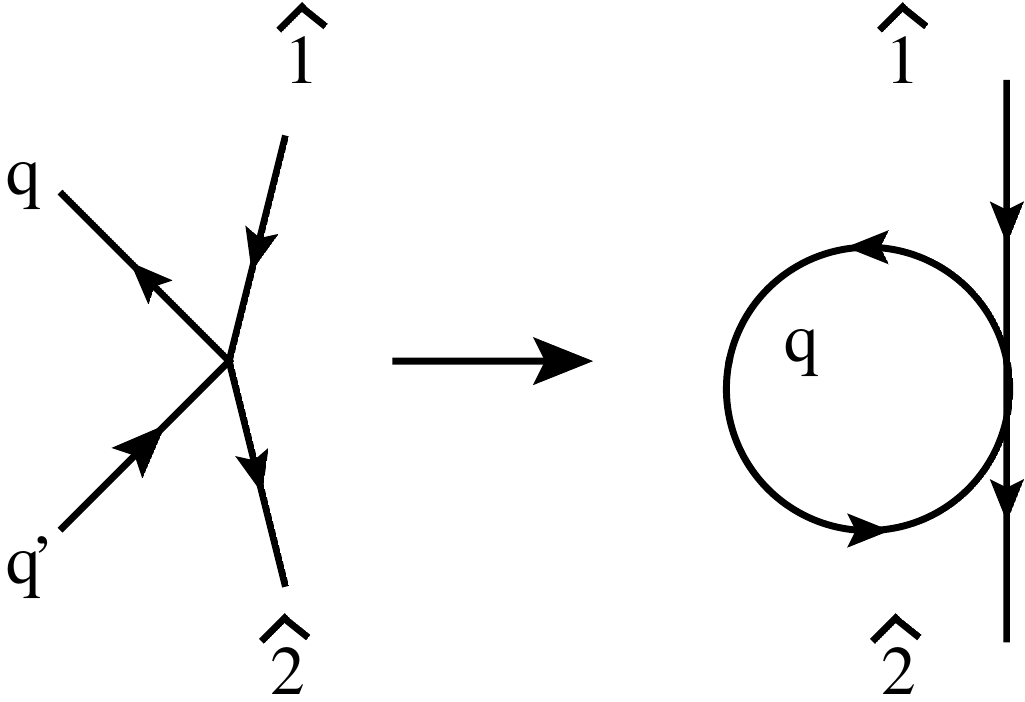}}}\quad\quad \begin{array}{c}\delta^{2|3}(\Lambda_{\hat{1}}+\Lambda_{\hat{2}}) \\\quad \delta^{2|3}(\Lambda_{q}+\frac{s}{1+c}\Lambda_{\hat{2}})\end{array}\,.
\eqe
In the above we've identified legs $q$ and $q'$. Note that the resulting kinematic constraint forces $p_{\hat{2}}=p_{\hat{1}}$, and $q||p_{\hat{1}}$, as promised. This fact makes evident by writing down the one-loop integrand as a BCFW shift acting on the forward-limit, 
\eqa
\nonumber A^{1-{\rm loop}} &=&\sum_{\alpha=\pm}\frac{1}{2}\int {d\log\tan\theta}\; \delta^{2|3}(C\cdot\Lambda)\;F_4\\
 &=&\delta^3 ( P_{full})\delta^6 ( Q_{full})\sum_{\alpha=\pm}\frac{1}{2}\int d^2 \lambda_{q} \int{ d \theta  \over s} 
{ \langle 3 \hat{4} \rangle   \over  \langle \hat{1} q\rangle \langle  q \hat{4} \rangle (q - p_{\hat{1}} - p_2)^2 } \, ,
\label{1loop4ptInt}
\eqae
with
\eq
\lambda_{\hat{4}} = -\alpha_4 (1/c_4)\lambda_4 + (s_4/c_4) \lambda_1 \,, \quad
\lambda_{\hat{1}} = (\alpha_4 s_4/c_4) \lambda_4 -(1/ c_4 )\lambda_1 \, .
\eqe
Above analysis also justifies the reason why we did not consider following singularity, 
\eq \label{lowerorderpole}
\includegraphics[scale=0.4]{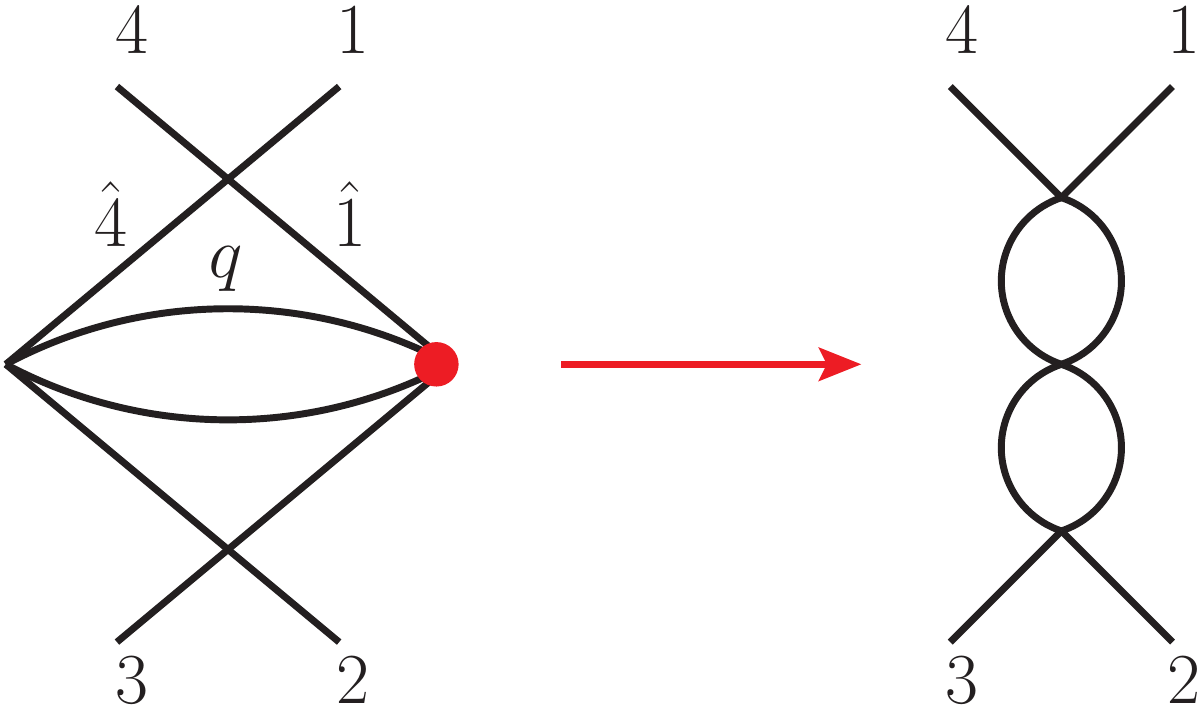}\, ,
\eqe
where one opens up internal vertex $3$ in a different way. That is because this singularity correspondences to a higher dimension comparing to those in (\ref{4ptSing}). In particular for the example given in (\ref{lowerorderpole}), it requires $p_{\hat{1}} + q =0$, instead of the collinear limit $p_{\hat{1}}||q$. So it is a higher-dimensional singularity. Since one-loop four-point diagram is obtained by taking forward-limit of six-point tree-level amplitude, so this singularity can be traced back to six-point tree-level diagram. It turns out that such singularity correspondences to so-called {\it soft singularity} discussed in \cite{HW}, which is also a higher-dimensional singularity there. We will encounter such singularities in higher-point and higher-loop amplitudes as well, but since they are higher-dimensional comparing to usual cut singularities, it's not necessary to consider them.

\subsubsection{Comparison to known integrand}
Finally, we can establish the equivalence of  eq.(\ref{looponshell}) with the known four-point one-loop integrand. We first note that the four-point one-loop amplitude in~\cite{BH}\cite{HW} can be rewritten as:
\eq\label{MomLog}
 A_4^{1-{\rm loop}} = A_4^{{\rm Tree}} \int d\log { (\ell-p_1)^2\over (\ell-p_1-p_2)^2 }\,d\log { \ell^2\over (\ell-p_1-p_2)^2}\,d\log{(\ell+p_4)^2\over (\ell-p_1-p_2)^2}\,.
\eqe
Note that in this form, it is clear that the integrand not only is a total derivative, but furthermore since there is no (complex)spurious singularity, the integral integrates to zero~\cite{SimonTalk}.  The validity of the above form can be straightforwardly derived by using dual coordinates in five-dimensional embedding space. The integrand is given as:
\eq\label{Embed}
\int_{X^2_0=0} \frac{\langle X_0dX_0dX_0dX_0dX_0\rangle}{X_0^2}\frac{\langle 0,1,2,3,4\rangle}{(0.1)(0.2)(0.3)(0.4)},\quad (i.j)\equiv X_i\cdot X_j
\eqe  
where $X_i$ lives on the five-dimensional null projective space in which the three-dimensional Minkowski space is embedded, $\langle \cdots\rangle$ indicate the contraction with the five-dimensional Levi-Cevita tensor, and the integration contour is understood to incircle the pole $X_0^2=0$. The dual positions are defined via $x_i-x_{i+1}=p_i$ and the labels are with respect to the following diagram:
$$\includegraphics[scale=0.6]{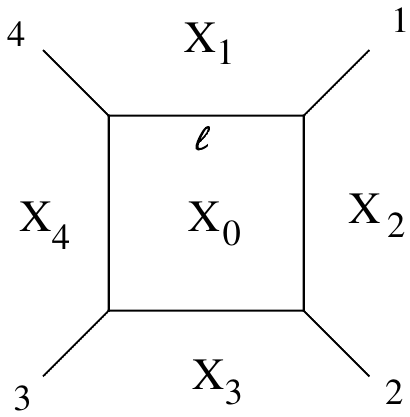}$$
We can parametrize the loop-region as:
\eq
X_0=a_1X_1+a_2X_2+a_3X_3+a_4X_4+a_\epsilon\langle *,1,2,3,4\rangle\,.
\eqe
Using this parameterization, we can use the projective nature of the integrand to fix $a_1=1$, and integrate localizing $a^2_\epsilon$ on the pole, we find 
\eq\label{aPara}
eq.(\ref{Embed}) =\int d\log a_2 d\log a_3 d\log a_4\,.
\eqe  
The $d\log$ form in Eq.(\ref{MomLog}) can be verified using the following identification\footnote{In Eq.(\ref{MomLog}) we have removed constant factors $s$ and $t$ in the differential form.}:
\eq
a_2=\frac{(0.4)(1.3)}{(0.3)(2.4)}=\frac{s(\ell+p_4)^2}{t(\ell-p_1-p_2)^2},\;\; a_3=\frac{\ell^2}{(\ell-p_1-p_2)^2},\;\;a_4=\frac{s(\ell-p_1)^2}{t(\ell-p_1-p_2)^2}\,.
\eqe
Armed with eq.(\ref{aPara}), we can verify that it is indeed equivalent to our on-shell $d\log$ form in eq.(\ref{One-Loop4ptIntegrand}).  To simplify our task, we can simply go to the boundary of each logarithmic singularity and compare the two-form. For example, consider the singularity $a_4=0$ in eq.(\ref{aPara}), whose residue is simply the two-form:
\eq  
F^{(1,2)}_4 = d\log a_2  d\log a_3 =d\log \frac{(\ell'+p_4+p_1)^2}{(\ell'-p_2)^2} d\log  \frac{(\ell'+p_1)^2}{(\ell'-p_2)^2}\, .
\eqe
where we defined $\ell' = \ell - p_1$ with $\ell'^2 =0$ under this cut. This is single cut corresponds to the on-shell diagram in eq.(\ref{4ptSing}), which, according to bubble reduction, can be written as, 
\eq
F^{(1,2)}_{4, {\rm on-shell}} =  d\log ({\rm csc}_3 -1 ) \ d\log {{\rm tan}_4  \over {\rm tan}_1 } \, .
\eqe
It is straightforward to check that these two $d\log$ forms match by expressing $\ell'$ in terms of on-shell variables
\eq
\lambda_{\ell'} =  {c_3 \over c_1(1 + s_3) } ( s_1 \lambda_3 + \lambda_2 ) \, .
\eqe
Similar analysis can be done for $a_2$ and $a_3$, thus verifying that the integrand in eq.(\ref{1loop4ptInt}) indeed matches with the known answer eq.(\ref{MomLog}).
\subsection{Six-point one-loop amplitude}
After extensive analysis of four-point amplitude at one loop, we now move on to the six-point case. The contribution from the forward-limit of the eight-point tree-level amplitude is given as:
\eq \label{8tforward}
\includegraphics[scale=0.5]{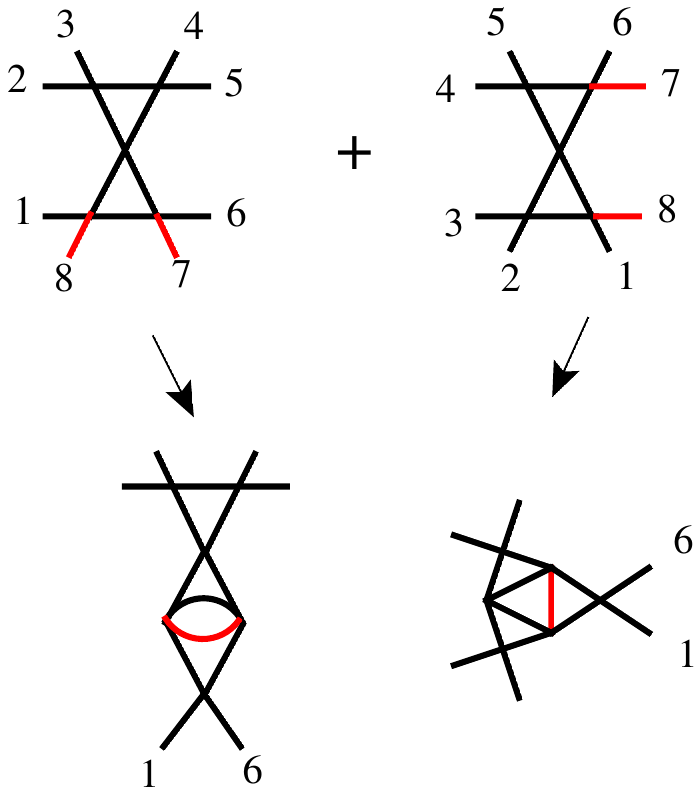}
\eqe
We also have two contributions that arise from the factorization of the six-point one-loop amplitude into a four-point one-loop amplitude multiplied with a four-point tree-level amplitude, given by:
\eq \label{8tfac}
\includegraphics[scale=0.6]{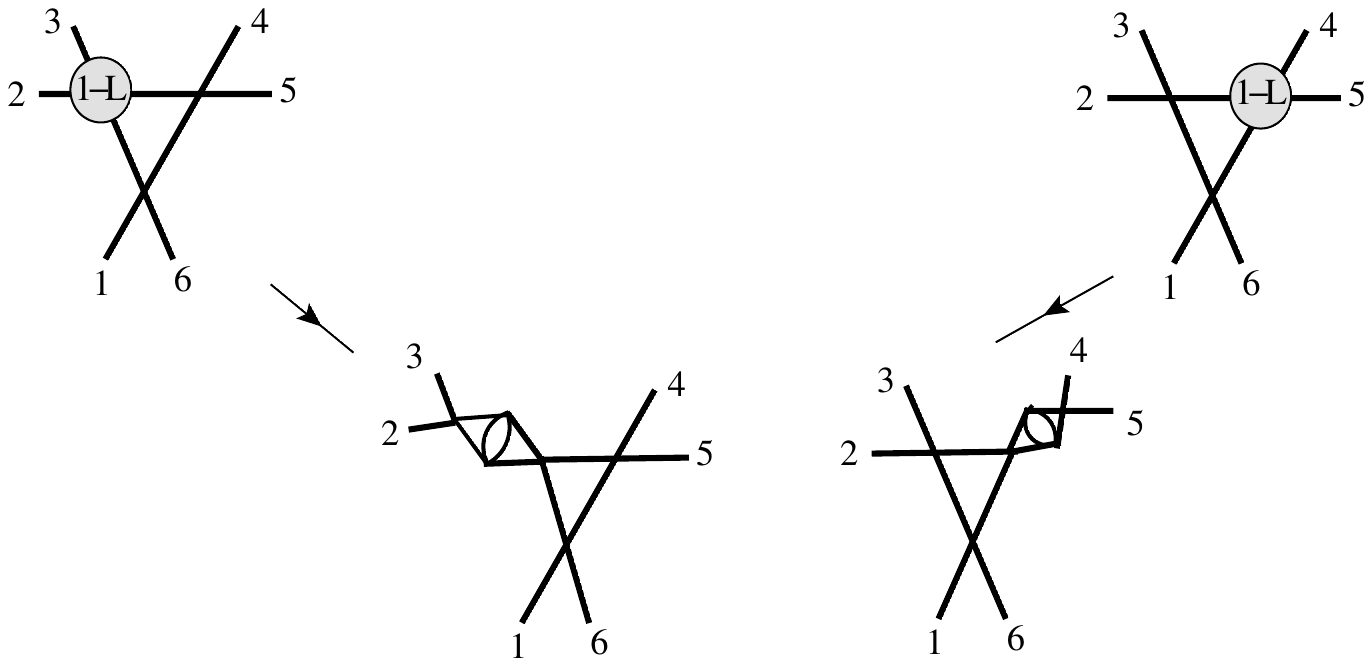}
\eqe
Thus the six-point one-loop amplitude is given as:
\eq                                                                                                                                                                                                                                                                                                                                                                                                                                                                                                                                                                                                                                                                                                                                                                                                                                                                                                                                                                                                                                                                                                                                                                                                                                                                                                                                                                                                                                       A^{1-{\rm loop}}_6=\vcenter{\hbox{\includegraphics[scale=0.6]{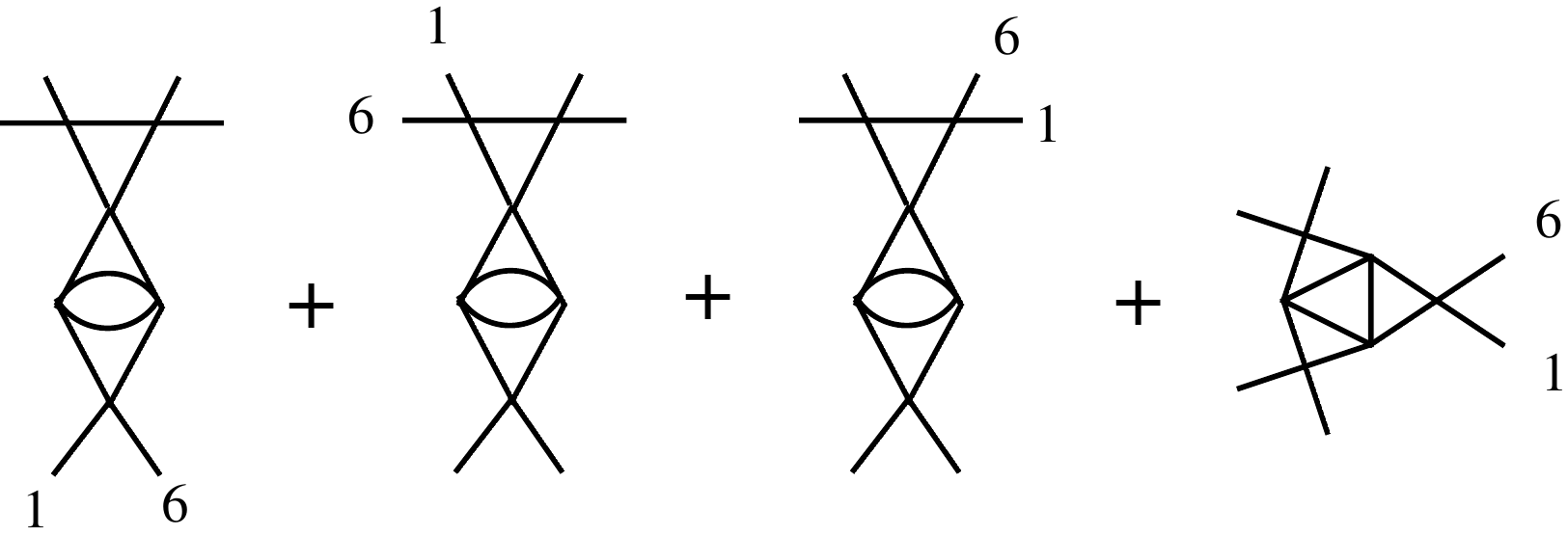}}}\,.
\label{6pt1Loop}
\eqe
Notably the above representation of six-point amplitude has manifest $i \rightarrow i+2$ cyclic symmetry. With this two-site cyclic symmetry, and the fact that only scattering amplitudes with even number of legs are non-vanishing in ABJM theory, it is guaranteed that all factorization poles are present. Like the case of four-point one-loop amplitude, actually only half of forward-limit singularities those are related by two-site cyclic rotations are manifest. So what we need to verify is the existence of the other half of single cuts that are related to the original BCFW bridge by a one-site shift. Again, like the case of four-point one-loop amplitude, these singularities are obtained by opening up the internal vertices. Following diagrams show the forward-limit singularities according BCFW bridge with legs $1$ and $2$, 
\eq
\includegraphics[scale=0.4]{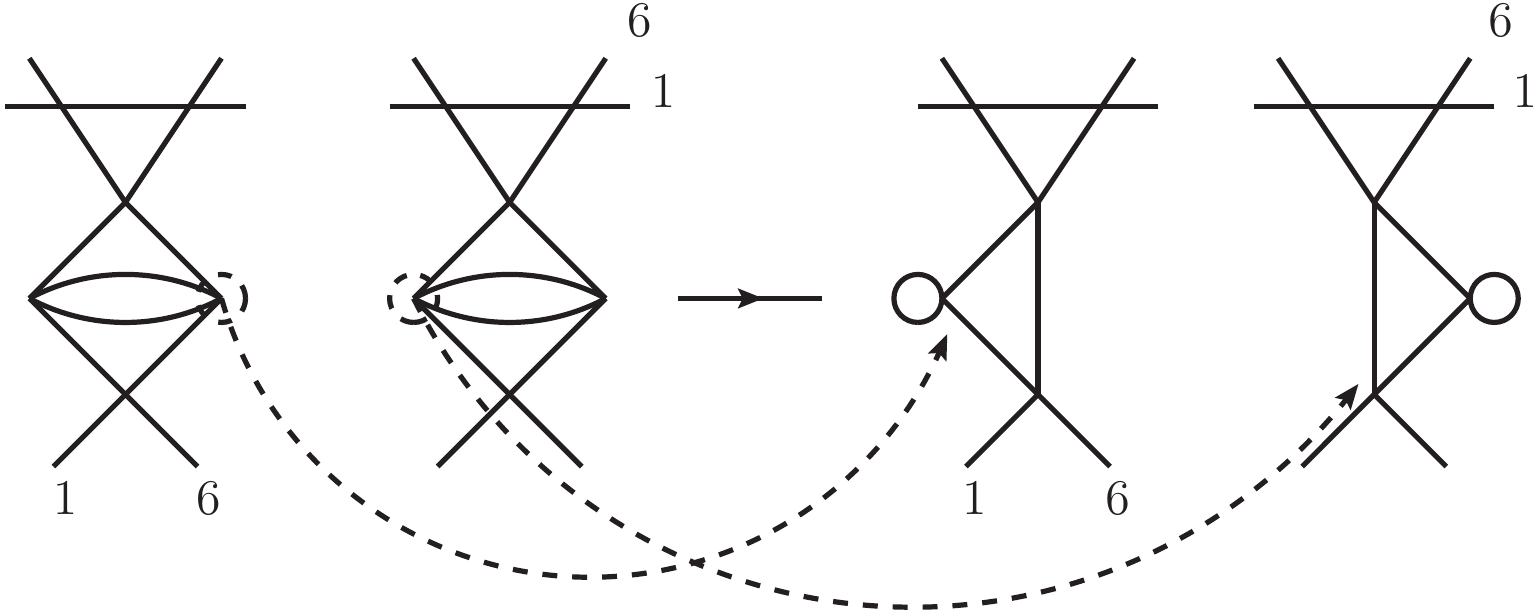} \, .
\eqe
On the other hand, unlike four-point one-loop amplitude, each term in eq.(\ref{6pt1Loop}) contains some unphysical poles, which we need to ensure those singularities are spurious, namely they cancel out each other in the sum. Indeed, all the spurious singularities appeared in pairs, as shown in following 
\eq
\includegraphics[scale=0.4]{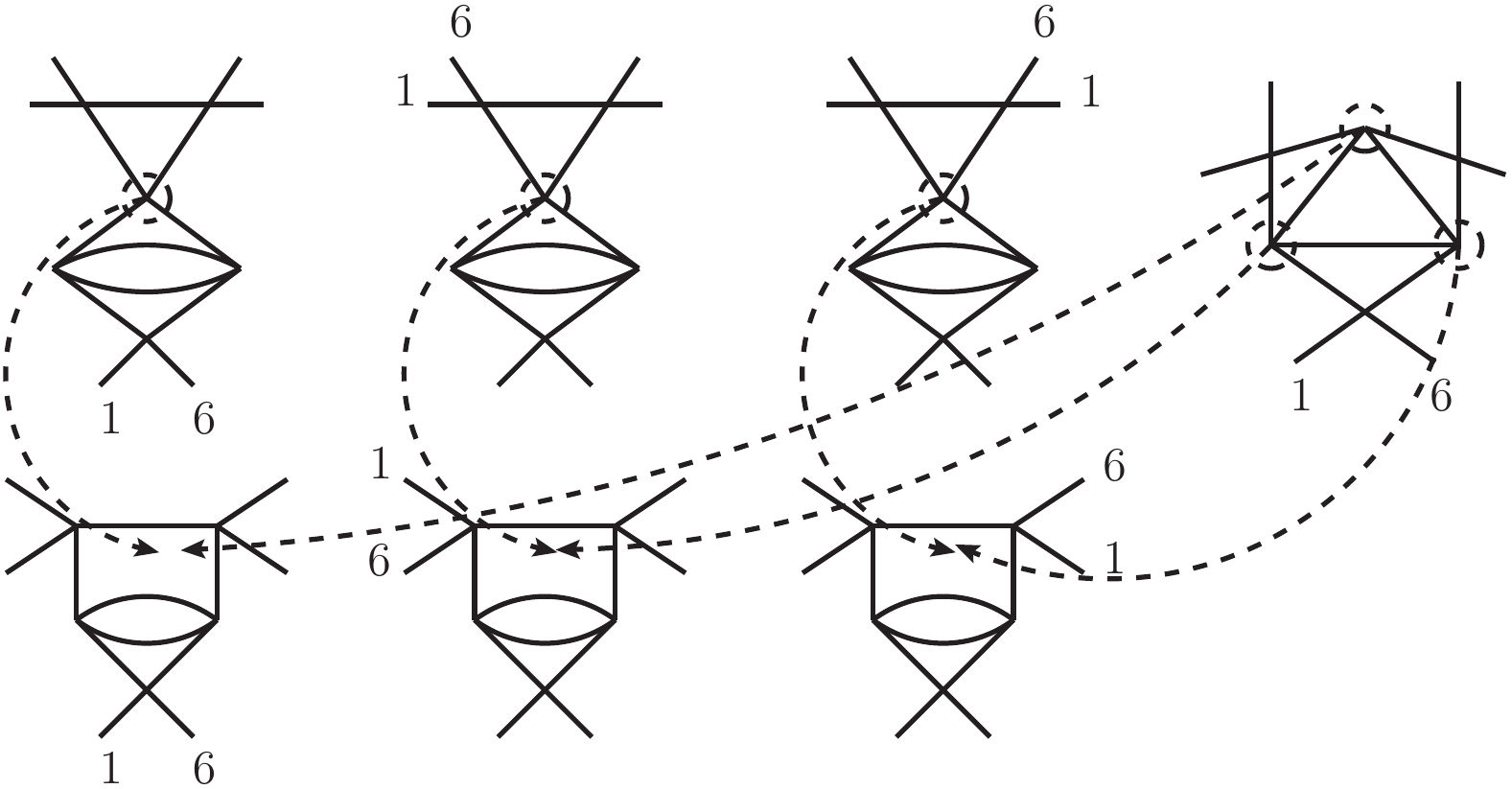} \, .
\eqe
Note actually the cancellation between the first and the last diagram can be traced back to the spurious singularities cancellation of tree-level eight-point amplitude. 

One can actually generally argue that spurious singularities should always cancel in pair in the loop recursion relations. As they can be easily associated with singularities arising from the forward limit of sub amplitudes in a factorization diagram, and factorization singularities arising from a forward limit diagrams, represented as:
\eq
\includegraphics[scale=0.45]{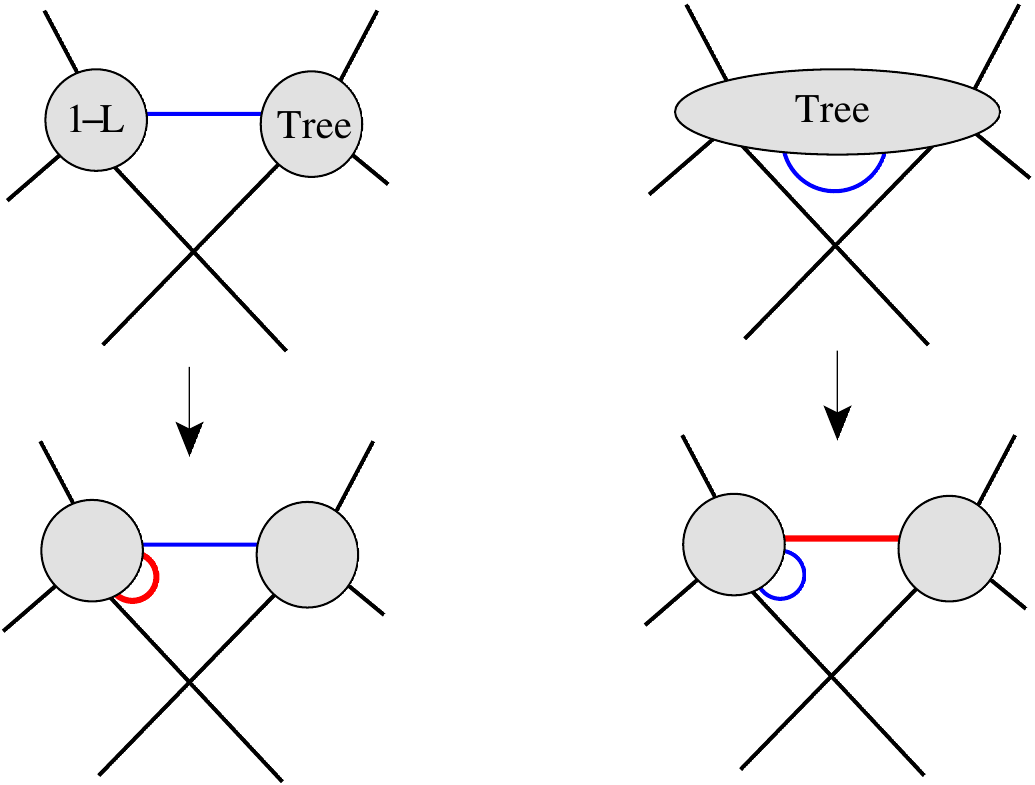} \, ,
\eqe
where we have used the red line to indicate the spurious singularity, and the blue line to represent the singularity that is used to construct the recursion. As one can see, the same spurious diagram can always appear in pair from two sources, exactly as discussed for $\mathcal{N}=4$ SYM~\cite{NimaBigBook}.

Before proceeding to next section, let us pause here to make a few comments on the result of six-point amplitude. First of all eq.(\ref{6pt1Loop}) is just one possible way of presenting six-point one-loop amplitude, however it is a special one by making of manifest two-site cyclic symmetry. As we argued previously, in fact that this symmetry of ABJM amplitudes is rather general due to the simplicity of the on-shell diagrams of $OG_k$. 

Apparently there are representations which do not have the manifest $i \rightarrow i+2 $ symmetry, and one may need all kinds of equivalence moves to see this symmetry. As already shown in \cite{HW}, by solving recursion relation in a particular way, all tree-level amplitudes can be presented in a form with manifest two-site cyclic symmetry.  We will refer such representation of amplitudes as the ``canonic" representation. In general such representation is obtained by solving the recursion relation, (\ref{looprecursion}), as follows: 
\begin{itemize}

\item Firstly, we compute the factorization diagram as proceed in \cite{HW}, where the BCFW bridge vertex, as a triangle, is connected to the rest diagrams through points only. 

\item Secondly, as for the forward-limit diagram, we always identify two legs on different (adjacent) vertices, then the BCFW vertex again forms a triangle. 

\item Finally lower-point or lower-loop amplitudes plugged in the recursion relations were obtained in the way described above. 

\end{itemize}
We will show explicitly that higher-point/higher-loop amplitudes obtained from recursion relation in such way does enjoy the symmetry manifestly.  

\subsection{Eight-point one-loop amplitude}
Before moving to the general one-loop amplitudes, let us take eight-point as one more example. It gets contributions from factorization diagrams with tree-level/one-loop four-point amplitude multiplied with one-loop/tree-level six-point amplitude, and the forward-limit diagram of ten-point tree-level amplitude. By solving the recursion in the canonic way we described in previous section, the result of eight-point one-loop amplitude can be represented in a form with manifest $i \rightarrow i+2$ symmetry, 
\eq \label{8ptloop}
\includegraphics[scale=0.45]{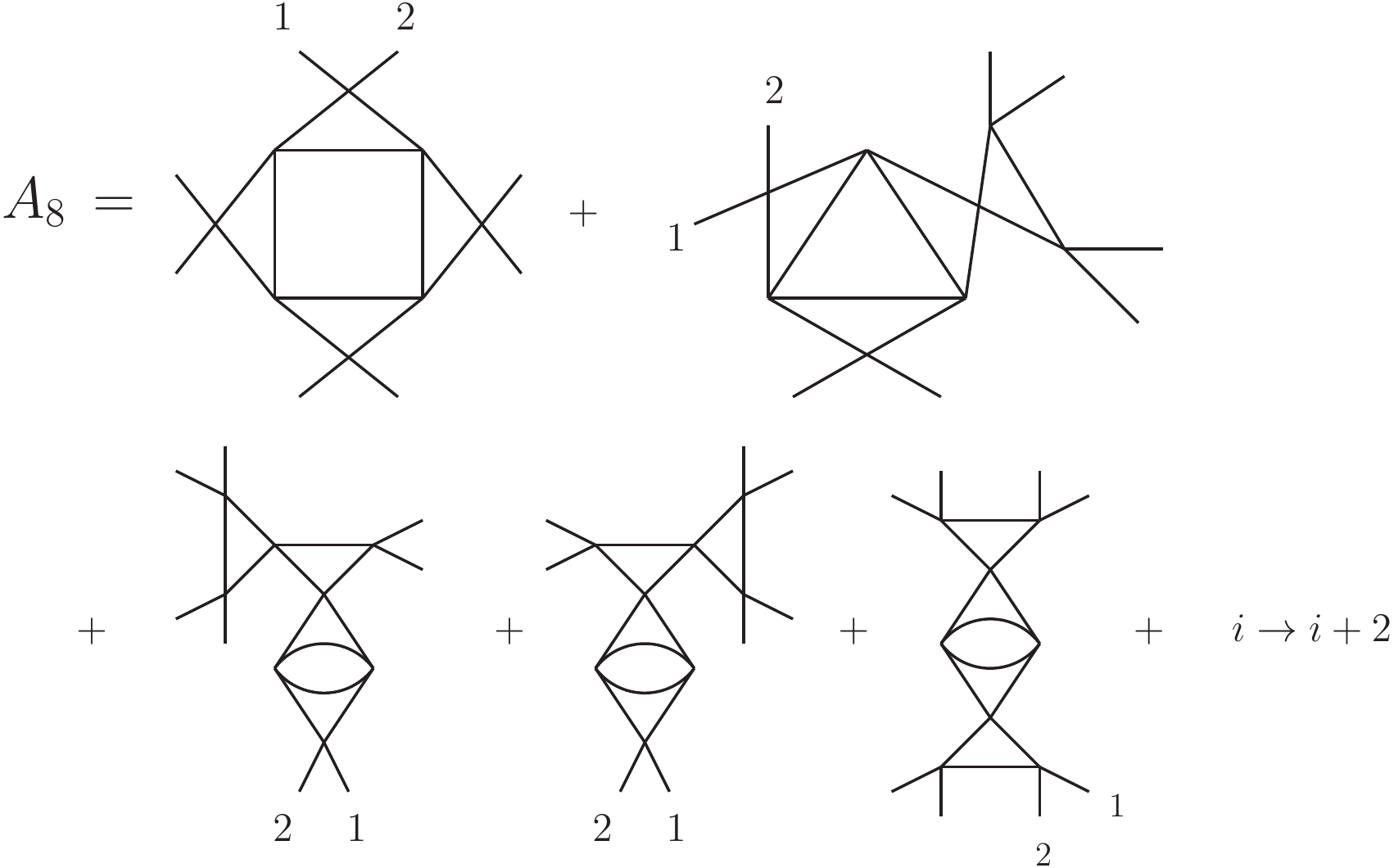} \, ,
\eqe
where $i \rightarrow i+2$ means summing over all other diagrams related by two-site cyclic symmetry. It is easy to see that forward-limit singularity corresponding to opening vertex with legs $8$ and $1$ (one-site shifted from the original BCFW bridge, vertex with legs $1$ and $2$) comes from the diagrams in the second line of eq.~(\ref{8ptloop}), again by opening up internal vertices, 
\eq
\includegraphics[scale=0.7]{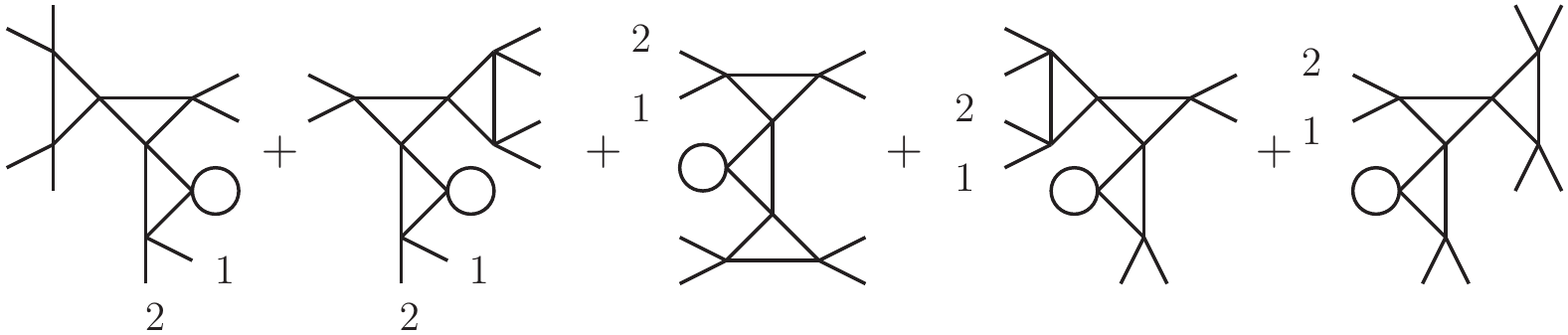} \, .
\eqe
Although we argued generally spurious singularities should cancel in pairs, it is still instructive to see this more explicitly. Actually the cancellation appears in a nice pattern: 
\begin{itemize}
\item Box diagram contains $4$ spurious singularities, where at each of the singularities the box reduces to a triangle. These singularities are cancelled out by the spurious singularities from the triangle diagrams--there are $4$ triangle diagrams after summing over all $i \rightarrow i+2$ cyclic diagrams. One example is shown as follows 
\eq
\includegraphics[scale=0.5]{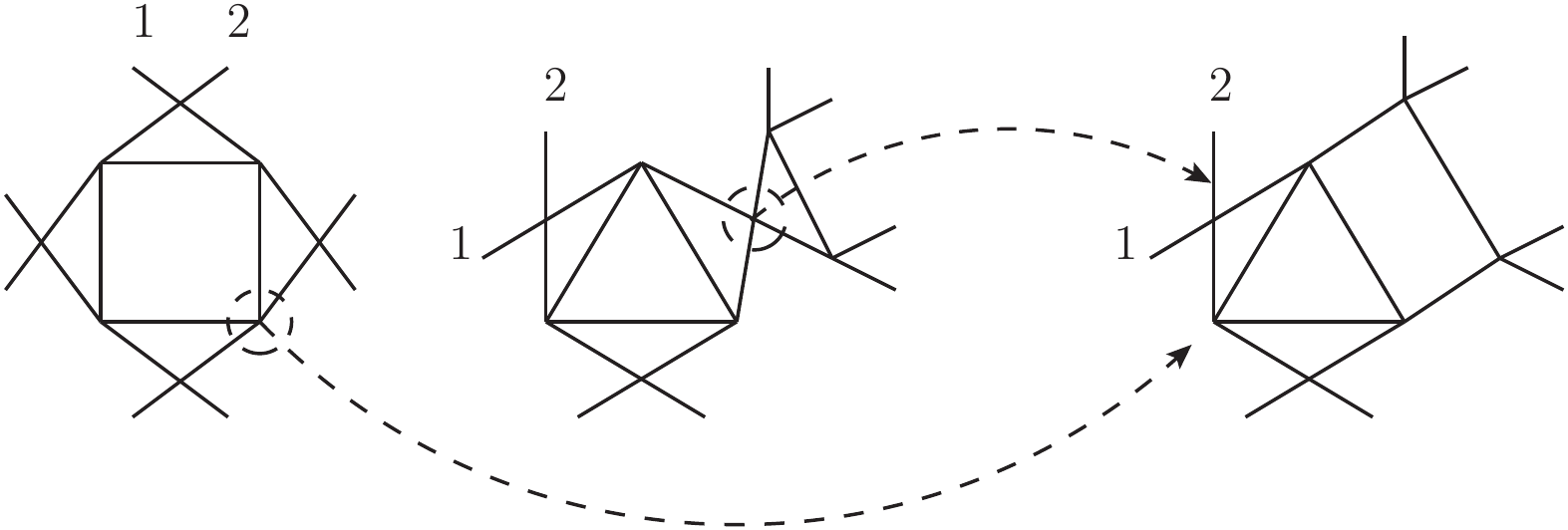} \, .
\eqe

\item Triangle diagrams contain another type spurious singularities, where at each of the singularities the triangle reduces to a bubble. Each triangle has $3$ such singularities (in total there are $4 \times 3=12$). They are all cancelled out by the spurious poles from the bubble diagrams, 
\eq
\includegraphics[scale=0.5]{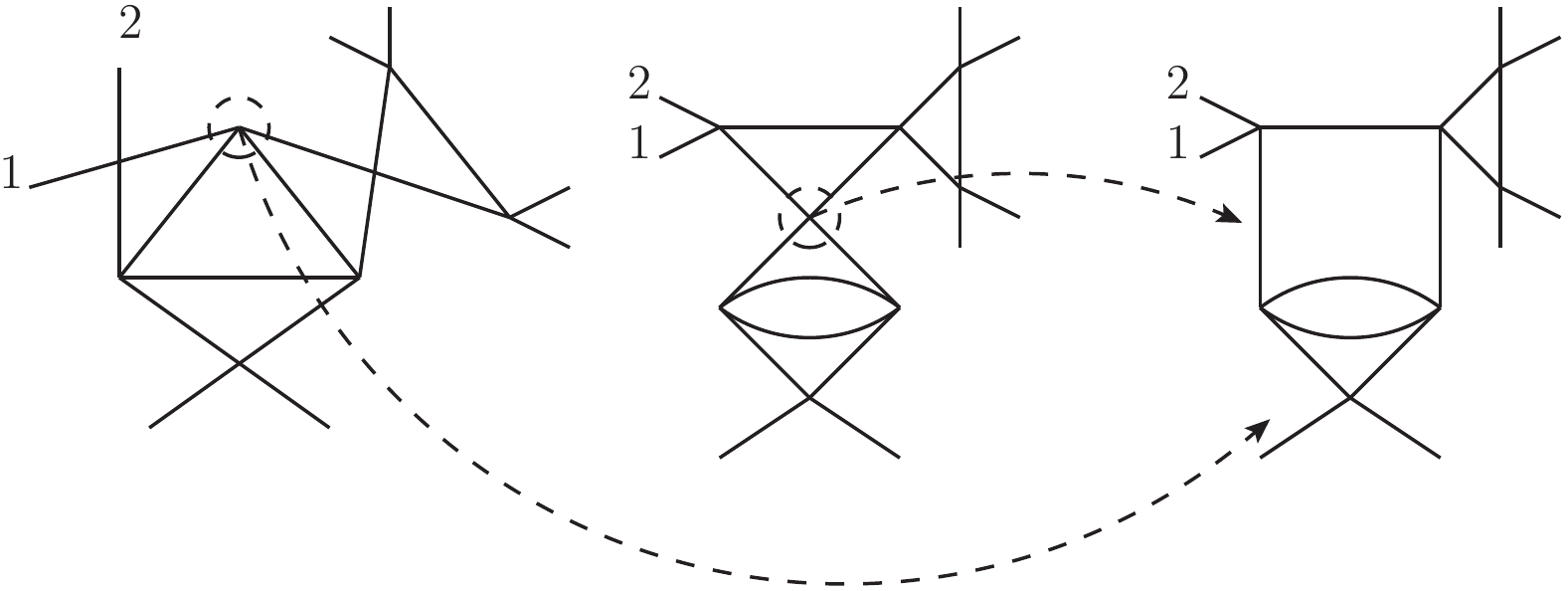} \, .
\eqe

\item Bubble diagrams have further spurious singularities, which are cancelled out between bubble diagrams themselves. For instance, 
\eq
\includegraphics[scale=0.5]{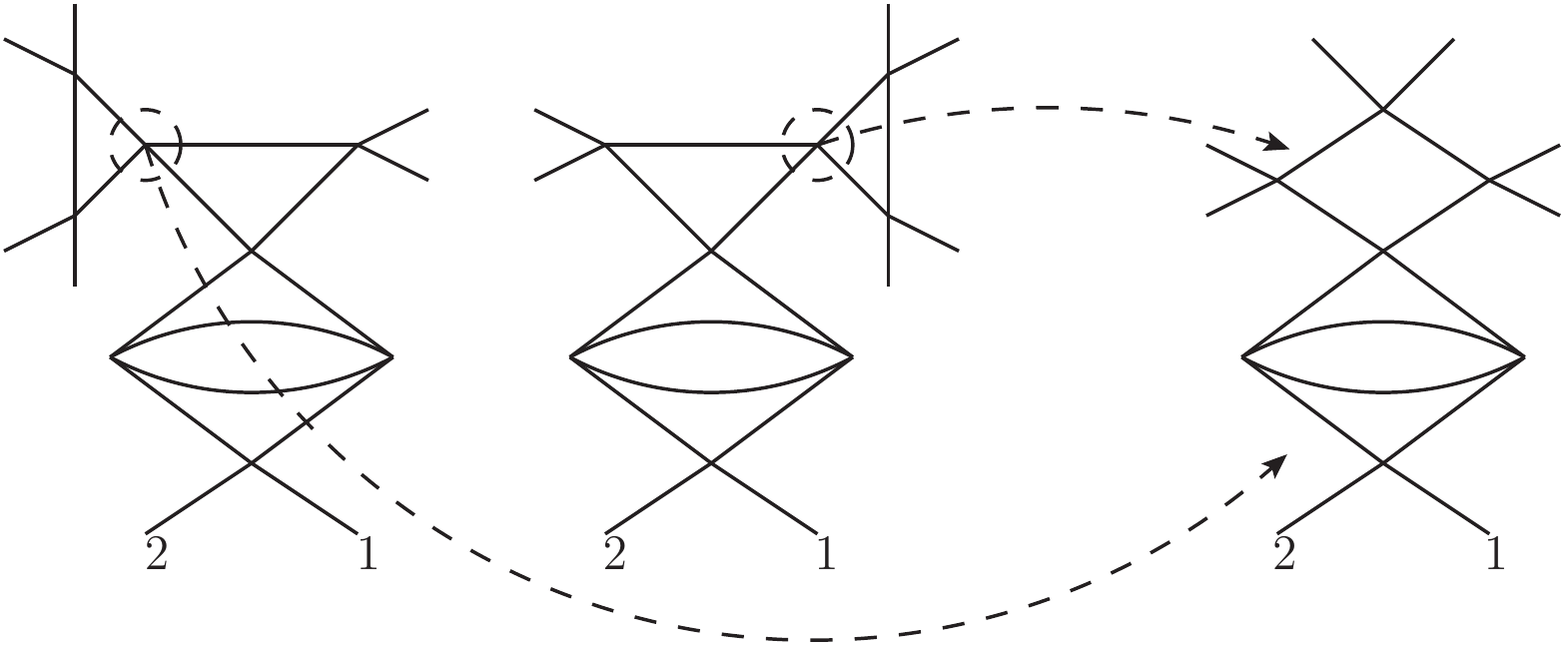} \, .
\eqe
which is essentially the spurious pole cancellation of eight-point amplitude at tree-level. 

\end{itemize}
Detailed analysis of six-point and eight-point examples makes the pattern of general one-loop amplitudes quite clear, as we will present shortly in following section. 

\subsection{General one-loop amplitudes}
As found in \cite{HW}, $(2p+4)$-point tree-level amplitude from recursion relation can be represented as a sum of $(2p)!/(p! (p+1)!)$ diagrams constructed all by triangles, which are connected with each other through points only. Similarly, the solution of one-loop recursion relation for $(2p+4)$-point amplitude contains $\binom{2p+2}{p}$ diagrams, which now are built up by $m$-gon's, with $m=2,\ldots, (p+2)$, and each edge of such a polygon is connected with a triangle through the edge. The rest of such a diagram are triangles connected through points only, just as the case of tree-level amplitudes. Finally we sum over all possible two-site cyclic permutations, which makes the result with manifest $i \rightarrow i+2$ cyclic symmetry. So, again, all the physical factorizations and half of forward-limit singularities are manifest because of this symmetry. Whereas the other half of forward-limit singularities can be seen by opening up the internal vertices of the diagrams with $2$-gon's, namely bubbles. 

As for the spurious singularities, they are cancelled out in order: spurious singularities from diagrams with $m$-gon cancel the spurious singularities from diagrams with $(m-1)$-gon; and diagrams with $2$-gon's contain further singularities, which are cancelled out between themselves. Clearly one can see all these structures from the six- and eight-point examples. 

It's not difficult to write down the results for higher-point one-loop amplitudes according to above description, here is another non-trivial example, ten-point one-loop amplitude, 
\eq \label{10pt1loop}
\includegraphics[scale=0.5]{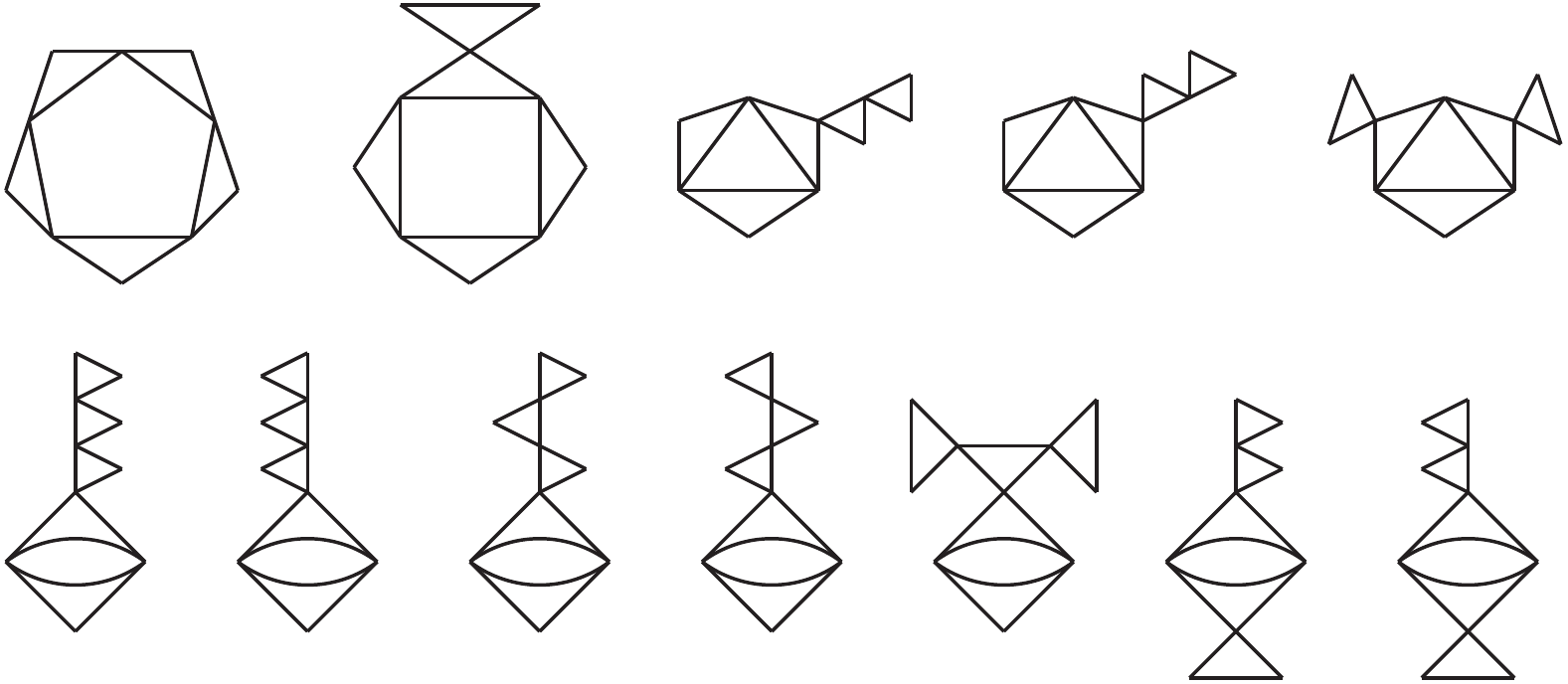} \, .
\eqe
As the case of tree-level amplitudes \cite{HW}, to simplify the notation we have omitted the external legs in the diagrams: the diagrams should be understood with two external legs attached to each external vertex, for instance the first diagram in eq.(\ref{10pt1loop}) is really
\eq 
\includegraphics[scale=0.2]{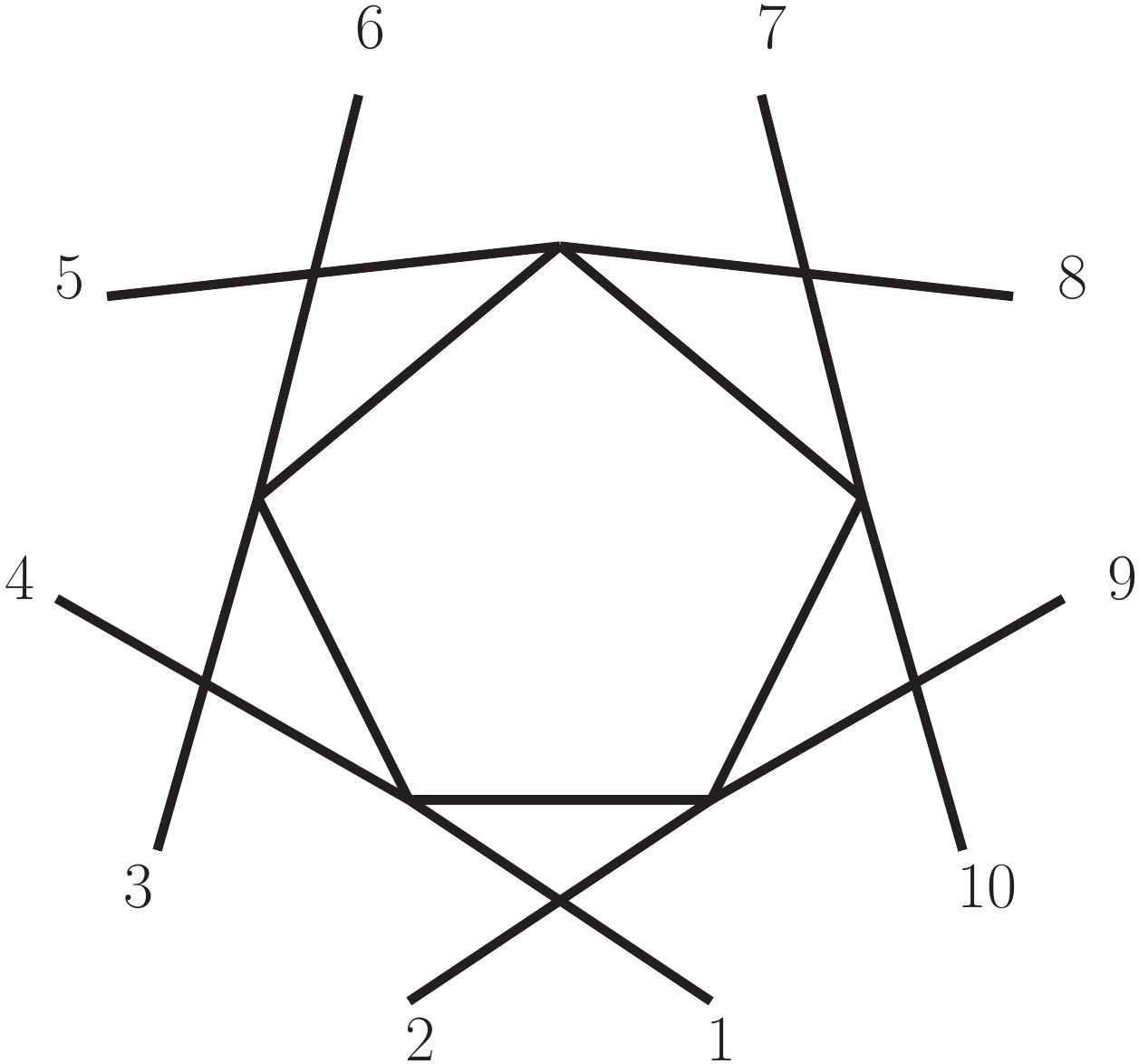} \, .
\eqe
Summing over all $i \rightarrow i+2$ cyclic rotations, we find $56$ different diagrams for ten-point amplitude, which is indeed the result of the recursion relation.

\subsection{Two-loop amplitudes}
We can proceed to solve the recursion relation for two-loop amplitudes. There are $4(p+1)\binom{2p+1}{p}$ diagrams for a $(2p+4)$-point amplitude at two loops. The simplest four-point amplitude at two loops can be obtained by attaching a BCFW bridge to forward-limit of six-point one-loop amplitude, eq.(\ref{6pt1Loop}). The result presented in the canonic way is given by
\eq
\label{Reduc2}
\includegraphics[scale=0.4]{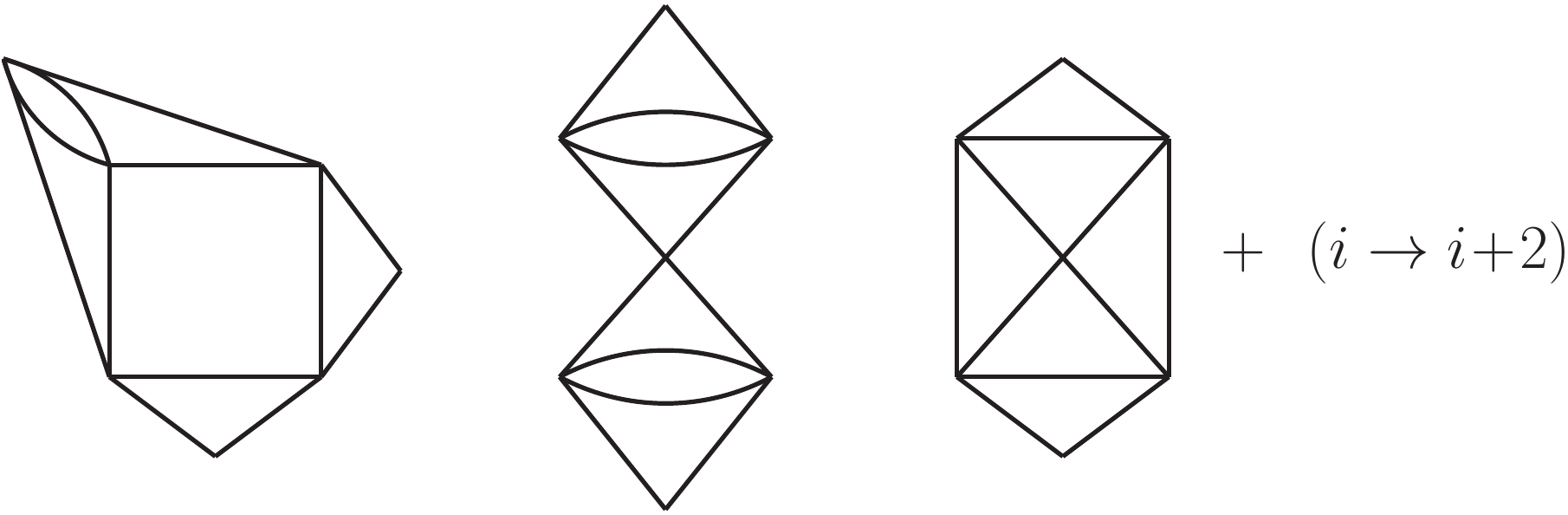} \, .
\eqe
we have also omitted the external legs to simplify the notation. In total there are four diagrams because the second and the last diagrams are symmetric under $i \rightarrow i+2$. Just as what we have found in the one-loop case, one half of physical forward-limit singularities can be trivially seen by opening up external vertices, while the other half is again hidden in the internal vertices. For instance the forward-limit singularity corresponding to opening up vertex with legs $2$ and $3$ can be shown as follows
\eq
\includegraphics[scale=0.6]{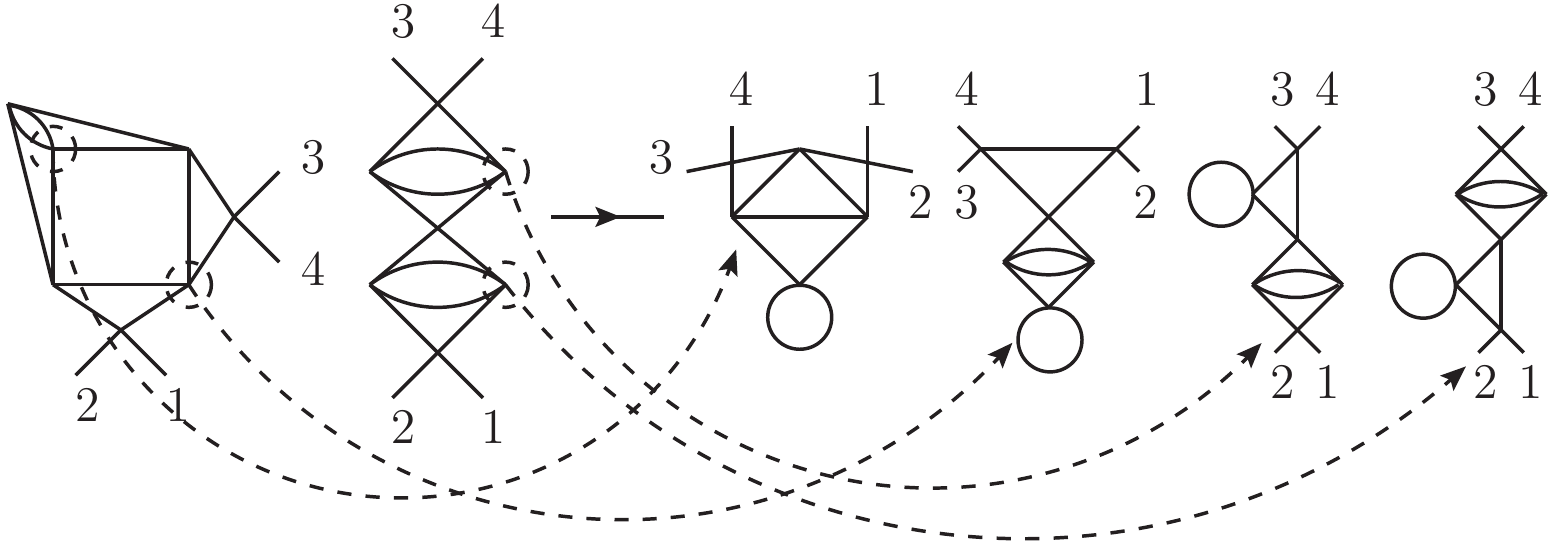} \, ,
\eqe
where a triangle move has been applied to obtain the second diagram of forward-limit singularity. Similarly the result of two-loop six-point amplitude can be presented in the canonic form with manifest two-site cyclic symmetry
\eq
\label{Reduc3}
\includegraphics[scale=0.6]{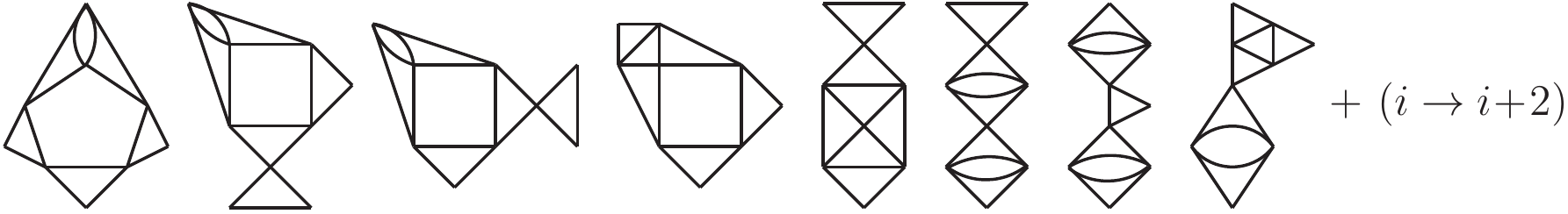} \, . 
\eqe
Again, besides those manifest physical poles, one half of forward-limit singularities can only be seen by opening up internal vertices, 
\eq \label{6pt2loopphy}
\includegraphics[scale=0.7]{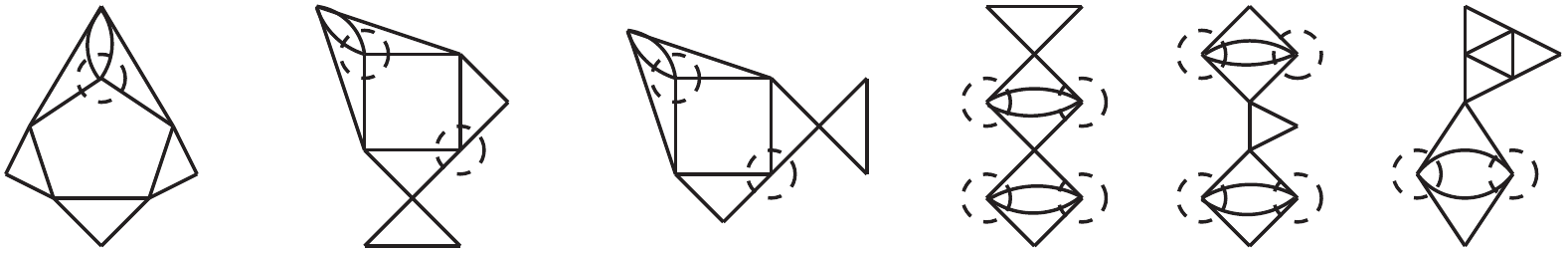} \, . 
\eqe
In the above diagram, we circle out the vertices whose boundary would lead to the physical singularities. It is straightforward to obtain all $15$ singularities in eq.(\ref{6pt2loopphy}), and the combination of these singularities is precisely the forward-limit of one-loop eight-point amplitude, eq.(\ref{8ptloop}). 

All those examples clearly show the advantage of the canonic representation, which not only has manifest of $i \rightarrow i+2$ cyclic symmetry, but also makes the hidden forward-limit singularities transparent: they all come from the diagrams with bubbles. 

\subsection{General structures of loop amplitudes in ABJM}
A remarkable property of the four-point, and six-point amplitudes displayed in eq.(\ref{Reduc1}), eq.(\ref{6pt1Loop}), eq.(\ref{Reduc2}), and eq.(\ref{Reduc3}), is that through equivalence moves, they can all be reduced to tree-level on-shell diagrams. For example, consider one of the four-point two-loop diagram given in eq.(\ref{Reduc2}):
\eq
\includegraphics[scale=0.7]{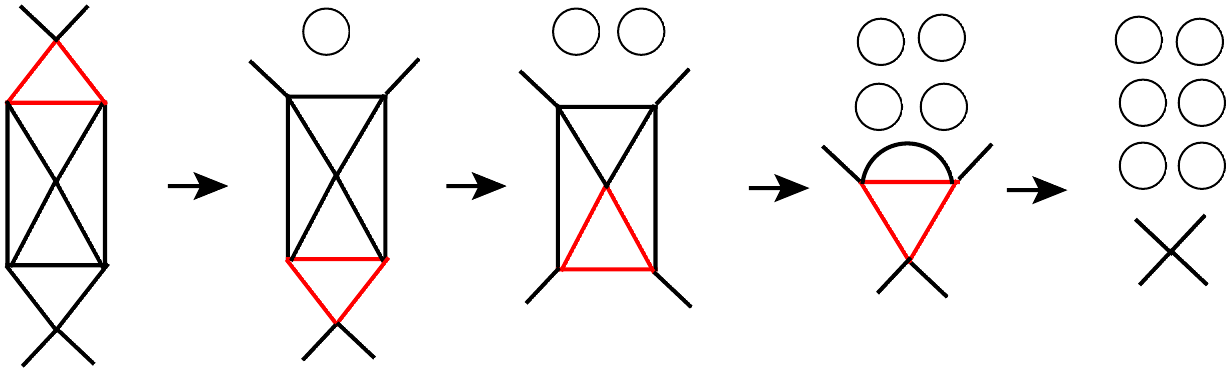}\,, 
\eqe
where we've demonstrated the reduction procedures by outlining, in red, the triangle subgraphs that undergoes the equivalence move. All four-point bubbles are automatically reduced, appearing as vacuum bubbles. The fact that all four-and six-point diagrams can be reduced to the tree-diagrams can be understood easily from the dimension of the top-cell in OG$_k$: $k(k-1)/2$. Since the number of kinematics constraint for any on-shell diagram is given by $2k-3$, for $k=2,3$ the dimension of the top-cell is exactly the same as the number of kinematic constraint. Thus all higher dimensional on-shell diagrams must be equivalent to the the top-cell, which is simply tree diagrams, with the extra degrees of freedom decoupled from the Grassmaniann. Thus we see for $k=2,3$ the integrands only has logarithmic singularities, just as the MHV amplitudes in $\mathcal{N}=4$ SYM.

We have seen that the on-shell diagram solution automatically gives an answer where uniform transcendentally is manifest. Indeed the known answer at of one-loop and two-loop amplitudes satisfy this result~\cite{Beisert1Loop, OneLoopABJM, 2LoopABJM, 2LoopABJM2}, as well as the all order conjecture for four-point amplitudes~\cite{BDSABJM}. Our result implies that $L$-loop four- and six-point amplitudes are given by uniform transcendental $L$ functions, where there is one two-particle cut discontinuity at each loop order.  

Finally, since we have verified that the solutions to the loop recursions indeed reproduce all physical singularities, this implies all reduced medial graphs correspond to the leading singularity of ABJM loop amplitudes. Since each medial graph is one-to-one correspondence to the cell in the $OG_{k+}$, this proves that the residues of the orthogonal Grassmannian integral~\cite{LeeOG}, which are evaluated on the boundaries of the top-cell  indeed correspond to the leading singularities.

\section{Acknowledgements}
It is a pleasure to thank Nima Arkani-Hamed, Andreas Brandhuber, Song He, Gabriele Travaglini and Jaroslav Trnka for discussions. We would especially like to thank Thomas Lam, for pointing out the relevance of the tetrahedron equation, Sangmin Lee and Joonho Kim for the many useful discussions as well as the sharing of their draft. Y.H. would like to thank Seoul National University for its hospitality, during which part of this work was completed. Y-t. H. and D. X. are supported by the Department
of Energy under contract DE-SC0009988. The work of C.W is supported by the Science and Technology Facilities Council Consolidated Grant ST/J000469/1 {\it String theory, gauge theory \& duality. }

\bibliographystyle{JHEP}

\end{document}